\documentclass[aps,prd,twocolumn,superscriptaddress,nofootinbib,nolongbibliography]{revtex4-2}
\usepackage{eurosym}
\usepackage{epsfig}
\usepackage{graphicx}
\usepackage{dcolumn}
\usepackage{bm}
\usepackage{longtable}
\usepackage{amssymb}
\usepackage{amsfonts}
\usepackage{amsmath}
\usepackage{hyperref}
\usepackage{slashed}
\usepackage{xcolor}

\newcommand{\be}{\begin{equation}}
\newcommand{\ee}{\end{equation}}
\newcommand{\ba}{\begin{align}}
\newcommand{\ea}{\end{align}}
\newcommand{\one}{{\rm 1\kern -.9mm l}}

\begin{document}

\title{Two Times for Freudenthal}

\author{Alexander~Kamenshchik}
\email{kamenshchik@bo.infn.it}
\affiliation{Dipartimento di Fisica e Astronomia, Universit\'{a} di Bologna, via Irnerio 46, 40126 Bologna, Italy}
\affiliation{INFN, Sezione di Bologna, viale Berti Pichat 6/2, 40127 Bologna, Italy}

\author{Alessio~Marrani}
\email{a.marrani@herts.ac.uk}
\affiliation{Center for Mathematics and Theoretical Physics, University of Hertfordshire, AL10 9AB Hatﬁeld, UK}

\author{Federica~Muscolino}
\email{federica.muscolino@gmail.com}
\affiliation{Dipartimento di Matematica e Applicazioni, Universit\'a di Milano Bicocca, via
Roberto Cozzi 55, 20125 Milano, Italy}
\affiliation{Gruppo Nazionale di Fisica Matematica, InDAM, Piazzale Aldo Moro 5, 00185 Roma, Italy}

\begin{abstract}
We investigate the algebraic structure of the two-time physics introduced
some time ago by I. Bars and his co-authors, clarifying its relations with
quadratic and cubic Jordan algebras, as well as with reduced Freudenthal triple
systems (FTS) based on them. In particular, the `extended' phase space
introduced by Bars can be endowed with the structure of a reduced FTS constructed over a
semi-simple cubic Jordan algebra (named Lorentzian spin factor), characterized by a primitive, invariant symmetric tensor of rank $4$. The $Sp(2,%
\mathbb{R})$-gauge fixing procedure typical of two-time physics yields
algebraic-differential constraints on the quartic polynomial associated to such a tensor, implying that
only two (isomorphic) nilpotent orbits of the non-transitive
action of the automorphism group of the Lorentzian spin factor are spanned by
the conjugated variables which coordinatize the `extended' phase space. We illustrate our results in relativistic, manifestly Lorentz-covariant physical systems, as well as in
non-relativistic systems (such as the non-relativistic massive particle, the
hydrogen atom, and the Carroll particle with non-vanishing energy).
\end{abstract}

\maketitle

\section{\protect\bigskip Introduction}

Two-time (2T) physics arose out during the '90s in the works of Itzhak Bars, and in recent years it has found applications into a large variety of physical fields and models, including classical particle theory, field theory, supersymmetric models, strings, branes and so on (see e.g. \cite{Bars-Terning} for a comprehensive review). Within this paradigm, different one-time (1T) physical systems can be ascertained to descend from a unique physical system in the '2T world', in which the physical world's dimensions are augmented by two additional (one timelike and one spacelike) dimensions.

More specifically, one of the foundational features of 2T physics is the gauging of the symplectic symmetry $Sp(2,\mathbb{R})$ of the (suitably enlarged) phase space (later referred to as \textit{enlarged} phase space), which mixes positions and momenta. The corresponding gauge fixing procedure gives rise to three first-class constraints, which, by means of three appropriate gauge choices solving the constraints, eventually lead to a dimensional reduction from the 2T world to the 1T, physical world.
Different gauge choices yield different 1T physical systems, which are only apparently unrelated; in fact, a remarkable outcome of the 2T paradigm is the discovery of new 'dualities' among physical systems, yet to be fully explored and understood, which are implemented in terms of (gauged) $Sp(2,\mathbb{R})$ transformations on the extended phase space coordinates. A number of examples within classical particle theory has been discussed in \cite{Bars-Gauged,Bars:DualityH,Bars-Emergent,KM}, whereas in \cite{Bars-Araya} the duality transformations between various classical systems have been investigated in detail. Within this framework, physical systems in curved spacetimes have been studied in \cite{Bars:Gravity,Bars:GravityII,Bars10}, and other applications have been considered in \cite{Bars:QFT,Bars:SM} for field theories, as well as in \cite{Bars:Super} and in \cite{Bars:Strings} for supersymmetric models and string theory, respectively.

In this work, we propose a systematic and algebraic method for the classification of the various possible gauge fixing procedures which can be implemented within 2T physics. We will point out that the extended phase space of the 2T physics can naturally be endowed with the structure of a reduced \textit{Freudenthal triple system} (FTS) $\mathcal{F}(\mathcal{J}_3)$, based on the rank-3 semisimple \textit{Jordan algebra} $\mathcal{J}_3\equiv\mathbb{R}\oplus\Gamma_{1,d-1}$, usually referred to as a (cubic) \textit{Lorentzian spin factor} in $d$ spacetime dimensions. The \textit{reduced} nature of this FTS implies it to inherit essentially all of its structures from the cubic Jordan algebra it is built upon \cite{small_orbits,small_orbits_maths,McCrimmon,McCrimmon:1969}; this fact plays a key role in the dimensional reduction, in which the 1T, physical world's phase space stems from the extended phase space of 2T physics. Relativistic systems provide a framework in which all this can be realized and appreciated in a clearer way, since the Lorentz algebra naturally emerges as the reduced structure algebra associated to the rank-2 Jordan algebra $\Gamma_{1,d-1}$, defined by the Minkowski space in $d$ dimensions (and the related quadratic norm).

As treated in Section \eqref{Sec:Freudenthal}, the Lie algebra of the global isometry group $Sp(2,\mathbb{R})\otimes SO(2,d)$ of the extended phase space is isomorphic to the Lie algebra of the automorphisms (i.e., to the derivation algebra) of the FTS $Aut(\mathcal{F}(\mathbb{R}\oplus\Gamma_{1,d-1}))$ (see, e.g., \cite{small_orbits_maths}). The non-transitive action of  $Sp(2,\mathbb{R})\otimes SO(2,d)$ onto $\mathcal{F}(\mathcal{J}_3)$ is characterized by a \textit{unique} primitive invariant homogeneous polynomial of degree four, $\mathcal{I}_4$, and it characterizes the whole FTS as a prehomogeneous vector space with a unique polynomial invariant function (cf. e.g. Prop. 19 (2) of \cite{Sato_Kimura_1977}), as well as an example of (irreducible) Vinberg theta group \cite{vinberg1976weyl} of type II (cf. e.g. Table III of \cite{KAC1980190}), whose ring of invariants is finitely generated by $\mathcal{I}_4$ only. Furthermore, the whole FTS vector space gets stratified into orbits, which are proper (generally homogeneous non-symmetric) submanifolds given by \textit{loci} defined by algebraic-differential $Sp(2,\mathbb{R})\otimes SO(2,d)$-invariant constraints on $\mathcal{I}_4$ itself. We will show that the constraints of the $Sp(2,\mathbb{R})$ gauge-fixing characterizing 2T physics enforce the coordinates of the phase space of all 1T physical systems to belong to a unique nilpotent orbit, with $\mathcal{I}_4=0$; as studied some time ago in \cite{small_orbits}, such a unique orbit further splits into two orbits, discriminated by the sign of a quadratic polynomial, $\mathcal{I}_2$. In Section \ref{Sec:Examples} we will revisit the embedding of a number of (relativistic or non-relativistic, massless or massive) 1T physical models into 2T physics, in the light of the algebraic machinery mentioned above. Among other things, we will investigate how the $Sp(2,\mathbb{R})$ gauge fixing and the subsequent resolution of further constraints reduce the symmetry group from $Sp(2,\mathbb{R})\otimes SO(2,d)$ to a certain, proper subgroup, which characterizes the structure and dynamics of the resulting 1T physical model. As we will see, the non-linear realization of some generators of $Sp(2,\mathbb{R})\otimes SO(2,d)$ (as isometry group of the extended phase space of the 2T physics) plays a crucial role in the above symmetry reduction, and can be related to the (reductive) symmetries of the aforementioned polynomial $\mathcal{I}_2$.
The vast landscape of gauge fixings and related 1T physical models which can be potentially discussed within 2T physics provides a strong motivation for constructing a systematic classification of all possible ways to gauge-fix (and non-linearly realize) some generators of the original (isometric) symmetry. The present work aims at this scope, which would also be instrumental in clarifying the intricate web of 'dualities' emerging among the various models of 1T physics. 


The plan of the paper is as follows.

Section \ref{Sec:2T} summarizes the main features of 2T physics, discussing how different 1T physical models can be obtained within the same \textit{extended} phase space. Then, in Section \ref{Sec:Freudenthal}, 2T physics is placed into the broad framework of \textit{Freudenthal triple systems}, whose algebraic structures can be exploited to characterize in an invariant (or covariant) manner each of the 1T physical systems. Next, in Section \ref{Sec:Examples}, a number of models already discussed by Bars in 2T physics is reconsidered by using the FTS's. A general discussion of the results (and a related conjecture) is given in Subsection \ref{sec:Results}. Section \ref{Sec:Conclusions} contains Conclusions and Outlooks.

Various appendices conclude the paper.

App. \ref{App-JA-etc} summarizes the main properties of Jordan algebras, focussing on the rank-3 case, and of (reduced) FTS constructed over them. Then, App. \ref{App-Frob} discusses how the Frobenius norm endows a (cubic) Lorentzian spin factor $\mathfrak{J}_{1,d-1}$ with a Lorentzian $(d+1)$-dimensional flat metric, which is the linearization of a quadratic norm with signature $(n_+,n_-)=(d,1)$. Finally, App. \ref{App:I2} contains various results omitted from the main text for the sake of clarity, namely the explicit forms of the infinitesimal transformations of the physical phase space, as well as the analysis of the symmetries of the polynomial $\mathcal{I}_2$.
 
\section{The 2T, `extended' phase space} \label{Sec:2T}
	The aim of the 2T physics is to describe the properties of different physical systems from a general point of view based on the consideration of the extended phase space with an additional timelike and an additional spacelike components. This phase space can be parameterized using the coordinates
	\begin{gather}\label{2TCoord}
		\begin{split}
		    X^M&=\left(X^{0'},X^{1'},X^\mu\right),\\ P^M&=\left(P^{0'},P^{1'},P^\mu\right),
		\end{split}
	\end{gather}
	indicating the $d+2$-dimensional position and momentum vectors. The indices $0'$ and $1'$ label an extra time and an extra space dimensions and the further index $\mu=0,\dots ,d-1$ labels usual spacetime coordinates. We can \textit{pack} these coordinates together by introducing the notation
	\begin{equation}
		X^M_i=\left(X^M,P^M\right),
	\end{equation}
	where $i=1,2$ labels the position and momentum respectively. The quantity $X^M_i$ transforms linearly as the $(\mathbf{2},\mathbf{2+d})$ of the group $G=Sp(2,\mathbb{R})\otimes SO(2,d)$. The simplest $G$-invariant worldline action for a particle moving in the 2T phase space is
	\begin{equation}\label{WorldAction}
		S=\frac{1}{2}\int d\tau\ \epsilon^{ij}g_{MN}\ \partial_\tau X_i^M X_j^N,
	\end{equation}
	where $g_{MN}$ represents the metric tensor, with signature $(2,d)$, $\epsilon^{ij}$ is the antisymmetric tensor with $\epsilon^{12}=1$ and $\tau$ is a proper time parameter. In this paper, we consider the simplest case with a flat metric, i.e. $g_{MN=}{\eta_{MN}=\text{Diag(-1,1,-1,1,\dots,1)}}$, but a more general metric can be considered \cite{Bars10}. The infinitesimal transformations under $G$ are given by
	\begin{equation}
		\delta_\omega X_i^M = \epsilon_{ij}\omega^{jk}X_k^M,
	\end{equation}
	where $\omega^{jk}$ are the infinitesimal parameters, symmetric in $j,k$.
	
	When the $Sp(2,\mathbb{R})$ symmetry is promoted to a local symmetry (in particular, when $\omega^{ij}\rightarrow\omega^{ij}(\tau)$), the derivative with respect to $\tau$ should be substituted with the covariant derivative
	\begin{equation}
		\partial_\tau X^M_i\rightarrow D_\tau X^M_i=\partial_\tau X^M_i-\epsilon_{ij} A^{jk}(\tau)X^M_k,
	\end{equation}
	where the gauge field $A^{jk}(\tau)$ is symmetric in the indices $i,j$ and belongs to the adjoint representation of the Lie algebra of $Sp(2,\mathbb{R})$ (that we call $\mathfrak{sp}(2,\mathbb{R})$). It transforms as a gauge field under the $Sp(2,\mathbb{R})$ group
	\begin{equation}
		\delta_\omega A^{ij}(\tau)=\partial_\tau\omega^{ij}+\omega^{ik}\epsilon_{kl}A^{lj}+\omega^{jk}\epsilon_{kl}A^{li}.
	\end{equation}
	The worldline action invariant under these gauge transformations is
	\begin{align}\label{WorldGaugedAction}
		S&=\frac{1}{2}\int d\tau\ \epsilon^{ij}\eta_{MN}D_\tau X_i^M X_j^N\cr
        &=\int d\tau\ \left[\eta^{MN} \partial_\tau X_MP_N-A^{ij}(\tau)Q_{ij}\right],
	\end{align}
	where
	\begin{gather}
		Q_{11}=\frac{1}{2}X\cdot X,\quad Q_{22}=\frac{1}{2}P\cdot P,\cr
        Q_{12}=Q_{21}=\frac{1}{2}X\cdot P
	\end{gather}
	are the $\mathfrak{sp}(2,\mathbb{R})$ conserved charges, in fact sitting in the adjoint representation $\mathbf{3}$ of $\mathfrak{sp%
}(2,\mathbb{R})$ itself, realized as a symmetric rank-$2$ tensor
representation on the fundamental (doublet) irrepr. $\mathbf{2}$ (spanned by
the indices $i,j=1,2$); the corresponding $\mathfrak{sp}(2,\mathbb{R})$
Poisson brackets follow from the Heisenberg algebra $\mathfrak{h}_{2d+5}$
(with trivial central extension given by the flat metric $\eta_{MN}$) obeyed by the
positions and momenta coordinatizing the extended phase space itself.
	
	The gauge fields $A^{ij}$ are not dynamical, since the kinetic terms are absent.  Thus, in the action \eqref{WorldGaugedAction} they play the role of Lagrange multipliers, introducing three first order constraints $Q_{ij}=0$. It is worth noticing that these constraints lead to a non-trivial parameterization of the one-time (1T) spacetime only when the starting theory has more than one timelike dimension (see, for instance, \cite{Bars-Gauged,Bars:DualityH,Bars-Terning,Bars-Araya,Gun-Bars} and references therein).
	Moreover, the gauge freedom allows us to choose three physical degrees of freedom. Then, when the gauge is fixed and the constraints $Q_{ij}=0$ are satisfied, one gets the right number of 1T variables. That means that $X^M_i(\tau)=X^M_i(\vec{x}(\tau),\vec{p}(\tau))$, where $\vec{x}$ and $\vec{p}$ are $(d-1)$-dimensional vectors. The action now looks like
	\begin{equation}
		S=\int d\tau \left(\dot{\vec{x}}\cdot\vec{p}-H\right),
	\end{equation}
	where $H$ is the Hamiltonian of the 1T theory. It is here worth remarking that the extended phase space is symplectic,
whereas the surface defined by the three first class constraint $Q_{ij}=0$
is \textit{pre-symplectic}, giving rise to a symplectic 1T, physical phase space
after further gauge fixing/constraints' resolution.
	
	The gauge fixing in the context of the 2T physics can be interpreted in a very intuitive way as follows. Different gauge fixings correspond to different choices of the Hamiltonian (and different choices of the time). This means that different systems in the 1T physics are described by a unique 2T model. In this sense, these systems are \textit{dual} to each other under local $Sp(2,\mathbb{R})$ transformations.
	
	In many cases, it is useful to fix the gauge \textit{partially}. For instance, one can make two gauge choices and solve two of the constraints $Q_{ij}=0$, in such a way that the remaining quantities are written in terms of an \textit{extended} phase space variables $(t,\vec{x},H,\vec{p})$ (or $(x^\mu,p_\mu)$ for the relativistic case). The last gauge fixing will set the physical time $t$ in terms of $\tau$ (this corresponds to a \textit{choice of the time}) and will define the Hamiltonian $H$. For example, we may fix $X\cdot X=X\cdot P=0$. At this stage, the action takes the form
	\begin{equation}\label{PartialFix}
		S=\int d\tau \left(\dot{\vec{x}}\cdot\vec{p}-\dot{t}H-\frac{A^{22}}{2}P\cdot P\right).
	\end{equation}
	Then, after the choice of two degrees of freedom, corresponding to the gauge fixing, the last constraint $P\cdot P=0$ generally characterizes the theory we want to describe. For instance, if we are describing a massless relativistic particle, the gauge can be chosen in such a way that $P\cdot P = p^2$, where $p$ is the 4-momentum. Conversely, if we want to describe a \textit{massive} relativistic particle, the gauge can be fixed in order to obtain $P\cdot P=p^2-m^2$ or, for a non-relativistic particle, $P\cdot P=\vec{p}^2-2mH$ and so on (see \cite{Bars-Gauged,Bars-Araya} for further information about different gauge choices).
	
	Let us take a closer look at the equations of motion and the symmetries of the system.
	
	The equations of motions for the action \eqref{WorldGaugedAction} are
	\begin{align}\label{EqI}
		\dot{X}^M&=A^{12}X^M+A^{22}P^M,\\\label{EqII}
		\dot{P}^M&=-A^{12}P^M-A^{22}X^M.
	\end{align}
	Thus, now we see how the choice of the gauge fields $A^{ij}$ affects the equations of motion. The equations and the action \eqref{WorldGaugedAction} are invariant under the gauge group $Sp(2,\mathbb{R})$ and the global transformations $SO(2,d)$, where $d$ is the  dimension of  our 1T spacetime. This group plays an  important role in both classical and quantum theory. The generators of the group $SO(2,d)$ are
	\begin{equation}\label{Lgenerator}
		L^{MN}=X^MP^N-X^NP^M,
	\end{equation}
	and are invariant under $Sp(2,{\mathbb R})$ transformations. When the gauge is fixed partially, these generators are written in terms of the 1T variables and a subset of them will provide the generators of the symmetries of the 1T subsystem.

\section{\label{Sec:Freudenthal}The `extended' phase space as the Freudenthal triple system $%
\mathfrak{F}(\mathfrak{J}_{1,d-1})$}

The Lie groups involved in the 2T physics \textit{\`{a} la Bars }%
\cite{Bars-Gauged,Bars-Emergent}\textit{\ }are\footnote{%
We will henceforth use the physicists' notation of symplectic groups:
namely, $Sp(2,\mathbb{R})$ is the split real form of the Lie group whose
algebra is (in the usual Cartan's notation) $\mathfrak{c}_{1}$, when
considered over the complex numbers.}:%
\begin{align}
G_{\text{Conf}} &\equiv Conf\left( \mathbb{R}\oplus \Gamma _{1,d-1}\right)\simeq Aut\left( \mathfrak{F}\left( \mathbb{R}\oplus \Gamma _{1,d-1}\right)
\right) \cr
&=Sp(2,\mathbb{R})\times SO(2,d);  \label{Gconf} \\
G_{\text{Lor}} &\equiv Str_{0}\left( \mathbb{R}\oplus \Gamma
_{1,d-1}\right)\cr
&=SO(1,1)_{II}\times SO(1,d-1),  \label{GLor}
\end{align}%
with%
\begin{equation}
G_{\text{Conf}}\supset SO(1,1)_{I}\times G_{\text{Lor}},  \label{decomp}
\end{equation}%
where the subscripts $I$ and $II$ discriminate between the two Abelian
non-compact factors. Here, the labels $Conf$, $Aut$ and $Str_0$ state for the conformal symmetry, the automorphisms and the reduced structure group respectively (see App. \ref{App-JA-etc}). As discussed in App. \ref{App-JA-etc}, $\left( \mathbb{R%
}\oplus \Gamma _{1,d-1}\right) \equiv \mathfrak{J}_{1,d-1}$ is a semisimple
Jordan algebra of rank 3 (i.e., endowed with a cubic norm), whereas $%
\mathfrak{F}\left( \mathbb{R}\oplus \Gamma _{1,d-1}\right) \equiv \mathfrak{F%
}(\mathfrak{J}_{1,d-1})$ is the primitive, reduced Freudenthal triple system
constructed on it, and the `$Conf$', `$Aut$' and `$Str$' symmetry groups are
related to such algebraic structures (for further detail and a list of
references, see below as well as App. \ref{App-JA-etc}). We will henceforth
make use of various types of indices, ranging as follows :
\begin{align}
    \begin{split}
        i &=1,2;\\
        M &=0^{\prime },0,1^{\prime },1,...,d-1;\\
        \mu &=0,I=0,1,...,d-1,\\
        &\ ~\text{such that~}\overrightarrow{X}\equiv \left\{
X^{I}\right\} =\left( X^{1},...,X^{d-1}\right) .
    \end{split}
\end{align}

The `extended' phase space, of (real) dimension $2\left( d+2\right) $) can
be identified with $\mathfrak{F}(\mathfrak{J}_{1,d-1})$. According to the
(maximal, symmetric) group embedding (\ref{decomp}), the smallest
non-trivial (\textit{aka} `fundamental' or `defining') representation of $G_{%
\text{Conf}}$, in which the coordinates of the `extended phase space' sit,
branches as follows\footnote{%
In Bars' papers (see e.g. \cite{Bars-Emergent}), the notation $%
X_{1}^{M}\equiv X^{M}$ and $X_{2}^{M}\equiv P^{M}$ is used.}%
\begin{widetext}
\begin{eqnarray}
\overset{X_{i}^{M}}{\underset{\mathfrak{F}(\mathfrak{J}_{1,d-1})}{(\mathbf{2}%
,\mathbf{2+d})}} &=&\underset{(2,d)}{\underbrace{\overset{X_{1}^{0^{\prime }}%
}{\mathbf{1}}_{-3,0}\oplus \underset{\text{Frob:}(1,d)}{\underbrace{\overset{%
X_{1}^{\mu }=X_{1}^{0},\overrightarrow{X}_{1}}{\mathbf{d}_{-1,-1}}\oplus
\overset{X_{1}^{1^{\prime }}}{\mathbf{1}}_{-1,2}}}}}\oplus \underset{(2,d)}{%
\underbrace{\underset{\text{Frob:}(1,d)}{\underbrace{\overset{X_{2}^{\mu
}=X_{2}^{0},\overrightarrow{X}_{2}}{\mathbf{d}_{1,1}}\oplus \overset{%
X_{2}^{1^{\prime }}}{\mathbf{1}}_{1,-2}}}\oplus \overset{X_{2}^{0^{\prime }}}%
{\mathbf{1}}_{3,0}}}  \notag \\
&=&\underset{(2,d)}{\underbrace{\overset{x^{0^{\prime }}}{\mathbf{1}}%
_{-3,0}\oplus \underset{\text{Frob:}(1,d)}{\underbrace{\overset{x^{\mu
}=x^{0},\overrightarrow{x}}{\mathbf{d}_{-1,-1}}\oplus \overset{x^{1^{\prime
}}}{\mathbf{1}}_{-1,2}}}}}\oplus \underset{(2,d)}{\underbrace{\underset{%
\text{Frob:}(1,d)}{\underbrace{\overset{p^{\mu }=p^{0},\vec{p}}{\mathbf{d}%
_{1,1}}\oplus \overset{p^{1^{\prime }}}{\mathbf{1}}_{1,-2}}}\oplus \overset{%
p^{0^{\prime }}}{\mathbf{1}}_{3,0}}}  \notag \\
&=&\underset{(2,d)}{\underbrace{\overset{x^{0^{\prime }}}{\mathbf{1}}%
_{-3,0}\oplus \underset{\text{Frob:}(1,d)}{\underbrace{\overset{x^{\mu }=t,%
\overrightarrow{x}}{\mathbf{d}_{-1,-1}}\oplus \overset{x^{1^{\prime }}}{%
\mathbf{1}}_{-1,2}}}}}\oplus \underset{(2,d)}{\underbrace{\underset{\text{%
Frob:}(1,d)}{\underbrace{\overset{p^{\mu }=E,\vec{p}}{\mathbf{d}_{1,1}}%
\oplus \overset{p^{1^{\prime }}}{\mathbf{1}}_{1,-2}}}\oplus \overset{%
p^{0^{\prime }}}{\mathbf{1}}_{3,0}}},  \label{br}
\end{eqnarray}%
\end{widetext}
and, at the algebraic level, it corresponds to the following decomposition
of the reduced Freudenthal triple system%
\begin{equation}
\mathfrak{F}\left( \mathbb{R}\oplus \Gamma _{1,d-1}\right) =\mathbb{R}\oplus
\left( \mathbb{R}\oplus \Gamma _{1,d-1}\right) \oplus \left( \mathbb{R}%
\oplus \Gamma _{1,d-1}\right) ^{(\prime )}\oplus \mathbb{R}^{(\prime )}.
\label{br-al}
\end{equation}

A number of observations and definitions concerning (\ref{br}) and (\ref%
{br-al}) are in order\footnote{%
Consistently with Bars' papers, we will henceforth use the `mostly plus'
signature convention for Lorentzian metrics, as well as for
pseudo-Riemannian metrics, in general.} (for further detail, see App. \ref%
{App-JA-etc}).

\begin{itemize}
\item $\mathbf{\Gamma }_{1,d-1}$ is the simplest algebraic model of a $d$%
-dimensional Minkowski space, namely a vector space $V$ over $\mathbb{R}$
with a non-degenerate quadratic form $Q(v)$, $v\in V$, of Lorentzian
signature $\left( 1,d-1\right) $ (containing a base point $c_{0}$ such that $%
Q(c_{0})=1$). Such a vector space actually is a (simple) rank-2 Jordan
algebra \cite{1,2,3,4,5}. An element of $\Gamma _{1,d-1}$ is, for instance,
the Lorentzian position $d$-dimensional vector\ (1-form)%
\begin{equation}
X_{1}^{\mu }\equiv x^{\mu }=\left( X_{1}^{0},\overrightarrow{X}_{1}\right)
\equiv \left( x^{0},\overrightarrow{x}\right) \overset{c=1}{=}\left( t,%
\overrightarrow{x}\right) ,
\end{equation}%
or the Lorentzian momentum $d$-dimensional vector (1-form)%
\begin{equation}
X_{2}^{\mu }\equiv p^{\mu }=\left( X_{2}^{0},\overrightarrow{X}_{2}\right)
\equiv \left( p^{0},\vec{p}\right) \overset{c=1}{=}\left( E,\vec{p}\right) .
\end{equation}

\item $\left( \mathbb{R}\oplus \Gamma _{1,d-1}\right) \overset{\text{\cite%
{small_orbits_maths}}}{\equiv }$ $\mathfrak{J}_{1,d-1}$ is the semisimple
rank-3 Jordan algebra \cite{6,7,8,9} constructed over $\Gamma _{1,d-1}$, and
it is named \textit{Lorentzian spin factor} (see e.g. \cite{GP, 9,
McCrimmon, small_orbits_maths}). It is defined as follows : $\mathfrak{J}%
_{1,d-1}:=\mathbb{R}\oplus \Gamma _{1,d-1}$ with base point $c=(1;c_{0})\in
\mathfrak{J}_{1,d-1}$ and cubic norm $N(A):=aQ(v)$ for any $A\equiv (a;v)\in
\mathfrak{J}_{1,d-1}$, where $a\in \mathbb{R}$ and $v\in \Gamma _{1,d-1}$.
By the use of the Frobenius (matrix) norm, besides the cubic norm $\mathcal{N%
}_{3}$ the rank-3 Jordan algebra can also be endowed with a Lorentzian
quadratic norm of signature $(1,d)$ : this is proved in App. \ref{App-Frob},
and it is denoted by `Frob:$\left( 1,d\right) $' in (\ref{br}).

\item The two extra one-dimensional vector spaces $\mathbb{R}\oplus \mathbb{R%
}^{(\prime )}$ are respectively spanned by the extra timelike coordinates of
$X$- and $P$- type, namely by $X_{1}^{0^{\prime }}\equiv x^{0^{\prime }}$
and $X_{2}^{0^{\prime }}\equiv p^{0^{\prime }}$. Light-cone coordinates can
then be defined as follows :%
\begin{eqnarray}\label{LC1}
X_{i}^{+^{\prime }} &:&=\frac{1}{\sqrt{2}}\left( X_{i}^{0^{\prime
}}+X_{i}^{1^{\prime }}\right) ; \\ \label{LC2}
X_{i}^{-^{\prime }} &:&=\frac{1}{\sqrt{2}}\left( X_{i}^{0^{\prime
}}-X_{i}^{1^{\prime }}\right) .
\end{eqnarray}%
Thus, within the `extended phase space', besides the frame $%
X_{i}^{M}=\left\{ X_{1}^{0^{\prime }},X_{1}^{\mu },X_{1}^{1^{\prime
}},X_{2}^{0^{\prime }},X_{2}^{\mu },X_{2}^{1^{\prime }}\right\} $, one can
also use the frame $X_{i}^{M}=\left\{ X_{1}^{+^{\prime }},X_{1}^{-^{\prime
}},X_{1}^{\mu },X_{2}^{+^{\prime }},X_{2}^{-^{\prime }},X_{2}^{\mu }\right\}
$; for instance, they are both used in \cite{Bars-Emergent}, respectively in
Tables 2 and 1 therein. Among the physical systems discussed in Sec. \ref{Sec:Examples}, the former frame is employed in Sec. \ref{Hydrogen}, whereas the
latter is used in Sec. \ref{nr-massive}.

\item $\mathfrak{F}\left( \mathbb{R}\oplus \Gamma _{1,d-1}\right) \equiv
\mathfrak{F}(\mathfrak{J}_{1,d-1})$ is the (reduced, non-degenerate)
Freudenthal triple system (see e.g. \cite{Brown}, and \cite{35,36,37,9}%
) constructed over the Lorentzian spin factor $\mathfrak{J}_{1,d-1}$. It is
coordinatized by $X_{i}^{M}$, and it is naturally endowed with a \textit{%
symplectic} rank-2 invariant structure $\Omega _{MN}^{ij}\equiv \epsilon
^{ij}\eta _{MN}$ (where $\eta _{MN}$ is a pseudo-Euclidean metric with
signature $\left( 2,d\right) $) and a rank-4 primitive invariant \textit{%
symmetric} (tensor) structure $K_{MNPQ}^{ijkl}$.
\end{itemize}

As one may expect, the cubic Jordan algebra $\mathfrak{J}_{1,d-1}$ and the
related reduced Freudenthal triple system $\mathfrak{F}\left( \mathfrak{J}%
_{1,d-1}\right) $ sit into (linear) representations of their reduced
structure resp. automorphism Lie groups:%
\begin{align}
\mathfrak{J}_{1,d-1} &\simeq \mathbf{d}_{-1}\oplus \mathbf{1}_{2}\cr
&~\text{of~}%
G_{\text{Lor}}\equiv Str_{0}\left( \mathfrak{J}_{1,d-1}\right) ; \\
\mathfrak{F}\left( \mathfrak{J}_{1,d-1}\right) &\simeq (\mathbf{2},\mathbf{%
2+d})\cr
&~\text{of~}G_{\text{Conf}}\simeq Aut\left( \mathfrak{F}\left( \mathfrak{%
J}_{1,d-1}\right) \right) .
\end{align}%
\bigskip

\paragraph{Remark}

Algebraically, the extension from a $d$-dimensional Minkowski space to the
corresponding `extended phase space', as occurring in the 2T physics
\textit{\`{a} la Bars}, corresponds to the following chain of progressive
enlargement of the algebraic structures :%
\begin{widetext}
\begin{equation}
\underset{d\text{-dim.~Minkowski~space}}{\underset{\text{rank-2~JA}}{\Gamma
_{1,d-1}}}~\rightarrow ~\underset{\text{rank-3~JA}}{\mathfrak{J}_{1,d-1}}%
~\rightarrow ~\underset{\text{`ext.~phase~space'~\textit{\`{a}~la~Bars}}}{%
\underset{\text{reduced~FTS}}{\mathfrak{F}\left( \mathfrak{J}_{1,d-1}\right)
}},
\end{equation}%
\end{widetext}
where \textquotedblleft JA\textquotedblright\ and \textquotedblleft
FTS\textquotedblright\ stand for \textit{Jordan algebra} and \textit{%
Freudenthal triple system}, respectively. It should be remarked that all
this has remarkable relations and implications within the generalized
spacetime and phase space formalism developed by G\"{u}naydin and Pavlyk in
\cite{GP}.

\subsection{Invariants and Orbits}

The unique, primitive, quartic, $G_{\text{Conf}}$-invariant homogeneous
polynomial constructed from $K_{MNPQ}^{ijkl}$ reads \cite{ADM}%
\begin{align}\label{I4}
I_{4}
:&=K_{MNPQ}^{ijkl}X_{i}^{M}X_{j}^{N}X_{k}^{P}X_{l}^{Q}=X_{1}^{2}X_{2}^{2}-%
\left( X_{1}\cdot X_{2}\right) ^{2} \notag\\
&=\frac{\eta _{MP}\eta _{NQ}}{2}( X_{1}^{M}X_{2}^{N}-X_{1}^{N}X_{2}^{M}) (
X_{1}^{P}X_{2}^{Q}-X_{1}^{P}X_{2}^{Q})\notag\\
&=\ :\frac{1}{2%
}T_{MN}T^{MN},
\end{align}%
where%
\begin{align}
    \begin{split}
        X_{i}^{2} :&=X_{i}^{M}X_{i}^{N}\eta _{MN};  \\
        X_{1}\cdot X_{2} :&=X_{1}^{M}X_{2}^{N}\eta _{MN};  \\
        T^{MN} :&=X_{1}^{M}X_{2}^{N}-X_{1}^{N}X_{2}^{M}=T^{[MN]}.
    \end{split}
\end{align}%
The $Sp(2,\mathbb{R})$ gauge-fixing conditions \cite{Bars-Gauged,Bars:DualityH,
Bars-Terning}, which are crucial within the 2T physics \textit{\`{a}
la Bars}, read%
\begin{equation}
X_{1}^{2}=0,~X_{2}^{2}=0,~X_{1}\cdot X_{2}=0,  \label{gf}
\end{equation}%
and they imply%
\begin{eqnarray}
I_{4} &=&0; \\
\frac{\partial I_{4}}{\partial X_{i}^{M}} &=&0,~\forall M,i,
\end{eqnarray}%
where, by the homogeneity of the polynomial $I_{4}$,%
\begin{equation}
\frac{\partial I_{4}}{\partial X_{i}^{M}}=0\Rightarrow I_{4}=0.
\end{equation}%
In particular, after the analyses carried out in a number of papers in
literature (cf. e.g. \cite{ADM}, \cite{small_orbits}, \cite{small_orbits_maths} and \cite{CFMZ1-D=5}, as well as Refs. cited therein), the $Sp(2,%
\mathbb{R})$ gauge-fixing conditions (\ref{gf}) can be proven to select one
of the two `rank'\footnote{%
Within Freudenthal triple systems, a notion of `rank' of any element can be
introduced in an invariant manner with respect to the action of the
automorphism Lie group of the triple system itself \cite{9}.}-2 orbits%
\footnote{%
These have been named critical orbits `of type \textbf{B}' in the analysis
of \cite{ADM}.} $\mathcal{O}_{2c^{+}}$ and $\mathcal{O}_{2c^{-}}$ given in
the classification provided by Table VIII of \cite{small_orbits}, which are
discriminated by the sign of the quantity%
\begin{align}
\mathcal{I}_{2}:&=\frac{1}{2}\left[ \left( X_{1}^{0^{\prime }}\right)
^{2}+\left( X_{2}^{0^{\prime }}\right) ^{2}-\left( X_{1}^{1^{\prime
}}\right) ^{2}-\left( X_{2}^{1^{\prime }}\right) ^{2}\right]\notag\\
&=X_{1}^{+^{\prime }}X_{1}^{-^{\prime }}+X_{2}^{+^{\prime }}X_{2}^{-^{\prime
}},  \label{I2}
\end{align}%
which is nothing but a sum of the quadratic forms pertaining to the
lightcones in the extra timelike and spacelike coordinates.

Thus, the $Sp(2,\mathbb{R})$ gauge-fixing conditions (\ref{gf}) imply the
(coordinates describing the) resulting physical systems to belong to either $%
\mathcal{O}_{2c^{+}}$ or $\mathcal{O}_{2c^{-}}$, which are orbits of the
(non-transitive) action of $G_{\text{Conf}}$ onto its representation space
provided by the `extended phase space' $\mathfrak{F}\left( \mathfrak{J}%
_{1,d-1}\right) \simeq (\mathbf{2},\mathbf{2+d})$. Such orbits are
respectively defined by the following $G_{\text{Conf}}$-invariant
constraints :
\begin{eqnarray}
\mathcal{O}_{2c^{+}} &:&\left\{
\begin{array}{l}
X_{1}^{2}=0,~X_{2}^{2}=0,~X_{1}\cdot X_{2}=0; \\
T^{MN}\neq 0; \\
\mathcal{I}_{2}>0;%
\end{array}%
\right.  \label{plus} \\
\mathcal{O}_{2c^{-}} &:&\left\{
\begin{array}{l}
X_{1}^{2}=0,~X_{2}^{2}=0,~X_{1}\cdot X_{2}=0; \\
T^{MN}\neq 0; \\
\mathcal{I}_{2}<0.%
\end{array}%
\right.  \label{minus}
\end{eqnarray}%
The subgroup(s) of $G_{\text{Conf}}$ stabilizing $\mathcal{O}_{2c^{+}}$ and $%
\mathcal{O}_{2c^{-}}$ were computed in \cite{small_orbits} : interestingly,
both $\mathcal{O}_{2c^{+}}$ and $\mathcal{O}_{2c^{-}}$ enjoy isomorphic
realizations as (non-symmetric) coset spaces, with such a subgroup acting as
isotropy group, namely%
\begin{gather}
\mathcal{O}_{2c^{+}}\simeq \mathcal{O}_{2c^{-}}\\ \nonumber
\simeq \frac{Sp(2,\mathbb{R}%
)\times SO(2,d)}{\left( SO(2,1)\ltimes \mathbb{R}\right) \times \left(
SO\left( d-2\right) \ltimes \left( \mathbb{R}^{d-2}\oplus \mathbb{R}%
^{d-2}\right) \right) }.\bigskip
\end{gather}

\section{\label{Sec:Examples}2T physics: Examples}

We will now consider various (relativistic and non-relativistic) physical
systems in $d$ Lorentzian spacetime dimensions, along with their
coordinatization in the `extended phase space' $\mathfrak{F}\left( \mathfrak{%
J}_{1,d-1}\right) $ within the 2T physics \textit{\`{a} la Bars}. In
each system, the symmetry Lie algebra of $\mathfrak{F}\left( \mathfrak{J}%
_{1,d-1}\right) $, which is the conformal algebra $\mathfrak{so}(2,d)$ in $d$
Lorentzian dimensions, generated by\footnote{%
The square brackets denote ($1/n!$ times the) antisymmetrization of the $n$
enclosed indices throughout.}%
\begin{align}
\begin{split}
    L^{MN}:&=X^{M}P^{N}-X^{N}P^{M}=2X^{[M}P^{N]}\\
    &=L^{[MN]},
\end{split}  \label{L^MN}
\end{align}%
gets broken by the $Sp(2,\mathbb{R})$-gauge fixing procedure (as well as by
the resolution of the physical constraints) to a proper (not necessarily
semisimple nor reductive) real subalgebra $\mathfrak{g}$ :%
\begin{equation}
\mathfrak{so}(2,d)\rightarrow \mathfrak{g},  \label{red}
\end{equation}%
which can be characterized as the maximal manifest (i.e., linearly realized)
Lie algebra of symmetries of the physical system itself. All other
generators of $\mathfrak{so}(2,d)$, namely those of the (not necessarily
maximal nor symmetric, and generally pseudo-Riemannian) coset space\footnote{%
$\mathfrak{so}(2,d)$ and $\mathfrak{g}$ are the Lie algebras of the Lie
groups $SO(2,d)$ and $G$, respectively. We will use an analogous notation
throughout this paper.} $SO(2,d)/G$, are \textit{non-linearly} realized, in
a way which can be regarded as a typical feature of the parametrization of
the system in the 2T physics \textit{\`{a} la Bars}, which is
particularly interesting in non-relativistic systems (see Secs. \ref{nr-massive}, \ref{Hydrogen}, \ref{Carroll-E0}).

From the definition (\ref{L^MN}), the Poisson brackets of the generators of $%
\mathfrak{so}(2,d)$ consistently read%
\begin{align}\label{conf}
&\{L^{MN},L^{PQ}\}\\ \notag
&\quad=\eta ^{MP}L^{NQ}+\eta ^{NQ}L^{MP}-\eta ^{MQ}L^{NP}-\eta
^{NP}L^{MQ},
\end{align}%
where $\eta ^{MP}$ is a diagonal metric with signature $\left(
-,-,+,...,+\right) $ :%
\begin{equation*}
\eta ^{MN}=\eta _{MN}=\text{diag}\left( -1,-1,\underset{d~\text{times}}{%
\underbrace{1,...,1}}\right) .
\end{equation*}

We should also here recall that the $L^{MN}$'s generate the infinitesimal
transformations of the `extended phase space' $\mathfrak{F}\left( \mathfrak{J%
}_{1,d-1}\right) $ through Poisson brackets, henceforth denoted by $\left\{
\cdot ,\cdot \right\} $: for any function $f=f\left( X,P\right)$%
\begin{align}
\delta_{\Lambda^{(MN)}}f :&=\frac{1}{2}\Lambda _{MN}\left\{ L^{MN},f\right\} \equiv %
\frac{1}{2}\Lambda _{MN}\left\langle \nabla _{\mathcal{X}}L^{MN},\nabla _{%
\mathcal{X}}f\right\rangle \notag \\
:&=\frac{1}{2}\Lambda _{MN}\left( \nabla _{%
\mathcal{X}}L^{MN}\right) ^{T}\Omega\ \nabla _{\mathcal{X}}f  \notag \\
&=\Lambda _{MN}\eta^{MR}\left( P^{N}%
\frac{\partial f}{\partial P^{R}}+X^{N}\frac{\partial f}{\partial X^{R}}%
\right) ,
\end{align}%
where
$\Lambda _{MN}=\Lambda _{\lbrack MN]}$ is the tensor of \textit{%
infinitesimal }conformal transformations' parameters, and $\left\langle
\cdot ,\cdot \right\rangle $ denotes the symplectic product in $\mathfrak{F}%
\left( \mathfrak{J}_{1,d-1}\right) $ itself, determined by the $2\left(
d+2\right) \times 2\left( d+2\right) $ symplectic metric%
\begin{equation}
\Omega :=\left(
\begin{array}{cc}
0_{d+2} & -I_{d+2} \\
I_{d+2} & 0_{d+2}%
\end{array}%
\right) ,
\end{equation}%
where $0_{d+2}$ and $I_{d+2}$ respectively denote the null and identity $%
\left( d+2\right) \times \left( d+2\right) $ matrices. Moreover, $\mathcal{X}
$ collectively denote the coordinates of $\mathfrak{F}\left( \mathfrak{J}%
_{1,d-1}\right) $ :%
\begin{equation}
\mathcal{X}:=\left( X^{M},P^{M}\right) ^{T}.
\end{equation}

\subsection{\label{Sec-rel-m0}Relativistic \textit{massless} particle}

The relativistic massless particle in flat (Minkowski) space is
characterized by%
\begin{equation}
p^{2}=p^{\mu }p^{\nu }\eta _{\mu \nu }=-\left( p^{0}\right) ^{2}+\left\vert
\vec{p}\right\vert ^{2}=0.
\end{equation}%

Renaming $X_1^M\equiv X^M$ and $X_2^M\equiv P^M$, where $M=+,-,0,1,\dots,(d-1)$, the particles coordinates can be parametrized in $\mathfrak{F}\left( \mathfrak{J}%
_{1,d-1}\right) $ as follows (cf. e.g. \cite{Bars-Emergent}, Table 1) :%

\begin{equation}
    \begin{array}{l}
        X^{+^{\prime }}=1, \\ X^{-^{\prime }}=\frac{1}{2}x^{2}, \\ X^{\mu }=x^{\mu }, \\  [8pt]
        P^{+^{\prime }}=0, \\ P^{-^{\prime }}=x\cdot p, \\ P^{\mu }=p^{\mu },
    \end{array}
    \label{par-1}
\end{equation}
where $\left\vert \cdot \right\vert $ denotes the Lorentzian norm in $1+(d-1)$
(spatial) dimensions, and $x^2=x\cdot x$. Furthermore, $\mu,\nu=0,1,\dots,(d-1)$ and the $+,-$ indexes denote the light-cone coordinates introduced in (\ref{LC1}) and (\ref{LC2}). This parametrization implies%
\begin{equation}
\mathcal{I}_{2}=X^{+^{\prime }}X^{-^{\prime }}+P^{+^{\prime }}P^{-^{\prime
}}=\frac{1}{2}x^{2}  \label{res}
\end{equation}%
to be manifestly invariant under the Lorentz group $SO(1,d-1)$. Its Lie
algebra $\mathfrak{so}(1,d-1)$, which is the symmetry of a relativistic
massless particle in flat (Minkowski) space, is fully manifest and linearly
realized within the parametrization (\ref{par-1}).

Indeed, using the parameterization \eqref{par-1}, the $SO(2,d)$ generators become
\begin{equation}
\begin{split}
L^{\mu\nu}&=x^{\mu}p^{\nu}-x^{\nu}p^\mu;\\
L^{+\mu}&=p^\mu;\\
L^{+-}&=x\cdot p;\\
L^{-\mu}&=\frac{1}{2}x^2p^\mu-x\cdot p\ x^\mu,
\end{split}
\label{MasslessGen}
\end{equation}
generating the \textit{conformal transformations} of the massless relativistic particles (rotations, translations, special conformal transformations and dilatation respectively). Their Poisson brackets can be easily computed from the general equation \eqref{conf}. Indeed, as discussed for instance in \cite{Bars-Emergent}, under equation (3.4), the Poisson brackets between the $L^{MN}$'s do not change after the gauge fixing and the constraints has been solved. For instance, using the definition of the Poisson brackets,
\begin{align}
    \{L^{+\mu},L^{-\nu}\}&=\sum_{\rho,\sigma=0,1}^{d-1}\eta^{\rho\sigma}\left(\frac{\partial L^{+\mu}}{\partial x^\rho}\frac{\partial L^{-\nu}}{\partial p^\sigma}-\frac{\partial L^{+\mu}}{\partial p^\rho}\frac{\partial L^{-\nu}}{\partial x^\sigma}\right)\cr
    &=x^\nu p^\mu-p^\nu x^\mu + \eta^{\mu\nu}x\cdot p\notag \\
    &=-L^{\mu\nu}+\eta^{\mu\nu}L^{+-}.
\end{align}
This is equivalent to using the equation \eqref{conf}. Indeed,
\begin{align}
    \{L^{+\mu},L^{-\nu}\}&=\eta^{+-}L^{\mu\nu}+\eta^{\mu\nu}L^{+-}-\eta^{+\nu}L^{\mu-}-\eta^{\mu-}L^{+\nu}\notag \\
    &=-L^{\mu\nu}+\eta^{\mu\nu}L^{+-}.
\end{align}
In this way, all the Poisson brackets can be easily computed
\begin{equation}
\begin{split}
\{L^{\mu\nu},L^{\rho\sigma}\}&=\eta^{\mu\rho}L^{\nu\sigma}+\eta^{\nu\sigma}L^{\mu\rho}-\eta^{\mu\sigma}L^{\nu\rho}-\eta^{\nu\rho}L^{\mu\sigma};\\
\{L^{+\mu},L^{+\nu}\}&=\{L^{-\mu},L^{-\nu}\}=0;\\
\{L^{+\mu},L^{-\nu}\}&=-L^{\mu\nu}+\eta^{\mu\nu}L^{+-};\\
\{L^{+\mu},L^{+-}\}&=L^{\mu+};\\
\{L^{-\mu},L^{+-}\}&=L^{\mu-};\\
\{L^{\mu\nu},L^{+\rho}\}&=\eta^{\nu\rho}L^{\mu+}-\eta^{\mu\rho}L^{\nu+};\\
\{L^{\mu\nu},L^{-\rho}\}&=\eta^{\nu\rho}L^{\mu-}-\eta^{\mu\rho}L^{\nu};\\
\{L^{\mu\nu},L^{+-}\}&=0.
\end{split}
\label{PB}
\end{equation}
These Poisson brackets are valid also for the examples in the next sections.

\subsubsection{\label{Sec-rel}The maximal linearly realized algebra: $%
\mathfrak{so}\left( 1,d-1\right)\oplus\mathfrak{so}(1,1)$}

The infinitesimal transformations of the physical phase space coordinates $(x^\mu,p^\mu)$ are
\begin{equation}
\begin{split}
\delta_{\Lambda^{(\mu\nu)}} x^\rho&=\frac{1}{2}\Lambda_{\mu\nu}\{L^{\mu\nu},x^\rho\}=\frac{1}{2}\Lambda_{\mu\nu}(\eta^{\mu\rho}x^\nu-\eta^{\nu\rho}x^\mu);\\
\delta_{\Lambda^{(+\mu)}} x^\nu&=b_\mu\{L^{+\mu},x^\nu\}=-b^\nu;\\
\delta_{\Lambda^{(-\mu)}} x^\nu&=c_\mu\{L^{-\mu},x^\nu\}=-c^\nu\frac{x^2}{2}+c\cdot x x^\nu;\\
\delta_{\Lambda^{(+-)}} x^\mu&=\Lambda_{+-}\{L^{+-},x^\mu\}=-\alpha x^\mu,\\
\delta_{\Lambda^{(\mu\nu)}} p^\rho&=\frac{1}{2}\Lambda_{\mu\nu}\{L^{\mu\nu},p^\rho\}=\frac{1}{2}\Lambda_{\mu\nu}(\eta^{\mu\rho}p^\nu-\eta^{\nu\rho}p^\mu);\\
\delta_{\Lambda^{(+\mu)}} p^\nu&=b_\mu\{L^{+\mu},p^\nu\}=0;\\
\delta_{\Lambda^{(-\mu)}} p^\nu&=c_\mu\{L^{-\mu},p^\nu\}=c\cdot p x^\nu-c\cdot x p^\nu-c^\nu x\cdot p;\\
\delta_{\Lambda^{(+-)}} p^\mu&=\alpha\{L^{+-},p^\mu\}=\alpha p^\mu,
\end{split}
\label{CSTrasformations}
\end{equation}
where $b_\mu=\Lambda_{+\mu}$, $c_\mu=\Lambda_{-\mu}$ and $\alpha=\Lambda_{+-}$. Thus, in this case, the formula (\ref{red}) reads%
\begin{equation}
\mathfrak{so}(2,d)\rightarrow\mathfrak{g}=\mathfrak{so}\left( 1,d-1\right)\oplus\mathfrak{so}(1,1),
\label{red-2}
\end{equation}
generated by the rotations $L^{\mu\nu}$ and the dilatation $L^{+-}$.

\subsubsection{The sign of $\mathcal{I}_2$}

The quantity $\mathcal{I}_2$ is manifestly invariant under rotations. Furthermore, under dilatation, it transforms as
\begin{equation}\label{I2Massless}
    \mathcal{I}'_2=e^{-2\alpha}\mathcal{I}_2,
\end{equation}
leaving the sign invariant.

On the other hand, under translations one gets
\begin{equation}
    \mathcal{I}'_2=(x^\mu-a^\mu)(x_\mu-a_\mu).
\end{equation}
If we only consider the time translation,
\begin{equation}\label{IIMasslessTTransl}
    \mathcal{I}'_2=-(x_0-a_0)^2+|\vec{x}|^2,
\end{equation}
or space translations,
\begin{equation}\label{IIMasslessXTransl}
    \mathcal{I}'_2=-x_0^2+|\vec{x}-\vec{a}|^2,
\end{equation}
it becomes evident that the sign of $\mathcal{I}_2$ is not invariant. The same happens for special conformal transformations, for which
\begin{align}
    \mathcal{I}_2'=(f_c(x)x^\mu+c^\mu g_c(x))(f_c(x)x_\mu+c_\mu g_c(x)),
\end{align}
and we can transform the \textit{timelike} ($\mu=0$) component differently from the \textit{spacelike} component.

The relativistic massless particle belongs to the degenerate orbit with $X^2P^2-(X\cdot P)^2=0$. Under the action of $\mathfrak{g}\equiv\mathfrak{so}\left( 1,d-1\right)\oplus\mathfrak{so}(1,1)$, the orbit splits into two suborbits $\mathcal{O}%
_{2c^{+}} $ resp. $\mathcal{O}_{2c^{-}}$ in $\mathfrak{F}\left( \mathfrak{J}%
_{1,d-1}\right) $ depending on whether\footnote{%
Note that each sheet of $T_{d-1}^{\text{time}}$ or $T_{d-1}^{\text{space}}$
is an orbit of the non-transitive action of the Lorentz group $SO(1,d-1)$,
which acts as the (reduced) structure group on the rank-2 Jordan algebra $%
\mathbf{\Gamma }_{1,d-1}$. \ To complete the resulting stratification of $%
\mathbf{\Gamma }_{1,d-1}$, the two cone branches (overlapping in the origin)
of the lightcone $x^{2}=0$, respectively defined as forward ($x^{0}>0$) or
backward ($x^{0}<0$) branches, should be considered; however, these latter
are not relevant to our analysis, because they pertain \cite{small_orbits}
to other orbits in the Freudenthal triple system, which are not consistent
with the $Sp(2,\mathbb{R})$ gauge-fixing conditions (\ref{gf}).} the Lorentz
vector $x^{\mu }$ belongs to the two-sheeted hyperboloids%

\begin{align}
T_{d-1}^{\text{time}} &\equiv \left. \frac{SO\left( 1,d-1\right)\otimes SO(1,1)}{SO(d-1)}%
\right\vert _{x^{2}<0}\notag\\
&=T_{d-1}^{\text{time},~x^{0}>0}\cup T_{d-1}^{\text{time%
},~x^{0}<0},  \label{hyperb-1} \\
&\text{resp.}  \notag \\
T_{d-1}^{\text{space}} &\equiv \left. \frac{SO\left( 1,d-1\right) \otimes SO(1,1)}{SO(d-1)}%
\right\vert _{x^{2}>0}\notag \\
&=T_{d-1}^{\text{space},~x^{0}>0}\cup T_{d-1}^{\text{%
space},~x^{0}<0}.  \label{hyperb-2}
\end{align}

\subsection{\label{Sec-rel-m0-max}Relativistic \textit{massless} particle in
maximally symmetric space}

The relativistic massless particle in a maximally symmetric space of
constant curvature $K$ is characterized by%
\begin{equation}
p^{2}-\frac{K\left( x\cdot p\right) }{1-Kx^{2}}=0,
\end{equation}%
where $x\cdot p:=-x^{0}p^{0}+\vec{x}\cdot \vec{p}$ denotes the Lorentzian
scalar product in $d$ dimensions (and thus, for instance, $x^{2}\equiv
x\cdot x$). This physical system can be parametrized in $\mathfrak{F}\left(
\mathfrak{J}_{1,d-1}\right) $ as follows\footnote{%
In the limit of vanishing curvature $K\rightarrow 0$, the parametrization (%
\ref{par-2}) consistently reduces to (\ref{par-1}).} (cf. e.g. \cite%
{Bars-Emergent}, Table 1):%
\begin{equation}
    \begin{array}{l}
        X^{+^{\prime }}=1+\sqrt{1-Kx^{2}}, \\ X^{-^{\prime }}=\frac{1}{2}\frac{x^{2}}{1+\sqrt{1-Kx^{2}}}, \\ X^{\mu }=x^{\mu }, \\ [8pt]
        P^{+^{\prime }}=0, \\ P^{-^{\prime }}=\frac{\sqrt{1-Kx^{2}}}{1+\sqrt{1-Kx^{2}}}x\cdot p, \\ P^{\mu }=p^{\mu }-\frac{Kx\cdot p}{1+\sqrt{1-Kx^{2}}}x^{\mu },
    \end{array}
    \label{par-2}
\end{equation}
which again yields to (\ref{res}). In this case, the generators become
\begin{equation}
\begin{split}
L^{\mu\nu}&=x^{\mu}p^{\nu}-x^{\nu}p^\mu;\\
L^{+\mu}&=(1+\sqrt{1-Kx^{2}})p^\mu-Kx\cdot p x^\mu;\\
L^{+-}&=\sqrt{1-Kx^{2}}\ x\cdot p;\\
L^{-\mu}&=\frac{1}{1+\sqrt{1-Kx^2}}\left[\frac{1}{2}x^2p^\mu-x\cdot p\ x^\mu\right]\\
&\qquad+\frac{1}{2}\frac{Kx^2\ x\cdot p\ x^\mu}{(1+\sqrt{1-Kx^2})^2}.
\end{split}
\label{SSpaceGen}
\end{equation}
In addition, $\mathcal{I}_{2}$ is the same as in \eqref{res}. 

\subsubsection{\label{Sec-relMax}The maximal linearly realized algebra: $%
\mathfrak{so}\left( 1,d-1\right)$}

The only \textit{linearly realized} infinitesimal transformations of the phase space coordinates under $SO(2,d)$ are
\begin{equation}
\begin{split}
\delta_{\Lambda^{(\mu\nu)}} x^\rho&=\frac{1}{2}\Lambda_{\mu\nu}\{L^{\mu\nu},x^\rho\}\\
&=\frac{1}{2}\Lambda_{\mu\nu}(\eta^{\mu\rho}x^\nu-\eta^{\nu\rho}x^\mu);\\
\delta_{\Lambda^{(\mu\nu)}} p^\rho&=\frac{1}{2}\Lambda_{\mu\nu}\{L^{\mu\nu},p^\rho\}\\
&=\frac{1}{2}\Lambda_{\mu\nu}(\eta^{\mu\rho}p^\nu-\eta^{\nu\rho}p^\mu),
\end{split}
\label{CSTrasformationsLinearlyRMMS}
\end{equation}
which generate the set of Lorentz transformations (see Section \ref{App:RMMS} an the Appendix \ref{App:I2} for the whole set of transformations). Therefore, the reduction formula (\ref{red}) specializes to
\begin{equation}
\mathfrak{so}(2,d)\rightarrow\mathfrak{g}\equiv\mathfrak{so}\left( 1,d-1\right).
\label{red-2s}
\end{equation}

\subsubsection{The sign of $\mathcal{I}_2$}
We can also observe that the transformations generated by $L^{+-}$, for which the infinitesimal form is
\begin{align}
    \begin{split}
        \delta_{\Lambda^{(+-)}} x^\mu&=\alpha\{L^{+-},x^\mu\}\\
        &=-\alpha\ (1+\sqrt{1-Kx^{2}})\ x^\mu;\\
        \delta_{\Lambda^{(+-)}} p^\mu&=\alpha\{L^{+-},p^\mu\}\\
        &=\frac{\alpha}{\sqrt{1-Kx^2}}\left[p^\nu-K(x^2p^\nu+x\cdot p)\right],
    \end{split}
\end{align}
are not linearly realized, in contrast with the relativistic massless particle. On the other hand, the full transformation of $x^\mu$ under $L^{+-}$ can be written as
\begin{equation}
    {x'}^\mu = f_\alpha(x)\ x^\mu,
\end{equation}
where $f_\alpha(x)$ is a scalar function of $x^2=x\cdot x$. Thus,
\begin{equation}
    \mathcal{I}_2'=f_\alpha^2(x)\ \mathcal{I}_2.
\end{equation}
We can conclude that the sign of $\mathcal{I}_2$ is invariant under the action of the whole $\mathfrak{h}\equiv\mathfrak{so}\left( 1,d-1\right)\oplus\mathfrak{so}(1,1)\supset\mathfrak{g}$ (the sign of $\mathcal{I}_2$ is analyzed in more depth in Subsection \ref{App:RMMS} of the Appendix \ref{App:I2}). When only the transformations generated by $\mathfrak{h}$ are considered, this physical system belongs to the orbit $\mathcal{O}_{2c^{+}}$
resp. $\mathcal{O}_{2c^{-}}$ in $\mathfrak{F}\left( \mathfrak{J}%
_{1,d-1}\right) $ depending on whether $x^{\mu }$ belongs to the two-sheeted
hyperboloids (\ref{hyperb-1}) resp. (\ref{hyperb-2}).

\subsection{ \label{Sec-rel-m}Relativistic \textit{massive} particle in
Minkowski space}

The relativistic massive particle in flat (Minkowski) space is characterized
by%
\begin{equation}
p^{2}=p^{\mu }p^{\nu }\eta _{\mu \nu }=-m^{2}.
\end{equation}%
By introducing%
\begin{equation}
a:=\sqrt{1+\frac{m^{2}x^{2}}{\left( x\cdot p\right) ^{2}}},
\end{equation}%
it can be parametrized in $\mathfrak{F}\left( \mathfrak{J}_{1,d-1}\right) $
as follows (cf. e.g. \cite{Bars-Emergent}, Table 1):%
\begin{equation}
    \begin{array}{l}
        X^{+^{\prime }}=\frac{1+a}{2a},  \\X^{-^{\prime }}=\frac{a}{1+a}x^{2}, \\ X^{\mu }=x^{\mu }, \\ [8pt]
        P^{+^{\prime }}=-\frac{m^{2}}{2ax\cdot p}, \\ P^{-^{\prime }}=ax\cdot p, \\ P^{\mu }=p^{\mu },
    \end{array}
    \label{par-3}
\end{equation}
which implies%
\begin{equation}
\mathcal{I}_{2}=X^{+^{\prime }}X^{-^{\prime }}+P^{+^{\prime }}P^{-^{\prime
}}=\frac{1}{2}\left( x^{2}-m^{2}\right)
\end{equation}%
to be manifestly invariant under the Lorentz group $SO(1,d-1)$.

\subsubsection{\label{Sec-relmass}The maximal linearly realized algebra: $%
\mathfrak{so}(1,d-1)$}

The generators of $SO(2,d)$ for the relativistic massless particle become
\begin{equation}
\begin{split}
L^{\mu\nu}&=x^{\mu}p^{\nu}-x^{\nu}p^\mu;\\
L^{+\mu}&=\frac{1+a}{2a}p^\mu+\frac{m^2}{2a\ x\cdot p} x^\mu;\\
L^{+-}&=a\ x\cdot p;\\
L^{-\mu}&=\frac{ax^2}{1+a}p^\mu-a\ x\cdot p\ x^\mu,
\end{split}
\label{MassiveGen}
\end{equation}
giving the \textit{linearly realized} infinitesimal transformations (the whole set of transformations is reported in Section \ref{App:RMassive})
\begin{equation}
\begin{split}
\delta_{\Lambda^{(\mu\nu)}} x^\rho&=\frac{1}{2}\Lambda_{\mu\nu}\{L^{\mu\nu},x^\rho\}\\
&=\frac{1}{2}\Lambda_{\mu\nu}(\eta^{\mu\rho}x^\nu-\eta^{\nu\rho}x^\mu);\\
\delta_{\Lambda^{(\mu\nu)}} p^\rho&=\frac{1}{2}\Lambda_{\mu\nu}\{L^{\mu\nu},p^\rho\}\\
&=\frac{1}{2}\Lambda_{\mu\nu}(\eta^{\mu\rho}p^\nu-\eta^{\nu\rho}p^\mu).\\
\end{split}
\label{CSTrasformationsRMassiveLinear}
\end{equation}

As for the previous case, the only linearly realized transformations are generated by the set of Lorentz transformations, and (\ref{red})
specializes to
\begin{equation}
    \mathfrak{so}(2,d)\rightarrow\mathfrak{g}\equiv\mathfrak{so}(1,d-1).
\end{equation}

\subsubsection{The sign of $\mathcal{I}_2$}

In contrast to the previous case, the sign of $\mathcal{I}_2$ is not invariant under $L^{+-}$-transformations, which infinitesimal realization reads
\begin{align}
    \delta_{\Lambda^{(+-)}} x^\mu&=\alpha\{L^{+-},x^\mu\}=-\frac{\alpha x^\mu}{a},\\
    \delta_{\Lambda^{(+-)}} p^\mu&=\alpha\{L^{+-},p^\mu\}=\alpha ax^\mu\notag\\
    &\qquad\qquad+\frac{\alpha m^2}{ax\cdot p}\left(x^\mu-\frac{x^2}{x\cdot p}p^\mu\right).
\end{align}
Indeed, it transforms as
\begin{equation}
    \mathcal{I}_2'=f_\alpha^2\left(x,p\right)x^2-m^2,
\end{equation}
where $f_\alpha\left(x,p\right)$ is a scalar function of both $x$ and $p$. It becomes evident that, due to the presence of $m^2$, the function $f_\alpha\left(x,p\right)$ leads to a change in the sign of $\mathcal{I}_2$ (the other transformations of $\mathcal{I}_2$ are analyzed in Subsection \ref{App:RMassive}).

Therefore, the sign of $\mathcal{I}_2$ is invariant only under $\mathfrak{g}\equiv\mathfrak{so}(1,d-1)$ and, under its action, the relativistic massive particle in Minkowski spacetime belongs
to the orbit $\mathcal{O}_{2c^{+}}$ resp. $\mathcal{O}_{2c^{-}}$ in $%
\mathfrak{F}\left( \mathfrak{J}_{1,d-1}\right) $ depending on the sign of $%
x^{2}-m^{2}$, namely whether $x^{\mu }$ belongs to the two-sheeted
hyperboloids%
\begin{align}
\tilde{T}_{d-1}^{\text{time}} &\equiv \left. \frac{SO\left( 1,d-1\right) }{%
SO(d-1)}\right\vert _{x^{2}-m^{2}<0}\notag\\
&=\tilde{T}_{d-1}^{\text{time}%
,~x^{0}>0}\cup \tilde{T}_{d-1}^{\text{time},~x^{0}<0}, \\
&\text{resp.}  \notag \\
\tilde{T}_{d-1}^{\text{space}} &\equiv \left. \frac{SO\left( 1,d-1\right) }{%
SO(d-1)}\right\vert _{x^{2}-m^{2}>0}\notag\\
&=\tilde{T}_{d-1}^{\text{space}%
,~x^{0}>0}\cup \tilde{T}_{d-1}^{\text{space},~x^{0}<0}.
\end{align}%
Notice that in the limit $m\rightarrow 0^{+}\Leftrightarrow a\rightarrow
1^{+}$ the relativistic massless particle system (case 1) is retrieved.

\subsection{\label{nr-massive}Non-relativistic \textit{massive} particle}

In the non-relativistic framework, the Lorentz-covariant coordinates and
momenta split as follows :%
\begin{eqnarray}
x^{\mu } &\rightarrow &t,x^{I}; \\
p^{\mu } &\rightarrow &H,p^{I}.
\end{eqnarray}%
As a consequence, the Lorentz-covariant (`non-extended') phase space $\Gamma
_{1,d-1}^{\oplus 2}=\Gamma _{1,d-1}\oplus \Gamma _{1,d-1}$, coordinatized by
$x^{\mu }$ and $p^{\mu }$, becomes the non-relativistic (`non-extended')
phase space, coordinatized by $t$, $x^{I}$, $H$ and $p^{I}$, and denoted by $%
\mathcal{N}_{d}$ :%
\begin{equation}
\Gamma _{1,d-1}^{\oplus 2}\rightarrow \mathcal{N}_{d}:=\mathbb{R}^{2}\oplus
\Gamma _{d-1}^{\oplus 2}.
\end{equation}%
The non-vanishing Poisson brackets between the coordinates of $\mathcal{N}%
_{d}$ read
\begin{equation}
\{x^{I},p^{J}\}=\delta ^{IJ},\qquad \{t,H\}=-1.  \label{PSPoisson}
\end{equation}

The non-relativistic massive particle of mass $m\in \mathbb{R}_{0}^{+}$ is
characterized by the physical constraint between the components of the
momentum vector :%
\begin{equation}
H-\frac{\left\vert \vec{p}\right\vert ^{2}}{2m}=0,
\label{constr-nr-particle}
\end{equation}%
where $\left\vert \vec{p}\right\vert ^{2}:=\vec{p}\cdot \vec{p}%
=p^{I}p^{J}\delta _{IJ}$. This physical system can be coordinatized\footnote{%
Note the $\pm $-branching.} in the `extended' phase space $\mathfrak{F}%
\left( \mathfrak{J}_{1,d-1}\right) $ as follows (cf. e.g. \cite%
{Bars-Emergent}, Table 1):%
\begin{equation}
    \begin{array}{l}
        X^{+}=t, \\ X^{-}=\frac{1}{m}\left( \vec{x}\cdot\vec{p}-tH\right), \\ X^{0}_\pm=\pm \left\vert \vec{x}-\frac{t%
}{m}\vec{p}\right\vert, \\ X^{I}=x^{I},\\ [8pt]
        P^{+}=m, \\ P^{-}=H, \\ P^{0 }=0, \\ P^{I}=p^{I},
    \end{array}
    \label{par-4}
\end{equation}%
where $I=1,\dots,(1-d)$, implying%
\begin{align}
    \begin{split}
        \mathcal{I}_{2}&=X^{+^{\prime }}X^{-^{\prime }}+P^{+^{\prime }}P^{-^{\prime}}\\
        &=\frac{t}{m}\left( \vec{x}\cdot \vec{p}-tH\right) +mH\\
        &=\frac{t\vec{x}\cdot\vec{p}+\left( m^{2}-t^{2}\right) H}{m}
    \end{split}
    \label{I2-nr}
\end{align}%
to have \textit{at least} $\mathfrak{so}(d-1)$ as manifest (linearly
realized) symmetry Lie algebra.

Within the coordinatization (\ref{par-4}), the generators $L^{MN}$ (\ref%
{L^MN}) of $\mathfrak{so}(2,d)$ read \cite{Bars-Emergent}%
\begin{equation}
\begin{split}
L^{IJ}& =x^{I}p^{J}-x^{J}p^{I}; \\
L^{+-}& =2tH-\vec{x}\cdot \vec{p}; \\
L^{+I}& =tp^{I}-mx^{I}; \\
L^{-I}& =\frac{\vec{x}\cdot \vec{p}-tH}{m}p^{I}-Hx^{I}; \\
L^{+0}& =\mp m\left\vert \vec{x}-\frac{t}{m}\vec{p}\right\vert ; \\
L^{-0}& =\mp H\left\vert \vec{x}-\frac{t}{m}\vec{p}\right\vert ; \\
L^{0I}& =\pm p^{I}\left\vert \vec{x}-\frac{t}{m}\vec{p}\right\vert .
\end{split}
\label{GenGalilei}
\end{equation}
Once again, the Poisson brackets can be simply deduced from \eqref{conf}, with $\eta^{IJ}$ being the Euclidean metric and $\eta^{00}=-1$. Notably, the generators $L^{IJ}$ and $L^{+I}$ form a proper subalgebra of the so called \textit{Bargmann algebra} \cite{Bargmann:UnitaryReps}, whose properties are outlined in the following sections.

\subsubsection{\label{Gal}The Galilei algebra $\mathfrak{gal}\left(
1,d-1\right) $}

Since the symmetry of a non-relativistic particle of mass $m$ in $d$
Lorentzian dimensions is given by the Bargmann Lie algebra $\mathfrak{bar}%
_{m}\left( 1,d-1\right) $, which is a central extension of the Galilei Lie
algebra $\mathfrak{gal}\left( 1,d-1\right) $ \cite%
{Figueroa:Galilei,FigueroaGalileiCarrollBargmann}, one may intuitively
guess that the generators of such algebras might lie within the set of
generators (\ref{GenGalilei}) of the conformal algebra $\mathfrak{so}(2,d)$
realized on the coordinatization (\ref{par-4}) of the `extended' phase space
$\mathfrak{F}\left( \mathfrak{J}_{1,d-1}\right) $.

This is \textit{only partially} true. Indeed, $\mathfrak{gal}\left(
1,d-1\right) $ is generated by $H$ and $p^{I}$, respectively giving rise to
time and space translations, and $B^{I}$ and $J^{IJ}$, respectively
determining boosts and rotations, and defined as follows :%
\begin{eqnarray}
B^{I} &:&=tp^{I}\overset{\text{(\ref{GenGalilei})}}{=}\left.
L^{+I}\right\vert _{m=0};  \label{def-1} \\
J^{IJ} &:&=L^{IJ}\overset{\text{(\ref{GenGalilei})}}{=}x^{I}p^{J}-x^{J}p^{I}.
\label{def-2}
\end{eqnarray}%
Notice that only the generators $B^{I}$, $J^{IJ}$ are generators of $%
\mathfrak{so}(2,d)$ (as given by (\ref{GenGalilei}); in the former case, in
the limit $m=0$); in fact, $B^{I}$, $J^{IJ}$ are \textit{quadratic} in the
coordinates of $\mathfrak{F}\left( \mathfrak{J}_{1,d-1}\right) $, whereas $H$
and $p^{I}$ are coordinates of $\mathfrak{F}\left( \mathfrak{J}%
_{1,d-1}\right) $ themselves (cf. (\ref{par-4})), and therefore they do
\textit{not} belong to $\mathfrak{so}(2,d)$ within the coordinatization (\ref%
{GenGalilei}).

By recalling the expressions of the conformal Poisson brackets (\ref{conf}),
and using (\ref{GenGalilei}), (\ref{def-1}) and (\ref{def-2}), one obtains
the non-vanishing Poisson brackets of $\mathfrak{gal}\left( 1,d-1\right) $ :
\begin{equation}
\begin{split}
\{H,B^{I}\}& =p^{I}; \\
\{p^{I},J^{JK}\}& =-\delta ^{IJ}p^{K}+\delta ^{IK}p^{J}; \\
\{B^{I},J^{JK}\}& =-\delta ^{IJ}B^{K}+\delta ^{IK}B^{J}; \\
\{J^{IJ},J^{KL}\}& =\delta ^{IK}J^{JL}+\delta ^{JL}J^{IK}\\
&\qquad\qquad-\delta
^{IL}J^{JK}-\delta ^{JK}J^{IL}.
\end{split}
\label{PB Bar}
\end{equation}%
On the other hand, by recalling (\ref{PSPoisson}), (\ref{rot}), (\ref{+I})
and (\ref{def-1})-(\ref{def-2}), the non-vanishing Poisson brackets of $%
\mathfrak{gal}\left( 1,d-1\right) $ with the coordinates $t$ and $x^{I}$ of
the $d$-dim. (non-relativistic) Lorentzian spacetime read
\begin{equation}
\begin{split}
\{H,t\}& =1; \\
\{p^{I},x^{J}\}& =-\delta ^{IJ}; \\
\{B^{I},x^{J}\}& =-t\delta ^{IJ}; \\
\{J^{IJ},x^{K}\}& =-\delta ^{JK}x^{I}+\delta ^{IK}x^{J}.
\end{split}
\label{GalST}
\end{equation}

Thus, the non-trivial \textit{finite} transformations of the coordinates $t$%
, $x^{I}$, $H$ and $p^{I}$ of $\mathcal{N}_{d}$ under the action of the
Galilei Lie group $Gal\left( 1,d-1\right) $ (whose Lie algebra is $\mathfrak{%
gal}\left( 1,d-1\right) $) can be grouped into the following classes :

\begin{itemize}
\item Rotations in $d-1$ spatial dimensions\footnote{%
There is no difference between upper and lower $I$-indices, since they are
raised and lowered with the $(d-1)$-dim. Kronecker delta $\delta
^{IJ}=\delta _{IJ}$. We will henceforth restore the Einstein summation
convention on dummy indices.} (generated by $J^{IJ}$, with parameters $%
\Lambda _{IJ}$) :\textbf{\ }
\begin{equation}
\begin{split}
x^{I}& \mapsto e^{\Lambda _{JK}J^{JK}}x^{I}e^{-\Lambda _{JK}J^{JK}}={\left(
\exp \Lambda \right) ^{I}}_{J}x^{J}; \\
p^{I}& \mapsto e^{-\Lambda _{JK}J^{JK}}p^{I}e^{\Lambda _{JK}J^{JK}}={\left(
\exp (-\Lambda )\right) ^{I}}_{J}p^{J}.
\end{split}
\label{rotations}
\end{equation}%
Note that the rotations (and in general, any transformations) act in the
opposite (i.e., inverse, at finite level) way when applied simultaneously to
conjugate variables\footnote{%
Possibly, apart from central charges; see e.g. (\ref{Bar boosts}), in which
the central charge $m$ occurs.}.

\item Translations in $d-1$ spatial dimensions (generated by $p^{I}$, with
parameters $a_{I}$) :
\begin{equation}
x^{I}\mapsto e^{a_{J}p^{J}}x^{I}e^{-a_{J}p^{J}}=x^{I}+a^{I}.
\label{space transl}
\end{equation}

\item Time translations (generated by $H$, with parameter $s$) :
\begin{equation}
t\mapsto e^{sH}te^{-sH}=t+s.  \label{time transl}
\end{equation}

\item Galileian boosts (generated by $B^{I}$ (\ref{def-1}), with parameters $%
\beta _{I}$) :\textbf{\ }
\begin{equation}
\begin{split}
x^{I}& \mapsto e^{\beta _{J}B^{J}}x^{I}e^{-\beta _{J}B^{J}}=x^{I}-t\beta
^{I}; \\
H& \mapsto e^{\beta _{J}B^{J}}He^{-\beta _{J}B^{J}}=H-\beta _{I}p^{I}.
\end{split}
\label{Gal boosts}
\end{equation}
\end{itemize}

\subsubsection{The Bargmann algebra $\mathfrak{bar}_{m}\left( 1,d-1\right) $}

By centrally extending $\mathfrak{gal}\left( 1,d-1\right) $ with the central
charge $m$, one obtains the Lie algebra $\mathfrak{bar}_{m}\left(
1,d-1\right) $, which is therefore generated by $H$, $p^{I}$, $B^{I}$, $%
J^{IJ}$ and $m$, with non-vanishing Poisson brackets given by
\begin{equation}
\begin{split}
\{H,B^{I}\}& =p^{I}; \\
\{p^{I},B^{J}\}& =-m\delta ^{IJ}; \\
\{p^{I},J^{JK}\}& =-\delta ^{IJ}p^{K}+\delta ^{IK}p^{J}; \\
\{B^{I},J^{JK}\}& =-\delta ^{IJ}B^{K}+\delta ^{IK}B^{J}; \\
\{J^{IJ},J^{KL}\}& =\delta ^{IK}J^{JL}+\delta ^{JL}J^{IK}\\
&\qquad\qquad-\delta
^{IL}J^{JK}-\delta ^{JK}J^{IL}.
\end{split}%
\label{PB Barg}
\end{equation}%
Notice the important difference with respect to $\mathfrak{gal}\left(
1,d-1\right) $ : in $\mathfrak{bar}_{m}\left( 1,d-1\right) $, the generators
$p^{I}$ do \textit{not} commute with the generators $B^{I}$ anymore.
Moreover, the $B^{I}$'s of $\mathfrak{bar}_{m}\left( 1,d-1\right) $ coincide
with the generators $L^{+I}$ of $\mathfrak{so}(2,d)$, as given by (\ref%
{GenGalilei}) :
\begin{equation}
B^{I}:=tp^{I}-mx^{I}\overset{\text{(\ref{GenGalilei})}}{=}L^{+I}.
\label{B
Barg}
\end{equation}%
On the other hand, the Poisson brackets of $\mathfrak{bar}_{m}\left(
1,d-1\right) $ with the coordinates $t$ and $x^{I}$ are the same of the ones
of $\mathfrak{gal}\left( 1,d-1\right) $, given by (\ref{GalST}).

Thus, the non-trivial \textit{finite} transformations of the coordinates $t$%
, $x^{I}$, $H$ and $p^{I}$ of $\mathcal{N}_{d}$ under the action of the
Bargmann Lie group $Bar_{m}\left( 1,d-1\right) $ (whose Lie algebra is $%
\mathfrak{bar}_{m}\left( 1,d-1\right) $) are still given by (\ref{rotations}%
), (\ref{space transl}) and (\ref{time transl}), with only the finite boosts
changing : namely, the Galileian boosts (\ref{Gal boosts}) are replaced by
the so-called

\begin{itemize}
\item Bargmann boosts (generated by $B^{I}$ (\ref{B Barg}), with parameters $%
\beta _{I}$) :
\begin{equation}
\begin{split}
x^{I}& \mapsto e^{\beta _{J}B^{J}}x^{I}e^{-\beta _{J}B^{J}}=x^{I}-t\beta
^{I}; \\
p^{I}& \mapsto e^{\beta _{J}B^{J}}p^{I}e^{-\beta _{J}B^{J}}=p^{I}-m\beta
^{I}; \\
H& \mapsto e^{\beta _{J}B^{J}}He^{-\beta _{J}B^{J}}=H-\beta _{I}p^{I}+\frac{1%
}{2}m\beta ^{2},
\end{split}
\label{Bar boosts}
\end{equation}%
where $\beta ^{2}:=\beta _{I}\beta _{J}\delta ^{IJ}\equiv \left\vert \vec{%
\beta}\right\vert ^{2}$. Notice that (\ref{Bar boosts}) consistently reduce
to (\ref{Gal boosts}) for $m\rightarrow 0^{+}$. Moreover, it can be checked
that under finite Bargmann boosts (\ref{Bar boosts}) the physical constraint
(\ref{constr-nr-particle}) stays invariant; in fact, (\ref{Bar boosts})
implies%
\begin{equation}
    \begin{split}
        H&\mapsto H-\vec{\beta}\cdot \vec{p}+\frac{1}{2}m\left\vert \vec{\beta}\right\vert ^{2}; \\
        \vec{p}&\mapsto \vec{p}-m\vec{\beta}\\
        \Rightarrow \frac{\left\vert \vec{p}%
        \right\vert ^{2}}{2m}&\mapsto \frac{1}{2m}\left( \vec{p}-m\vec{\beta}\right)\cdot \left( \vec{p}-m\vec{\beta}\right)\\ &=\frac{\left\vert \vec{p}%
        \right\vert ^{2}}{2m}-\vec{\beta}\cdot \vec{p}+\frac{1}{2}m\left\vert \vec{%
        \beta}\right\vert ^{2},%
    \end{split}
\end{equation}%
such that%
\begin{equation}
H-\frac{\left\vert \vec{p}\right\vert ^{2}}{2m}=0\ \mapsto\ H-\frac{\left\vert
\vec{p}\right\vert ^{2}}{2m}=0.  \label{inv-constr}
\end{equation}%
Since the constraint (\ref{constr-nr-particle}), defining the Hamiltonian of
a non-relativistic particle of mass $m$ in $d-1$ spatial (Euclidean)
dimensions, is manifestly invariant under spatial rotations (\ref{rotations}%
), and space (\ref{space transl}) and time (\ref{time transl}) translations,
the result (\ref{inv-constr}) implies that the Hamiltonian of a
non-relativistic particle of mass $m$ in $d$ Lorentzian spacetime
dimensions is invariant under the whole Bargmann Lie algebra $\mathfrak{bar}%
_{m}\left( 1,d-1\right) $ (group $Bar_{m}\left( 1,d-1\right) $), as mentioned
at the start of Subsection \ref{Gal}.
\end{itemize}

\subsubsection{\label{Sec-bars}The maximal linearly realized algebra: $%
\widehat{\mathfrak{bars}}_{m}(1,d-1)$}

The whole set of infinitesimal transformations of the non-relativistic phase space under $SO(2,d)$ are reported in Subsection \ref{App:NRM} of the Appendix \ref{App:I2}. The only linearly realized transformations read
\begin{eqnarray}
&&%
\begin{array}{l}
\delta_{\Lambda^{(IJ)}}t:=\frac{1}{2}\Lambda _{IJ}\left\{ L^{IJ},t\right\} =0; \\
\delta _{\Lambda^{(IJ)}}x^{K}:=\frac{1}{2}\Lambda _{IJ}\left\{ L^{IJ},x^{K}\right\}
=-\Lambda _{I}^{~K}x^{I}; \\
\delta_{\Lambda^{(IJ)}}H:=-\frac{1}{2}\Lambda _{IJ}\left\{ L^{IJ},H\right\} =0;\\
\delta_{\Lambda^{(IJ)}}p^{K}:=\frac{1}{2}\Lambda _{IJ}\left\{ L^{IJ},p^{K}\right\}
=\Lambda ^{K}_{~I}p^{I};
\end{array}
\label{rot}\\
&&%
\begin{array}{l}
\delta_{\Lambda^{(+-)}}t:=\alpha\left\{ L^{+-},t\right\} = 2\alpha t;\\
\delta_{\Lambda^{(+-)}}x^{I}:=\alpha\left\{ L^{+-},x^{I}\right\} = \alpha x^I;\\
\delta_{\Lambda^{(+-)}}H:=\alpha\left\{ L^{+-},H\right\} = -2\alpha H;\\
\delta_{\Lambda^{(+-)}}p^{I}:=\alpha\left\{ L^{+-},p^{I}\right\} = -\alpha p^I;%
\end{array}
\label{+-} \\
&&%
\begin{array}{l}
\delta_{\Lambda^{(+I)}}t:=b_I\left\{ L^{+I},t\right\} = 0;\\
\delta_{\Lambda^{(+I)}}x^J:=b_I\left\{ L^{+I},x^J\right\} = -b^Jt;\\
\delta_{\Lambda^{(+I)}}H:=b_I\left\{ L^{+I},H\right\} = -b\cdot p;\\
\delta_{\Lambda^{(+I)}}p^{J}:=b_I\left\{ L^{+I},p^{J}\right\} = -mb^J.%
\end{array}
\label{+I}
\end{eqnarray}%

The generators $L^{IJ}$ and $L^{+I}$ of $\mathfrak{so}(2,d)$ (\ref%
{GenGalilei}), respectively generating the rotations $\mathfrak{so}(d-1)$ and
the Bargmann boosts, determine a proper subalgebra of $\mathfrak{%
bar}_{m}\left( 1,d-1\right) $ itself, as enlightened by their Poisson brackets (\ref{Bars Gen}), denoted as $%
\mathfrak{bars}_{m}\left( 1,d-1\right) $:%
\begin{equation}
    \begin{split}
        \mathfrak{bars}_{m}\left( 1,d-1\right) :&=\mathfrak{so}(d-1)\oplus _{s}\mathbb{R}^{(d-1)}\\
        &\subsetneq \mathfrak{bar}_{m}\left( 1,d-1\right),
    \end{split}
\label{deff-1}
\end{equation}%
with $\mathbb{R}^{(d-1)}$ generated by the Bargmann boosts $B^{I}$, $I=1,\dots,d-1$, given in (\ref{B Barg}). More rigorously, $\mathfrak{bars}%
_{m}\left( 1,d-1\right) $ can be defined as the intersection of the Bargmann
algebra $\mathfrak{bar}_{m}\left( 1,d-1\right) $ with the conformal algebra $%
\mathfrak{so}(2,d)$ (\ref{GenGalilei}) :%
\begin{equation}
\mathfrak{bars}_{m}\left( 1,d-1\right) =\mathfrak{bar}_{m}\left(
1,d-1\right) \cap \mathfrak{so}(2,d).  \label{deff-2}
\end{equation}%
In this case, the linearly realized subalgebra of $\mathfrak{so}(2,d)$ is represented by a \textit{conformal extension} of $\mathfrak{bars}_m(1,d)$. Namely,
\begin{align}\label{deff-3}
    \widehat{\mathfrak{bars}}_m(1,d-1):=\mathfrak{bars}_{m}\left(1,d-1\right)\oplus_s\mathfrak{so}(1,1),
\end{align}
where $\mathfrak{so}(1,1)$ is generated by $L^{+-}$, representing a dilatation \footnote{It is worth noticing here that $\mathfrak{so}(1,1)$ commutates with $\mathfrak{so}(2,d)$, so that
\begin{align}
    \widehat{\mathfrak{bars}}_m(1,d-1)=(\mathfrak{so}(d-1)\oplus\mathfrak{so}(1,1))\oplus_s \mathbb{R}^{(d-1)}.
\end{align} }.
Thus, the formula (\ref{red}) specializes to%
\begin{equation}
\mathfrak{so}(2,d)\rightarrow\mathfrak{g}=\widehat{\mathfrak{bars}}_{m}\left( 1,d-1\right).
\label{red-3}
\end{equation}



From the observation on linearity \textit{versus} non-linearity made below (%
\ref{rot})-(\ref{+I}), one concludes that, within the 2T realization
\textit{\`{a} la Bars }\cite{Bars-Emergent}, the \textit{maximal}, manifest
(i.e., \textit{linearly} realized) symmetry of the non-relativistic massive
particle of mass $m$ is the non-compact, non-semisimple Lie algebra $%
\widehat{\mathfrak{bars}}_{m}\left( 1,d-1\right) $, defined by (\ref{deff-3}). All the remaining generators of $\mathfrak{so}(2,d)$
(namely, $L^{-I}$, $L^{+0}$, $L^{-0}$ and $L^{0I}$, which generate
the coset\footnote{$\widehat{Bars}_{m}\left( 1,d-1\right) $ is the Lie group whose Lie
algebra is $\widehat{\mathfrak{bars}}_{m}\left( 1,d-1\right) $.} $SO(2,d)/\widehat{Bars}_{m}%
\left( 1,d-1\right) $) are \textit{non-linearly} realized in this framework.

\subsubsection{Time and space translations}

The remaining set of $d$ generators of $\mathfrak{bar}_{m}\left(
1,d-1\right) $, namely%
\begin{equation*}
H,\left\{ p^{I}\right\} _{I=1,...,d-1}=\mathfrak{bar}_{m}\left( 1,d-1\right)
\ominus \mathfrak{bars}_{m}\left( 1,d-1\right) ,
\end{equation*}
does not belong to $\mathfrak{so}(2,d)$ (within the realization (\ref%
{GenGalilei}), stemming from the coordinatization (\ref{par-4}) of $%
\mathfrak{F}\left( \mathfrak{J}_{1,d-1}\right) $ under consideration). As
reminded above, $H$ and $p^{I}$ generate time and space translations,
respectively. It is interesting to notice that (apart from a $\pm $
branching) such generators appear in the expressions of the generators $%
L^{-0}$ resp. $L^{0I}$ of (\ref{GenGalilei}) as multiplied by the non-linear
factor%
\begin{equation}
    \begin{split}
        \pm X_{1,\pm }^{0}&\overset{\text{(\ref{par-4})}}{=}\left\vert \vec{x}-\frac{t%
        }{m}\vec{p}\right\vert\\
        &\quad=\sqrt{\left( \vec{x}-\frac{t}{m}\vec{p}\right) \cdot
        \left( \vec{x}-\frac{t}{m}\vec{p}\right) },
    \end{split}
\end{equation}
whose non-vanishing Poisson brackets with the coordinates of $\mathcal{N}%
_{d} $ read%
\begin{equation}
\begin{split}
& \left\{ \left\vert \vec{x}-\frac{t}{m}\vec{p}\right\vert ,x^{I}\right\} =%
\frac{t}{m}\frac{x^I-\frac{t}{m}p^I}{\left\vert \vec{x}-\frac{t}{m}\vec{p}\right\vert}; \\
& \left\{ \left\vert \vec{x}-\frac{t}{m}\vec{p}\right\vert ,p^{I}\right\} =%
\frac{x^I-\frac{t}{m}p^I}{\left\vert \vec{x}-\frac{t}{m}\vec{p}\right\vert}; \\
& \left\{ \left\vert \vec{x}-\frac{t}{m}\vec{p}\right\vert ,H\right\} =\frac{%
\vec{p}}{m}\cdot\frac{\vec x-\frac{t}{m}\vec p}{\left\vert \vec{x}-\frac{t}{m}\vec{p}\right\vert}.
\end{split}%
\end{equation}%

\subsubsection{The sign of $\mathcal{I}_{2}$}

As commented below (\ref{I2-nr}), $\mathcal{I}_{2}$ is manifestly invariant
under $\mathfrak{so}(d-1)$, which is the simple, reductive part of the
maximal linearly realized Lie algebra $\mathfrak{bars}_{m}\left(
1,d-1\right) $ given by (\ref{deff-1})-(\ref{deff-2}). On the other hand,
under the finite action of the other generators of $\mathfrak{bars}%
_{m}\left( 1,d-1\right) $, namely under finite Bargmann boosts (\ref{Bar
boosts}), it transforms as follows :
\begin{equation}
\mathcal{I}_{2}\rightarrow \mathcal{I}_{2}+\frac{1}{2}\left(
m^{2}+t^{2}\right) \beta ^{2}-(t\vec{x}+m\vec{p})\cdot \vec{\beta},
\label{tt}
\end{equation}%
and under the finite action of $\mathfrak{so}(1,1)$ transforms as:
\begin{equation}
\mathcal{I}_{2}\rightarrow\frac{e^{2\alpha}t\vec{x}\cdot
\vec{p}+\left( m^{2}-e^{4\alpha}t^{2}\right) e^{-2\alpha}H}{m},
\label{ttu11}
\end{equation}%
where $\alpha$ is the parameter of the transformations under $\mathfrak{so}(1,1)$. Thus, both boosts and dilatations cause a variation in the sign of $\mathcal{I}_2$. Interestingly, the presence of the mass \textit{breaks} the invariance of the sign under $\mathfrak{so}(1,1)$ explicitly. Indeed, if we ignore the mass at the denominator of \eqref{ttu11} and put $m=0$ at the numerator, then $\mathcal{I}_2\rightarrow e^{2\alpha}\mathcal{I}_2$. This behavior has been observed also for the \textit{massive} relativistic particle in Subsection \ref{Sec-rel-m}. Moreover, the \textit{non-relativistic behavior} breaks the boost invariance. As discussed in Appendix \ref{App:I2}, Subsection \ref{App:NRM}, also the other transformations of $SO(2,d)$ transforms the sign of $\mathcal{I}_2$. Therefore, the sign of $\mathcal{I}_2$ remains invariant only under $\mathfrak{h}=\mathfrak{so}(d-1)$.

\subsection{\label{Carroll-E0}Carroll particle with non-vanishing energy}

The Carroll particle has been introduced in 1965 by L´evy-Leblond \cite{LL} and, independently, in 1966 by Sen Gupta \cite{SenGupta:Carroll}. He noticed that applying the Wigner-Inönu contraction \cite{WI} to the Poincarè group, as the limit of the light speed $c$ going to zero, one can obtain a new group, named “Carroll group” as a tribute to Lewis Carroll
books about Alice. It is a sort of \textit{complementary group} to the Galilei group, which arises by sent the light speed to infinite. The Carroll group of transformations is widely used in different areas of physics (see \cite{Duval0,Duval,Duval1,Bergshoeff,Ciambelli,Donnay,Henneaux,Boer,Gomis,Hansen,Campoleoni,FigueroaGalileiCarrollBargmann,Figueroa:Fractons,Boer1,LL1} for some examples). In \cite{KM}, the Carroll massive particle has been studied in the context of the 2T physics. Below, we describe it using the formalism of the Freudenthal triple systems. Before that, let us summarize in the following section the main properties of the massive Carroll particles.

\subsubsection{The Carroll algebra $\mathfrak{carr}(1,d-1)$}
We call the generators of the \textbf{Carrollian transformations} as $%
(H,p^{I},C^{I},J^{IJ})$, where, the only non-trivial Poisson
brackets are
\begin{equation}\label{CarrollGen}
\begin{split}
\{C^{I},p^{J}\}& =H\delta ^{IJ}; \\
\{p^{I},J^{JK}\}& =-\delta ^{IJ}p^{K}+\delta ^{IK}p^{J}; \\
\{C^{I},J^{JK}\}& =-\delta ^{IJ}C^{K}+\delta ^{IK}C^{J}; \\
\{J^{IJ},J^{KL}\}& =\delta ^{IK}J^{JL}+\delta ^{JL}J^{IK}\\
&\qquad\qquad-\delta
^{IL}J^{JK}-\delta ^{JK}J^{IL}.
\end{split}%
\end{equation}%
Let us call the Carroll algebra in $d$-dimensions, generating the Carroll transformations, as
\begin{align}
    \mathfrak{carr}(1,d-1):=\mathfrak{so}(d-1)\oplus_s\mathbb{R}^{(d-1)},
\end{align}
where $\mathbb{R}^{(d-1)}$ is generated by $C^{I}$,
denoted as \textit{Carrollian boosts}, and $\mathfrak{so}(d-1)$ is generated by $J^{IJ}$, defining the algebra of rotations $SO(d-1)$. These generators can be represented as
\begin{equation}
C^{I}=Hx^{I}\qquad \mbox{and}\qquad J^{IJ}=x^{I}p^{J}-x^{J}p^{I}.
\end{equation}%
As usual, $H$ and $p^{I}$ generate the time and space translations. Then,
using the canonical Poisson brackets introduced in \eqref{PSPoisson}, the
non-zero Poisson brackets with the spacetime coordinates read,
\begin{equation}
\begin{split}
\{H,t\}& =1; \\
\{p^{I},x^{J}\}& =-\delta ^{IJ}; \\
\{C^{I},t\}& =p^{I}; \\
\{J^{IJ},x^{K}\}& =-\delta ^{JK}x^{I}+\delta ^{IK}x^{J}.
\end{split}%
\end{equation}%
It is noteworthy that the difference between the Carrollian algebras and the
Galilean algebras (and thus the Bargmann algebra) resides in the boosts.
Indeed, in the latter cases the boost transforms the space coordinates $%
x^{I} $; conversely, the Carroll boosts acts on the time coordinate $t$.

The full transformations of the phase space coordinates result

\begin{itemize}
\item Rotation in $d-1$ dimensions (generated by $J^{IJ}$ with parameters $\Lambda _{IJ}$):
\begin{equation*}
\begin{split}
x^{I}& \rightarrow e^{\Lambda _{JK}J^{JK}}x^{I}e^{-\Lambda _{JK}J^{JK}}={%
\left( \exp \Lambda \right) ^{I}}_{J}x^{J}; \\
p^{I}& \rightarrow e^{\Lambda _{JK}J^{JK}}p^{I}e^{-\Lambda _{JK}J^{JK}}={%
\left( \exp (-\Lambda )\right) ^{I}}_{J}p^{J}
\end{split}%
\end{equation*}

\item Space translations in $d-1$ dimensions (generated by $p^I$, with parameters $a_I$):
\begin{equation*}
x^{I}\rightarrow e^{a_{J}p^{J}}x^{I}e^{-a_{J}p^{J}}=x^{I}+a^{I}.
\end{equation*}%

\item Time translations in $d-1$ dimensions (generated by $H$, with parameter $s$):
\begin{equation*}
t\rightarrow e^{sH}te^{-sH}=t+s.
\end{equation*}

\item Boosts in $d-1$ dimensions (generated by $C^I$, with parameters $\beta_I$):
\begin{equation*}
\begin{split}
p^{I}& \rightarrow e^{\beta _{J}C^{J}}p^{I}e^{-\beta _{J}C^{J}}=p^{I}+H\beta
^{I}; \\
t& \rightarrow e^{\beta _{J}B^{J}}te^{-\beta _{J}B^{J}}=t+\beta _{I}x^{I}.
\end{split}%
\end{equation*}
\end{itemize}

\subsubsection{The Carroll particle in the $\mathfrak{F}\left( \mathfrak{J}_{1,d-1}\right) $ coordinates}

The Carroll particle with non-vanishing energy can be parametrized in $%
\mathfrak{F}\left( \mathfrak{J}_{1,d-1}\right) $ as follows (cf. Table II of
\cite{KM}):
\begin{equation}
    \begin{array}{l}
        X^{+}=E_{0}t, \\ X^{-}=\frac{\vec{p}\cdot
\vec{x}}{E_{0}}+\frac{t\left\vert
\vec{p}\right\vert ^{2}}{2E_{0}}, \\ X^0=\left\vert \vec{x}\right\vert , \\ X^{I}=x^{I}+tp^{I}, \\ [8pt]
        P^{+}=E_{0}, \\ P^{-}=\frac{\left\vert
\vec{p}\right\vert ^{2}}{2E_{0}}, \\ P^{0}=0, \\ P^{I}=p^{I},
    \end{array}
    \label{par-6}
\end{equation}
where $E_0$ is the \textit{rest energy} of the particle, also interpreted as the \textit{rest mass}. This implies that%
\begin{equation}
    \begin{split}
        \mathcal{I}_{2}&=X^{+^{\prime }}X^{-^{\prime }}+P^{+^{\prime }}P^{-^{\prime}}\\
        &=\left( \vec{p}\cdot \vec{x}\right) t+\frac{\left\vert \vec{p}\right\vert ^{2}}{2}\left( t^{2}+1\right) .
    \end{split}
\end{equation}%
Thus, the $SO(2,d)$ generators become
\begin{equation}\label{GenCarroll}
\begin{split}
L^{IJ}& =x^{I}p^{J}-x^{J}p^{I}, \\
L^{0I}& =p^{I}r, \\
L^{+I}& =-E_{0}x^{I} \\
L^{-I}& =-\frac{p^{2}}{2E_{0}}x^{I}+\frac{p\cdot x}{E_{0}}p^{I}, \\
L^{+-}& =-p\cdot x, \\
L^{-0}& =-r\frac{p^{2}}{2E_{0}}, \\
L^{+0}& =-E_{0}r,
\end{split}%
\end{equation}%
where $r=\sqrt{\vec x\cdot\vec x}$.

\subsubsection{\label{Sec-carr}The maximal linearly realized algebra: $%
\widehat{\mathfrak{carr}}(1,d-1)$}

Also in this case, we can compute the infinitesimal transformation of the phase space under $SO(2,d)$ explicitly, which are entirely reported in Subsection \ref{App:Carroll} of the Appendix \ref{App:I2}. The linearly realized transformations are
\begin{eqnarray}
&&%
\begin{array}{l}
\delta _{\Lambda^{(IJ)}}x^{K}:=\frac{1}{2}\Lambda _{IJ}\left\{ L^{IJ},x^{K}\right\}
=-\Lambda _{I}^{~K}x^{I}; \\
\delta_{\Lambda^{(IJ)}}p^{K}:=\frac{1}{2}\Lambda _{IJ}\left\{ L^{IJ},p^{K}\right\}
=\Lambda ^{K}_{~I}p^{I}; \\
\delta_{\Lambda^{(IJ)}}t:=\frac{1}{2}\Lambda _{IJ}\left\{ L^{IJ},t\right\} =0; \\
\delta_{\Lambda^{(IJ)}}H:=-\frac{1}{2}\Lambda _{IJ}\left\{ L^{IJ},H\right\} =0.%
\end{array}
\label{IJc}\\
&&%
\begin{array}{l}
\delta_{\Lambda^{(+-)}}t:=\alpha\left\{ L^{+-},t\right\} = 0;\\
\delta_{\Lambda^{(+-)}}x^{I}:=\alpha\left\{ L^{+-},x^{I}\right\} = \alpha x^I;\\
\delta_{\Lambda^{(+-)}}H:=\alpha\left\{ L^{+-},H\right\} = 0;\\
\delta_{\Lambda^{(+-)}}p^{I}:=\alpha\left\{ L^{+-},p^{I}\right\} = -\alpha p^I;%
\end{array}
\label{+-c} \\
&&%
\begin{array}{l}
\delta_{\Lambda^{(+I)}}t:=b_{I}\left\{ L^{+I},t\right\} = b_{I}x^I;\\
\delta_{\Lambda^{(+I)}}x^{J}:=b_{I}\left\{ L^{+I},x^{J}\right\} = 0;\\
\delta_{\Lambda^{(+I)}}H:=b_{I}\left\{ L^{+I},H\right\} = 0;\\
\delta_{\Lambda^{(+I)}}p^{J}:=b_{I}\left\{ L^{+I},p^{J}\right\} = -Hb^J.%
\end{array}
\label{+Ic}
\end{eqnarray}%

As for the non-relativistic particle of Subsection \ref{nr-massive}, and through the identification $%
E_{0}\leftrightarrow H$, the generators of rotations, boosts and dilatation can be associated to $L^{IJ}$, $L^{+I}$ and $L^{+-}$ of %
\eqref{GenCarroll} respectively (see also \eqref{CarrollGen}), forming the subalgebra $\mathfrak{carrs}(1,d-1)\equiv\mathfrak{so}(2,d)\cap\mathfrak{carr}(1,d-1)$ of $\mathfrak{so}(2,d)$ (see also Appendix \ref{App:I2}, Subsection \ref{App:Carroll}). Once again, the formula (\ref{red}) specializes to%
\begin{equation}
\mathfrak{so}(2,d)\rightarrow\mathfrak{g}=\widehat{\mathfrak{carrs}}\left( 1,d-1\right),
\end{equation}
where $\widehat{\mathfrak{carrs}}(d-1)$ is a \textit{conformal extension} of $\mathfrak{carrs}(d-1)$. Namely,
\begin{align}\label{DefCarrs}\notag
    \widehat{\mathfrak{carrs}}\left( 1,d-1\right):&=(\mathfrak{so}(d-1)\oplus\mathfrak{so}(1,1))\oplus_s\mathbb{R}^{(d-1)}\\
    &=\mathfrak{carrs}\left( 1,d-1\right)\oplus_s\mathfrak{so}(1,1).
\end{align}

\subsubsection{Time and space translations}

In analogy with the non-relativistic particle discussed in Subsection \ref{nr-massive}, the remaining set of $d$ generators of $\mathfrak{carr}\left(
1,d-1\right) $ representing space and time translations, namely%
\begin{equation}
H,\left\{ p^{I}\right\} _{I=1,...,d-1}=\mathfrak{carr}\left( 1,d-1\right)
\ominus \mathfrak{carrs}\left( 1,d-1\right) ,
\end{equation}
does not belong to $\mathfrak{so}(2,d)$. Interestingly enough, as for the non-relativistic particle, such generators appear in the expressions of the generators $%
L^{+0}$ resp. $L^{0I}$ of (\ref{GenCarroll}) as multiplied by the non-linear factor $r$.%


\subsubsection{The sign of $\mathcal{I}_2$}

The quantity $\mathcal{I}_{2}$ is manifestly invariant under rotations. On the other
hand, neither its value nor its sign is invariant under the rest of the $SO(2,d)$ transformations. As an example (see Appendix \ref{App:I2} for a more detailed discussion), it transforms under boosts as
\begin{equation*}
\begin{split}
\mathcal{I}_{2}\rightarrow & \mathcal{I}_{2}+\left[ tH+x\cdot p+(x\cdot
p)^{2}t+x\cdot pH(t^{2}+1)\right] (x\cdot \beta ) \\
&\quad +\left[ H+\frac{(x\cdot p)^{2}}{2}+\frac{H^{2}}{2}(t^{2}+1)+2tx\cdot p%
\right] (x\cdot \beta )^{2}\\
&\quad+\left[ Hx\cdot p+H^{2}t\right] (x\cdot \beta
)^{3}+\frac{H^{2}}{2}(x\cdot \beta )^{4},
\end{split}%
\end{equation*}
and under $\mathfrak{so}(1,1)$ as
\begin{equation*}
\begin{split}
\mathcal{I}_{2}\rightarrow &\left( \vec{p}\cdot \vec{x}\right) t+e^{-2\beta}\frac{\left\vert \vec{p}\right\vert ^{2}}{2} \left( t^{2}+1\right).
\end{split}%
\end{equation*}
Therefore, we can state that the sign of $\mathcal{I}_2$ is invariant only under $\mathfrak{h}=\mathfrak{so}(d-1)$.

In this case, the breaking of the invariance of the sign of $\mathcal{I}_2$ under $\mathfrak{so}(1,1)$ due to presence of a mass $E_0$ is not explicit as for the other massive cases (see Subsections \ref{Sec-relmass} and \ref{nr-massive}). In other words, we cannot restore the invariance under $\mathfrak{so}(1,1)$ by sending $E_0\rightarrow 0$. It could be a consequence of the \textit{intrinsic degeneracy} of the Carrollian particle (and the fact that the mass $E_0$ is also the Hamiltonian).

\subsection{\label{Hydrogen}Hydrogen atom}

In the hydrogen atom in $d$ Lorentzian spacetime dimensions, the components
of the momentum and the radial coordinate are related by the constraint%
\begin{equation}
H=\frac{\left\vert \vec{p}\right\vert ^{2}}{2m}-\frac{\alpha }{r}<0,
\label{H-H}
\end{equation}%
where $r\equiv \left\vert \vec{x}\right\vert $, $\alpha$ is the fine-structure constant, and negative values
of the energy are selected because, for simplicity's sake, we will only
consider bound states. By defining the (linearly time-dependent) function%
\begin{equation}
u(t):=\frac{\sqrt{-2mH}}{m\alpha }\left( \vec{x}\cdot \vec{p}-2mHt\right) ,
\label{u}
\end{equation}%
the hydrogen atom can be coordinatized in the `extended' phase space $%
\mathfrak{F}\left( \mathfrak{J}_{1,d-1}\right) $ as follows (\cite%
{Bars-Emergent}, Table 2):%
\begin{equation}
    \begin{array}{l}
        X^{0^{\prime }}=r\cos u, \\ X^{1^{\prime }}=-\frac{r}{m\alpha }\sqrt{-2mH}\vec{x}\cdot \vec{p}, \\ X^0=r\sin u , \\ X^{I}=x^{I}-\frac{r}{m\alpha }\vec{x}\cdot \vec{p}p^{I}, \\ [8pt]
        P^{0^{\prime }}=-\frac{m\alpha }{r\sqrt{-2mH}}\sin u, \\ P^{1^{\prime }}=\frac{1}{\sqrt{-2mH}}\left( \frac{m\alpha }{r}-\left\vert \vec{p}\right\vert ^{2}\right), \\ P^{0}=\frac{m\alpha }{r\sqrt{-2mH}}\cos u, \\ P^{I}=p^{I},
    \end{array}
    \label{par-5}
\end{equation}
thus implying%
\begin{align}
    \begin{split}
        \mathcal{I}_{2}&=\frac{1}{2} \left( r^{2}-\frac{m\alpha ^{2}}{2Hr^{2}}\right) \cos ^{2}u\\
        &\quad+\frac{2Hr^{2}}{2m\alpha ^{2}}\left( \vec{x}\cdot \vec{p}\right) ^{2}+\frac{\left\vert \vec{p}\right\vert ^{4}}{4mH}-\frac{\alpha \left\vert \vec{p}\right\vert ^{2}}{2Hr}
    \end{split}
\label{I2-H}
\end{align}%
to have \textit{at least} $\mathfrak{so}(d-1)$ as manifest (linearly
realized) symmetry Lie algebra (see further below for details). It is worth noticing here that in this example, the light-cone coordinates are not considered. Instead, the timelike and spacelike indices $0'$ and $1'$ have been used.

\subsubsection{The $SO(2,d)$ symmetry}

Within the coordinatization (\ref{par-5}), the generators $L^{MN}$ (\ref%
{L^MN}) of $\mathfrak{so}(2,d)$ read \cite{Bars-Emergent}
\begin{equation}
\begin{split}
L^{IJ}& =x^{I}p^{J}-x^{J}p^{I}; \\
L^{0I}& =\left( r\sin u+\frac{\vec{x}\cdot \vec{p}}{\sqrt{-2mH}}%
\cos u\right) p^{I}\\
&\qquad\qquad-\frac{m\alpha x^{I}}{r\sqrt{-2mH}}\cos u; \\
L^{0^{\prime }I}& =\left( r\cos u-\frac{\vec{x}\cdot \vec{p}}{\sqrt{%
-2mH}}\sin u\right) p^{I}\\
&\qquad\qquad+\frac{m\alpha x^{I}}{r\sqrt{-2mH}}\sin u;
\\
L^{01^{\prime }}& =\frac{r}{\sqrt{-2mH}}\left( \frac{m\alpha }{%
r}-\left\vert \vec{p}\right\vert ^{2}\right) \sin u\\
&\qquad\qquad+\vec{x}\cdot
\vec{p}\cos u; \\
L^{0^{\prime }1^{\prime }}& =\frac{r}{\sqrt{-2mH}}\left( \frac{%
m\alpha }{r}-\left\vert \vec{p}\right\vert ^{2}\right) \cos u\\
&\qquad\qquad-\vec{x%
}\cdot \vec{p}\sin u; \\
L^{I1^{\prime }}& =\left[ \frac{\sqrt{-2mH}}{m\alpha }-\frac{\left( \frac{%
m\alpha }{r}-\left\vert \vec{p}\right\vert ^{2}\right) }{m\alpha
\sqrt{-2mH}}\right] r\vec{x}\cdot \vec{p}p^{I}\\
&\qquad\qquad+\frac{\left( \frac{%
m\alpha }{r}-\left\vert \vec{p}\right\vert ^{2}\right) }{\sqrt{-2mH}%
}x^{I}; \\
L^{0^{\prime }0}& =\frac{m\alpha }{\sqrt{-2mH}}.
\end{split}
\label{GenH}
\end{equation}%

The expressions (\ref{GenH}) of the generators of the
conformal Lie algebra $\mathfrak{so}(2,d)$ implied by the coordinatization (%
\ref{par-5}) of the hydrogen atom in the `extended' phase space $\mathfrak{F}%
\left( \mathfrak{J}_{1,d-1}\right) $ look quite involved. Nevertheless, it is interesting observing that the full
symmetry of the hydrogen atom physical system actually amounts to the whole
$SO(2,d)$ transformations in $d$ Lorentzian spacetime dimensions \cite{Bars-Terning,Bars-Emergent,Bars-Gauged,Bars:DualityH}. In particular, when $d=4$, accounting for the physical hydrogen atom, the $SO(2,4)$ symmetry is generally known as \textit{the dynamical symmetry of the Hydrogen atom} \cite{Bars:DualityH,Bars-Terning,Maclay:DynamicalSymmetries}. The whole set of the generators of the group does not necessarily commute with the Hamiltonian (which is a characteristic of the dynamical symmetries), but it represents a symmetry of the action \cite{Bars-Terning}. From a more physical point of view, the generators have the following meaning:
\begin{itemize}
    \item $L^{I}=\frac{1}{2}\delta^{IJ}\epsilon_{JMN}L^{MN}$ are the generators of $SO(3)$;
    \item $A^I=\frac{1}{\sqrt{-2mH}}\delta^{IJ}\epsilon_{JMN}p^ML^N$ is the Runge-Lenz vector and describes the orientation and shape of the orbit. It forms (together with ${L}^I$) the $SO(4)$ group;
    \item ${L}^I$ and ${A}^I$ leave the Hamiltonian invariant;
    \item the rest of the generators of $SO(2,4)$ do not commutate with the Hamiltonian and determine the different energy levels of the orbits;
    \item the entire group $SO(2,4)$ describe the \textit{whole} spectrum of the hydrogen atom.
\end{itemize}
In the following section, we propose a $d$-dimensional generalization of the Runge-Lenz vector and the $SO(d)$ symmetry group of the hydrogen atom.

\subsubsection{From $\mathfrak{so}(d-1)$ to $\mathfrak{so}(d)$, and the
Runge-Lenz vector}

It is immediate to realize that
the $L^{IJ}$'s in (\ref{GenH}) generate the Lie algebra $\mathfrak{so}(d-1)$
of rotations in $d-1$ spatial dimensions. This algebra can be enhanced to $%
\mathfrak{so}(d)=$mcs$(\mathfrak{so}(2,d))\ominus \mathfrak{so}(2)$ (which
is the semisimple part of the maximal compact subalgebra, denotes by `mcs',
of $\mathfrak{so}(2,d)$ itself) in the following way :

\begin{enumerate}
\item Dualize the generators $L^{IJ}=L^{[IJ]}=x^{I}p^{J}-x^{J}p^{I}$ of $%
\mathfrak{so}(d-1)$ into the generators%
\begin{align}\label{dualized}
L^{I_{1}...I_{d-3}} :&=\frac{1}{2}\delta ^{I_{1}J_{1}}\delta
^{I_{2}J_{2}}...\delta ^{I_{d-3}J_{d-3}}\\ \notag
&\qquad\times\epsilon
_{J_{1}J_{2}...J_{d-3}J_{d-2}J_{d-1}}L^{J_{d-2}J_{d-1}}  \\ \notag
&=\delta ^{I_{1}J_{1}}\delta ^{I_{2}J_{2}}...\delta
^{I_{d-3}J_{d-3}}\\ \notag
&\qquad\times\epsilon
_{J_{1}J_{2}...J_{d-3}J_{d-2}J_{d-1}}x^{J_{d-2}}p^{J_{d-1}},
\end{align}%
where $\epsilon _{J_{1}J_{2}...J_{d-3}J_{d-2}J_{d-1}}$ is the Levi-Civita
completely anti-symmetric $\mathfrak{so}(d-1)$-invariant tensor of rank $d-1$%
. When $d=4$, (\ref{dualized}) reduces to%
\begin{equation}
L^{I}\equiv J^{I}=\frac{1}{2}\delta ^{IJ}\epsilon
_{KLM}L^{LM}=\delta ^{IJ}\epsilon _{KLM}x^{L}p^{M}.
\end{equation}

\item Define the ($(d-1)$-dim. generalization of the) Runge-Lenz vector :%
\begin{widetext}
\begin{align}
A^{I} :&=\frac{1}{\sqrt{-2mH}}\left({\frac{1}{(d-3)!}} \delta ^{IJ_{1}}\epsilon
_{J_{1}J_{2}...J_{d-3}J_{d-2}J_{d-1}}p^{J_{2}}L^{J_{3}...J_{d-1}}-\frac{%
m\alpha }{r}x^{I}\right)  \notag \\
&\overset{\text{(\ref{dualized})}}{=}-\frac{1}{\sqrt{-2mH}}\notag\\
&\times\left({\frac{1}{(d-3)!}} \delta
^{IJ_{1}}\delta ^{J_{3}K_{3}}\delta ^{J_{4}K_{4}}...\delta
^{J_{d-1}K_{d-1}}\epsilon _{(J_{2}|J_{1}J_{3}J_{4}...J_{d-1}}\epsilon
_{K_{3}K_{4}...K_{d-1}K_{d}|K_{d+1})}x^{K_{d}}p^{J_{2}}p^{K_{d+1}}
+\frac{%
m\alpha }{r}x^{I}\right)  \notag \\
&\overset{\text{(\ref{q2})}}{=}-\frac{1}{\sqrt{-2mH}}\left(
-x^{I}\left\vert \vec{p}\right\vert ^{2}+\vec{x}\cdot \vec{p}p^{I}+\frac{%
m\alpha }{r}x^{I}\right)  \notag \\
&\overset{\text{(\ref{GenH})}}{=}\frac{1}{\sqrt{-2mH}}\left( p_{J}L^{IJ}-%
\frac{m\alpha }{r}x^{I}\right) ,  \label{q1}
\end{align}%
\end{widetext}
which for $d=4$ allows to retrieve the usual expression of the Runge-Lenz
vector in three spatial dimensions:%
\begin{align}
    \begin{split}
        A^{I}&=\frac{1}{\sqrt{-2mH}}\left( \delta ^{IJ}\left( \vec{p%
        }\times \vec{J}\right) _{J}-\frac{m\alpha }{r}x^{I}\right)
        \\
        &\overset{\text{(\ref{dualized})~with~}d=4}{=}-\frac{1}{\sqrt{-2mH}}\\
        &\times\left(
        \delta ^{IJ}\delta ^{LM}\epsilon _{JL(K}\epsilon _{P)MN}p^{K}x^{N}p^{P}+%
        \frac{m\alpha }{r}x^{I}\right)   \\
        &=-\frac{1}{\sqrt{-2mH}}\left( -x^{I}\left\vert \vec{p}\right\vert ^{2}+%
        \vec{x}\cdot \vec{p}p^{I}+\frac{m\alpha }{r}x^{I}\right)  \\
        &\overset{\text{(\ref{q2-d=4})}}{=}\frac{1}{\sqrt{-2mH}}\left( p_{J}L^{IJ}-%
        \frac{m\alpha }{r}x^{I}\right) ,
    \end{split}
\label{q1-d=4}
\end{align}%
where in the first line \textquotedblleft $\times $\textquotedblright\
denotes the vector cross product in three spatial dimensions. In order to
obtain (\ref{q1}) and (\ref{q1-d=4}), we have used the following result%
\begin{align}\label{q2}
&\delta ^{IJ_{1}}\delta ^{J_{3}K_{3}}\delta ^{J_{4}K_{4}}...\delta
^{J_{d-1}K_{d-1}}\\ \notag
&\qquad\quad\times\epsilon _{(J_{2}|J_{1}J_{3}J_{4}...J_{d-1}}\epsilon
_{K_{3}K_{4}...K_{d-1}K_{d}|K_{d+1})}\\ \notag
&={(d-3)!}[-\delta _{K_{d}}^{I}\delta
_{J_{2}K_{d+1}}+\delta _{(K_{d+1}}^{I}\delta _{J_{2})K_{d}}],
\end{align}%
and its specialization for $d=4$:%
\begin{align}\label{q2-d=4}
    \begin{split}
        d=4&\Rightarrow \delta ^{IJ}\delta ^{LM}\epsilon _{JL(K}\epsilon_{P)MN}\\
        &=\delta ^{IJ}\delta ^{LM}\epsilon _{(K|JL}\epsilon _{MN|P)}\\
        &=-\delta_{N}^{I}\delta _{KP}+\delta _{(P}^{I}\delta _{K)N}.
    \end{split}
\end{align}%
It can be proved that $A^{I}$ is in one-to-one correspondence with the $d-1$
generators of the set $\mathfrak{so}(d)\ominus \mathfrak{so}(d-1)$ \cite{Bars:DualityH,Maclay:DynamicalSymmetries}; more specifically, by glancing at (\ref{GenH}), one realizes that%
\begin{equation}
L^{I1^{\prime }}\overset{\text{(\ref{q1})}}{=}-A^{I}.
\end{equation}%
Physically speaking, $A^{I}$ characterizes the shape and orientation of the
orbit in a two-body system \cite{Maclay:DynamicalSymmetries}.

\item Thus, $L^{I_{1}...I_{d-3}}$ (or, equivalently, $L^{IJ}$) and $A^{I}$
generate $\mathfrak{so}(d)$, with non-vanishing Poisson brackets given by%
\begin{align}  \label{2}
\{L_{IJ},L_{KL}\}& =\delta _{IK}L_{JL}+\delta _{JL}L_{IK}\\ \notag
&\qquad\qquad-\delta
_{IL}L_{JK}-\delta _{JK}L_{IL};  \label{1} \\ \notag
\{A_{I},A_{J}\}& ={\frac{p^2-\frac{2m\alpha}{r}}{2mH}}\frac{1}{(d-3!)}\\ \notag
&\qquad\times\epsilon
_{IJK_{1}...K_{d-3}}L^{K_{1}...K_{d-3}}\\
&={\frac{p^2-\frac{2m\alpha}{r}}{2mH}}L_{IJ};\\   \label{3}
\{A_{I},L_{J_{1}...J_{d-3}}\}&=\epsilon _{IJ_{1}...J_{d-3}J_{d-2}}A^{J_{d-2}};
\end{align}%
or, equivalently,
\begin{align*}
    \{A_{I},L_{JK}\}&=\{A_{I},x_{J}p_{K}-x_{K}p_{J}\}\\
    &=-\delta_{IJ}A_K+\delta_{IK}A_J.%
\end{align*}
We may notice that when the constraint $H=p^2/(2m)-\alpha/r$ is used, then the coefficient in \eqref{2} simplifies to 1. In this situation and when $d=4$, the Poisson brackets (\ref{1})-(\ref{3}) reduce to
\begin{align}
\{J_{I},J_{J}\}& =\epsilon _{IJK}J^{K}; \\
\{A_{I},A_{J}\}& =\epsilon _{IJK}J^{K}; \\
\{A_{I},J_{J}\}& =\epsilon _{IJK}A^{K}.
\end{align}
Thus, the generators $J_I$ and $A_J$ forms the algebra of $SO(4)$.

\item By recalling (\ref{H-H}), one can compute%
\begin{align}
\{H,L^{IJ}\}&=\{H,L^{I_{1}...I_{d-3}}\}\\ \notag
&=\{H,A^{I}\}=0,
\end{align}%
implying that the $\mathfrak{so}(d)$ generated by such generators is a
symmetry of the (Hamiltonian of the) hydrogen atom.
\end{enumerate}

\subsubsection{The maximal linearly realized algebra: $\mathfrak{so}\left(
d-1\right)$}
From the whole set of infinitesimal transformations collected in Appendix \ref{App:I2}, Subsection \ref{App:HA}, it is immediate to observe that the linearly realized infinitesimal transformation given by $SO(2,d)$ on the phase space after the gauge fixing are represented by the only set of rotations $L^{IJ}$, which acts as usual:
\begin{eqnarray}
&&%
\begin{array}{l}
\delta _{\Lambda^{(IJ)}}x^{K}:=\frac{1}{2}\Lambda _{IJ}\left\{ L^{IJ},x^{K}\right\}
=-\Lambda _{I}^{~K}x^{I}; \\
\delta_{\Lambda^{(IJ)}}p^{K}:=\frac{1}{2}\Lambda _{IJ}\left\{ L^{IJ},p^{K}\right\}
=\Lambda ^{K}_{~I}p^{I}; \\
\delta_{\Lambda^{(IJ)}}t:=\frac{1}{2}\Lambda _{IJ}\left\{ L^{IJ},t\right\} =0; \\
\delta_{\Lambda^{(IJ)}}H:=-\frac{1}{2}\Lambda _{IJ}\left\{ L^{IJ},H\right\} =0.%
\end{array}
\label{IJH}
\end{eqnarray}%
One can conclude that, within
the 2T realization \textit{\`{a} la Bars }\cite{Bars-Emergent}, the
\textit{maximal}, manifest (i.e., \textit{linearly} realized) symmetry of
the hydrogen atom (with $H<0$) is the Lie algebra $\mathfrak{so}\left(
d-1\right) $, which is also the manifest (linearly realized) symmetry of the
invariant\footnote{%
In this case, the rule of thumb on the symmetry of $\mathcal{I}_{2}$ stated
below (\ref{tt}) is respected once again; in particular, the manifest
symmetry of $\mathcal{I}_{2}$ is once again $\mathfrak{so}(d-1)$.} $\mathcal{%
I}_{2}$ (\ref{I2-H}). Within the above definitions, this implies that (\ref{red}) specializes to%
\begin{equation}
\mathfrak{so}(2,d)\rightarrow\mathfrak{g}=\mathfrak{so}(d-1).
\end{equation}%
All the remaining generators of $\mathfrak{so}(2,d)$ (namely, $L^{0I}$, $%
L^{0^{\prime }I}$, $L^{10^{\prime }}$, $L^{1^{\prime }0^{\prime }}$, $%
L^{I1^{\prime }}$ and $L^{0^{\prime }0}$, which generate the coset $%
SO(2,d)/SO\left( d-1\right) $) are \textit{non-linearly} realized in this
framework.

\subsubsection{The sign of $\mathcal{I}_2$}
As outlined in Subsection \ref{App:HA}, $\mathcal{I}_2$ is manifestly invariant under the group of rotations $SO(d-1)$ generated by $L^{IJ}$, but it is not invariant, nor is its sign, under the rest of the transformations of $SO(2,d)$. Thus, in this case, the sign of $\mathcal{I}_2$ is conserved only by the maximal linearly realized subgroup of $SO(2,d)$, namely by $SO(d-1)$. This result is consistent with analogous ones obtained in the previous examples : also in the other massive, non-relativistic models, the sign of $\mathcal{I}_2$ is invariant (at most) under rotations.

\subsection{Summary of the results} \label{sec:Results}
Table \ref{Tab:summary} summarizes all the results obtained in this section;
\begin{table*}[tbp]
\begin{tabular}{|p{8.3cm}||p{3.6cm}|p{3.6cm}|}
\hline
Gauge choice & $\mathfrak{g}$ & $\mathfrak{h}$ \\
\hline\hline
Massless relativistic particle & $\mathfrak{so}(1,d-1)\otimes_s\mathfrak{so}(1,1)$  & $\mathfrak{so}(1,d-1)\otimes_s\mathfrak{so}(1,1)$ \\
\hline
Massless rel. particle in
maximally symmetric space & $\mathfrak{so}(1,d-1)$  & $\mathfrak{so}(1,d-1)\otimes_s\mathfrak{so}(1,1)$ \\
\hline
Massive relativistic particle & $\mathfrak{so}(1,d-1)$  & $\mathfrak{so}(1,d-1)$ \\
\hline
Massive non-relativistic & $\widehat{\mathfrak{bars}}\left( 1,d-1\right)$  & $\mathfrak{so}(d-1)$ \\
\hline
Carrollian particle & $\widehat{\mathfrak{carrs}}\left( 1,d-1\right)$  & $\mathfrak{so}(d-1)$ \\
\hline
Hydrogen atom & $\mathfrak{so}(d-1)$ & $\mathfrak{so}(d-1)$  \\
\hline
\end{tabular}
\caption{\label{Tab:summary} Various gauge(-fixing) choices in 2T physics yielding different 1T physical models. $\mathfrak{g}$ is the largest proper subalgebra of the conformal Lie algebra in $d$ Lorentzian dimensions, which is still linearly realized after the gauge fixing (and constraints' resolution) procedure. $\mathfrak{h}$ is its (generally proper) subalgebra which is a symmetry of the sign of $\mathfrak{I}_2$. For the definition of the Lie algebras $\widehat{\mathfrak{bars}}$ and $\widehat{\mathfrak{carrs}}$, see \eqref{deff-3} and \eqref{DefCarrs}, respectively.}
\end{table*}
$\mathfrak{g}$ is the linearly realized subalgebra of $\mathfrak{so}(2,d)$, whereas $\mathfrak{h}$ denotes its (generally proper) subalgebra which leaves the sign of $\mathcal{I}_2$ invariant.

Such results provide a substantial hint for the formulation of the following conjectures :

\bigskip

\textbf{Conjecture 1}. In a massless relativistic system, the sign of $\mathcal{I}_2$ is invariant under the whole $SO(1,d-1)\otimes SO(1,1)$ subgroup of $SO(2,d)$, \textit{even when some of its generators are non-linearly realized}. A paradigmatic example is provided by the relativistic massless particle in a maximally symmetric spacetime, in which the linearly realized subgroup of $SO(2,d)$ is only the Lorentz group $SO(1,d-1)$, but the sign of $\mathcal{I}_2$ is invariant under the whole $SO(1,d-1)\otimes SO(1,1)$, where this latter Abelian factor is non-linearly realized.

\bigskip

\textbf{Conjecture 2}. In presence of a non-vanishing mass, the $SO(1,1)$ factor does not belong anymore to the symmetry group of the sign of $\mathcal{I}_2$. Since the \textit{non-relativistic} behavious breaks the Lorentz group to the rotational group only, the resulting symmetry group of the sign of $\mathcal{I}_2$ in a massive non-relativistic model is only $SO(d-1)$.



\section{Conclusions and Outlook}\label{Sec:Conclusions}
In this paper, we unveiled the Jordan and Freudenthal algebraic structures underlying the 2T physics \textit{à la Bars}. The global isometry group $Sp(2,\mathbb{R})\otimes SO(2,d)$ of the extended phase space of the 2T physics is recognized as the automorphism group of the reduced Freudenthal triple system $\mathfrak{F}\left( \mathbb{R}\oplus \Gamma _{1,d-1}\right)$ constructed over the (cubic) Lorentzian spin factor $\mathcal{J}_{1,d-1}\equiv \mathbb{R}\oplus \Gamma _{1,d-1}$ (which is a rank-3 semi-simple Jordan algebra), where $\Gamma _{1,d-1}$ is the rank-2 Jordan algebra given by the Minkowski space and its quadratic norm in $d$ dimensions. Since the automorphism group of a (reduced) FTS is isomorphic to the conformal group of the underlying (cubic) Jordan algebra, one can also regard $Sp(2,\mathbb{R})\otimes SO(2,d)$ as the conformal group of $\mathcal{J}_{1,d-1}$; in this regard, it is worth noticing how the semi-simple, rank-3 nature of the Jordan algebra $\mathcal{J}_{1,d-1}$ entails the enlargement of the conformal group $SO(2,d)$ of $\Gamma _{1,d-1}$ through the commuting factor $Sp(2,\mathbb{R}$, which endows the enlarged phase space (which, as a vector space, is $\mathfrak{F}\left( \mathbb{R}\oplus \Gamma _{1,d-1}\right)$) with a symplectic structure.  The action of $Aut(\mathfrak{F}\left( \mathbb{R}\oplus \Gamma _{1,d-1}\right))$ on the FTS $\mathfrak{F}\left( \mathbb{R}\oplus \Gamma _{1,d-1}\right)$ is non-transitive, and each orbit of the resulting stratification can be defined in terms of algebraic-differential constraints on the unique primitive $Aut(\mathfrak{F}\left( \mathbb{R}\oplus \Gamma _{1,d-1}\right))$-invariant degree-4 homogeneous polynomial $\mathcal{I}_4$ \cite{small_orbits}.

We have shown that each physical model obtained by means of a gauge fixing within the 2T physics \textit{á la Bars} has the coordinates of its enlarged phase space lying into a nilpotent orbit (also known as \textit{small} orbit) with $\mathcal{I}_4=0$, which further splits into two sub-orbits labeled by the sign of a degree-2 homogeneous polynomial $\mathcal{I}_2$ \cite{small_orbits}. Studying in some detail the retrieval of a number of well known physical systems within the 2T paradigm, and highlighting the role of the Freudenthal triple system which can be associated to the enlarged phase space, we observed that the sign of $\mathcal{I}_2$ is generally not invariant under the transformations of the whole Lie group $Sp(2,\mathbb{R})\otimes SO(2,d)$, but instead its symmetry is restricted to a proper (not necessarily semisimple) subgroup thereof, $\mathcal{H}$.

In this framework, we put forward a conjecture on the explicit structure of $\mathcal{H}$ in relation to the general properties of the corresponding physical model. Specifically, we have observed that the structure of $\mathcal{H}$, and thus of its Lie algebra $\mathfrak{h}$, depends on the massive/massless and on the relativistic/non-relativistic nature of the model itself (for further discussion, see Subsec. \ref{sec:Results}). Moreover, for each of the models we studied, we also determined the maximal \textit{linearly realized} subalgebra $\mathfrak{g}$ of the Lie algebra of $Sp(2,\mathbb{R})\otimes SO(2,d)$; interestingly, such a subalgebra $\mathfrak{g}$ (reported in Table \ref{Tab:summary} for the specific models under consideration) is the manifest symmetry algebra of the resulting physical model in the 1T physical world.

\bigskip

This work lays the foundations of a systematic classification of the physical models and related gauge-fixings within the 2T physics, exploiting the invariant structures and stratification of (Jordan and Freudenthal algebraic) triple systems which can be associated to the enlarged phase space of 2T physics. A number of directions and topics is urging us for further investigation, since tantalizing evidences of very interesting results can already be foreseen.

Firstly, we would like to attempt at turning the aforementioned conjecture on the symmetry of (the sign of) $\mathcal{I}_2$ into a theorem, possibly highlighting the role of various non-relativistic limits and contractions (such as the Galilean and Carrollian ones). In relativistic physical models, a \textit{partial} resolution of the constraints (as discussed above \eqref{PartialFix}), fixing only two (out of three) gauge degrees of freedom (and correspondingly solving two - out of three - related constraints) yields the following restriction of the underlying algebraic structures : 

\begin{equation}
\underset{\text{extended~phase~space~\textit{\`{a}~la~Bars}}}{%
\underset{\text{reduced~FTS}}{\mathfrak{F}\left( \mathfrak{J}_{1,d-1}\right)
}}~\rightarrow~\underset{d\text{-dim.~Minkowski~space}}{\underset{\text{rank-2~JA}}{\Gamma
_{1,d-1}}}.
\end{equation}
In this regard, non-relativistic models enjoy a less manifest symmetry, since - at least for the models we investigated in this work - the partial gauge-fixing procedure projects the coordinates of the extended phase space onto to a subspace thereof, involving coordinates that do not pertain to $\Gamma_{1,d-1}$.

Secondly, a deeper investigation is warranted in order to elucidate the structure (and realization) of the Lie algebra $\mathfrak{so}(2,d)$ after the gauge-fixing and the resolution of the corresponding constraints \eqref{gf}. As it can be easily realized by glancing Table \ref{Tab:summary}, space-time translations cannot be included into $\mathfrak{so}(2,d)$ for all the models we considered, but the massless relativistic particle, for which $\mathfrak{so}(2,d)$ is the \textit{conformal} algebra. However, translations are symmetries of the models under consideration (with the exclusion of the hydrogen atom); this would prompt us at including a (translational) non-semisimple component in the isometry algebra of the starting, extended phase space. As far as this point is concerned, we are investigating the possibility to extend the symmetry group of the 2T physics to the semi-direct product $(Sp(2,\mathbb{R})\otimes SO(2,d))\ltimes\mathbb{R}^{2(d+2)+1}$, where $\mathbb{R}^{2(d+2)+1}$ encodes a central extension of the Abelian translational group in $2(d+2)$ dimensions, thereby realizing `generalized translations' in the extended phase space of the 2T physics. A study of the corresponding coadjoint orbits stratification is currently underway.

\textit{Dulcis in fundo}, the (Jordan and Freudenthal) algebraic point of view introduced and investigated in the present work may potentially shed new light onto the issue of quantization, in particular of 2T physics. Within this (not yet thoroughly explored) framework, the non-vanishing of the commutation relations involving $X^M$'s and $P^M$'s, as well as related ordering prescriptions, would play a crucial role \cite{Bars-Gauged,KM}. Intriguingly, $I_4$ (given by \eqref{I4}) would turn out to be non-zero (and expectedly proportional to $\hbar$) also after the gauge-fixing, thus \textit{placing} any quantized system within a non-nilpotent (i.e., generic, open) orbit. As far as the algebraic classification of 1T physical models retrieved within Bars' 2T paradigm is concerned, prior to any gauge-fixing, the `quantum transition' from the relevant \textit{classical} nilpotent orbit (determined in this paper) to the corresponding generic, open \textit{quantum} orbit can be implemented by a suitable transformation of the non-symmetric coset $Sp(4+2d,\mathbb{R})/(Sp(2,\mathbb{R})\otimes SO(2,d))$. We cannot help but remark the potential similarities of such an implementation to the `symplectic deformations' of gauged supergravity, as firstly discussed in \cite{DallAgata:2014tph}. We leave this intriguing topic to further future work.

\section*{Acknowledgments}

This article is based upon work from COST Action CaLISTA CA21109 supported
by COST (European Cooperation in Science and Technology).

F. M. acknowledges the support of the Next Generation EU - Prin 2022 project "Singular Interactions and Effective Models in Mathematical Physics- 2022CHELC7".

\appendix

\section{\label{App-JA-etc}Jordan Algebras and Freudenthal Triple Systems}

\subsection{\label{sec:J}Jordan Algebras}

A Jordan algebra $\mathfrak{J}$ \cite%
{1,2,3,4,5} is a
vector space defined over a ground field $\mathbb{F}$ equipped with a
bilinear product which is commutative but non-associative, namely satisfying
\begin{equation}
\begin{split}
X\circ Y& =Y\circ X, \\
X^{2}\circ (X\circ Y)& =X\circ (X^{2}\circ Y),\quad \forall \ X,Y\in
\mathfrak{J}.
\end{split}
\label{eq:Jid}
\end{equation}%
For the treatment given in the present investigation, the relevant Jordan
algebras are examples of the class of \textit{cubic} Jordan algebras over $%
\mathbb{F}=\mathbb{R}$ \cite{6,7,
8,9}. A cubic Jordan algebra is endowed with a
cubic form $N:\mathfrak{J}\rightarrow \mathbb{R}$, such that $N(\lambda
X)=\lambda ^{3}N(X),\quad \forall \ \lambda \in \mathbb{R},\ X\in \mathfrak{J%
}$. Moreover, an element $c\in \mathfrak{J}$ exists, satisfying $N(c)=1$
(usually named \textit{base point}). A general procedure for constructing
cubic Jordan algebras, due to Freudenthal, Springer and Tits \cite%
{Springer:1962, McCrimmon:1969, McCrimmon}, exists, in which all
properties of the Jordan algebra are determined by the cubic form itself.

Let $V$ be a vector space equipped with a cubic norm $\mathcal{N}%
_{3}:V\rightarrow \mathbb{R}$ such that $\mathcal{N}_{3}(\lambda X)=\lambda
^{3}\mathcal{N}_{3}(X),\ \forall \ \lambda \in \mathbb{R},\ X\in V$, and
with a base point $c\in V$ satisfying $\mathcal{N}_{3}(c)=1$. Then, if the
full linearization of the cubic norm, denoted by $\mathcal{N}_{3}(X,Y,Z)$
and defined as%
\begin{align}
6\mathcal{N}_{3}(X,Y,Z):&=\mathcal{N}_{3}\left( X+Y+Z\right) -\mathcal{N}%
_{3}(X+Y)\\ \notag
&-\mathcal{N}_{3}(Y+Z)-\mathcal{N}_{3}(X+Z)\\ \notag
&+\mathcal{N}_{3}(X)+%
\mathcal{N}_{3}(Y)+\mathcal{N}_{3}(Z),
\end{align}%
is trilinear, the following four maps can be introduced :

\begin{enumerate}
\item The trace
\begin{equation}
Tr(X):=3\mathcal{N}_{3}(c,c,X);  \label{eq:cubicdefs}
\end{equation}

\item A quadratic map
\begin{equation}
S(X):=3\mathcal{N}_{3}(X,X,c),
\end{equation}

\item A bilinear map
\begin{equation}
S(X,Y):=6\mathcal{N}_{3}(X,Y,c),
\end{equation}

\item A trace bilinear form
\begin{equation}
Tr(X,Y)=Tr(X)Tr(Y)-S(X,Y).  \label{eq:tracebilinearform}
\end{equation}
\end{enumerate}

A cubic Jordan algebra $\mathfrak{J}$ with multiplicative identity $Id=c$
can be obtained starting from the vector space $V$ above \textit{iff} $%
\mathcal{N}_{3}$ is \textit{Jordan cubic}, namely \textit{iff}: \textbf{I]}
The trace bilinear form \eqref{eq:tracebilinearform} is non-degenerate, and
\textbf{II]} the quadratic adjoint map, $\sharp \colon \mathfrak{J}%
\rightarrow \mathfrak{J}$, uniquely defined by $Tr(X^{\sharp },Y):=3\mathcal{%
N}_{3}(X,X,Y)$, satisfies

\begin{equation}
(X^{\sharp })^{\sharp }=\mathcal{N}_{3}(X)X,\quad \forall X\in \mathfrak{J}.
\label{eq:Jcubic}
\end{equation}

In a cubic Jordan algebra, the so-called \textit{Jordan product} can be
introduced in the following way :
\begin{align}
X\circ Y:&=\tfrac{1}{2}( X\times Y+Tr(X)Y+Tr(Y)X\\ \notag
&\quad\qquad\qquad\qquad\qquad\qquad-S(X,Y)Id) ,
\end{align}%
where $X\times Y$ denotes the linearization of the quadratic adjoint map :
\begin{equation}
X\times Y:=(X+Y)^{\sharp }-X^{\sharp }-Y^{\sharp }.  \label{eq:FreuProduct}
\end{equation}%
Another related map is the \textit{Jordan triple product} :
\begin{equation}
\{X,Y,Z\}:=(X\circ Y)\circ Z+X\circ (Y\circ Z)-(X\circ Z)\circ Y.
\label{eq:Jtripleproduct}
\end{equation}

Jordan algebras were introduced and completely classified in \cite%
{3} in an attempt to generalize quantum mechanics beyond the
complex numbers $\mathbb{C}$. Below, we list all allowed possibilities of
cubic Jordan algebras \cite%
{4,5,9,McCrimmon:1969,Baez:2001dm}:

\begin{enumerate}
\item the simplest case: $\mathfrak{J}$ $=\mathbb{R}$, $\mathcal{N}%
_{3}(X):=X^{3}$;

\item the infinite sequence of semi-simple Jordan algebras given by $%
\mathfrak{J}=\mathbb{R}\oplus {\Gamma }_{m,n}$ (named \textit{%
pseudo-Euclidean} \textit{spin factors}), where $\Gamma _{m,n}$ is an $%
\left( m+n\right) $-dimensional vector space over $\mathbb{R}$, with a cubic
norm $\mathcal{N}_{3}(X=\xi \oplus \gamma ):=\xi \gamma ^{a}\gamma ^{b}\eta
_{ab}$;

\item Four exceptional\ and simple cases, given by $\mathfrak{J}=J_{3}^{%
\mathbb{A}}$ or $\mathfrak{J}=J_{3}^{\mathbb{A}_{s}}$, the algebra of $%
3\times 3$ Hermitian matrices over the four division algebras $\mathbb{A=R}$
(real numbers)$,\mathbb{C}$ (complex numbers)$,\mathbb{H}$ (quaternions)$,%
\mathbb{O}$ (octonions), or their split versions $\mathbb{A}_{s}=\mathbb{C}%
_{s},\mathbb{H}_{s},\mathbb{O}_{s}$ :
\begin{gather}
X=\left(
\begin{array}{ccc}
\alpha _{1} & x_{3} & \overline{x}_{2} \\
\overline{x}_{3} & \alpha _{2} & x_{1} \\
x_{2} & \overline{x}_{1} & \alpha _{3}%
\end{array}%
\right) ,\\ \notag
\alpha _{1},\alpha _{2},\alpha _{3}\in \mathbb{R}%
,~x_{1},x_{2},x_{3}\in \mathbb{A}\text{ (or }\mathbb{A}_{s}\text{)},
\end{gather}%
with conjugation (denoted by bar) pertaining to the relevant (division or
split) algebra. In these cases, the cubic norm is given by%
\begin{align}
\mathcal{N}_{3}\left( X\right) :&=\alpha _{1}\alpha _{2}\alpha _{3}-\alpha
_{1}x_{1}\overline{x}_{1}\\ \notag
&-\alpha _{2}x_{2}\overline{x}_{2}-\alpha _{3}x_{3}%
\overline{x}_{3}+2Re\left( x_{1}x_{2}x_{3}\right) .
\end{align}%
This reproduces the usual determinant\footnote{%
For explicit constructions of $N(X)$, see \textit{e.g.} \cite{F-Gimon-K} and
\cite{Wissanji}.} for $\mathbb{A}=\mathbb{R\ }$\ and $\mathbb{C}$. In these
cases, the Jordan product simply reads $X\circ Y=\tfrac{1}{2}(XY+YX)$, where
$XY$ is just the conventional $3\times 3$ matrix product. See \textit{e.g.}
\cite{5} for a comprehensive account.
\end{enumerate}

Moreover, we should recall that the following Jordan-algebraic isomorphisms
hold :%
\begin{eqnarray}
J_{2}^{\mathbb{A}} &\cong &\Gamma _{1,q+1}\left( \cong \Gamma
_{q+1,1}\right) ; \\
J_{2}^{\mathbb{A}_{s}} &\cong &\Gamma _{q+2+1,q/2+1},
\end{eqnarray}%
with $q:=$dim$_{R}\mathbb{A}=8,4,2,1$ for $\mathbb{A}=\mathbb{O},\mathbb{H},%
\mathbb{C},\mathbb{R}$, and $q:=$dim$_{R}\mathbb{A}_{s}=8,4,2$ for $\mathbb{A%
}_{s}=\mathbb{O}_{s},\mathbb{H}_{s},\mathbb{C}_{s}$ (see \textit{e.g.} App.
A of \cite{Gun-2} - and Refs. therein - for an introduction to division and
split algebras), implying the following (maximal, rank-preserving)
Jordan-algebraic embeddings :%
\begin{eqnarray}
J_{3}^{\mathbb{A}} &\supset &\mathbb{R}\oplus J_{2}^{\mathbb{A}}\cong
\mathbb{R}\oplus \Gamma_{1,q+1}; \\
J_{3}^{\mathbb{A}_{s}} &\supset &\mathbb{R}\oplus J_{2}^{\mathbb{A}%
_{s}}\cong \mathbb{R}\oplus \Gamma _{q/2+1,q/2+1}.
\end{eqnarray}

\subsection{\label{Symms-JAs}Symmetries of Jordan Algebras}

To each cubic Jordan algebra, a number of symmetry groups can be associated :

\begin{itemize}
\item $Aut(\mathfrak{J})$, the group of automorphisms of $\mathfrak{J}$,
which leaves invariant the structure constants of the Jordan product (the
Lie algebra of $Aut(\mathfrak{J})$ is given by the derivations of $\mathfrak{%
J}$ : $\mathfrak{Aut}(\mathfrak{J})=Der(\mathfrak{J})$).

\item $Str(\mathfrak{J})$, the \textit{structure} group, which leaves the
cubic norm $\mathcal{N}_{3}$ \textit{invariant} up to a rescaling:%
\begin{gather}\notag
\mathcal{N}_{3}(g(X))=\lambda \mathcal{N}_{3}(X),\text{ }\lambda \in \mathbb{%
R},\\
\forall \ g\in Str(\mathfrak{J});
\end{gather}%
the \textit{reduced structure} group $Str_{0}(\mathfrak{J})$ is obtained
from $Str(\mathfrak{J})$ by modding it out by its center \cite{Schafer:1966,
5, Brown:1969}:%
\begin{equation}
\mathcal{N}_{3}(g(X))=\mathcal{N}_{3}(X),\quad \forall \ g\in Str_{0}(%
\mathfrak{J}).
\end{equation}

\item $Conf(\mathfrak{J})$, the \textit{conformal} group, defined as
\begin{gather}\notag
\mathcal{N}_{3}(\sigma (X-X^{\prime }))=f_{\sigma }(X)f_{\sigma }(X^{\prime
})\mathcal{N}_{3}(X-X^{\prime }),\\ \forall \ \sigma \in Conf(\mathfrak{J}%
),
\end{gather}%
where $f$ is a $\sigma $-dependent, real function defined over $\mathfrak{J}$
:consequently, the \textit{cubic light-cone} defined by the vanishing $%
\mathcal{N}_{3}(X-X^{\prime })=0$ of the cubic norm of a difference of
Jordan elements $X$ and $X^{\prime }$ is invariant under $Conf(\mathfrak{J})$%
. This group can also be characterized as the automorphism group of the
Freudenthal triple system defined over $\mathfrak{J}$; see below.

\item $QConf(\mathfrak{J})$, the \textit{quasi-conformal} group, which can
be defined by introducing Freudenthal triple systems and their further
extension named \textit{extended} Freudenthal triple system \cite%
{Gunaydin:2000xr}; see below.
\end{itemize}

All symmetry groups of (simple and semi-simple) cubic Jordan algebras over $%
\mathbb{R}$ are listed\footnote{%
Besides cubic Jordan algebras and their symmetries, another remarkable
\textit{Jordan triple system} is given by $2$-dimensional octonionic
vectors, and denoted by $M_{2,1}\left( \mathbb{O}\right) $. Its relevant
symmetries are $QConf\left( M_{2,1}\left( \mathbb{O}\right) \right) \cong
E_{6(-14)}$, and $Conf\left( M_{2,1}\left( \mathbb{O}\right) \right) \cong
SU(5,1)$ (\textit{cfr.} \cite{Koecher,Loos,Gun-Bars} and \cite{GST,GST2}).} in
Table 1.

\begin{table*}[tbp]
\begin{equation*}
\begin{array}{|c||c|c|c|c|}
\hline
\mathfrak{J} & Aut(\mathfrak{J}) & Str_{0}(\mathfrak{J}) & Conf(\mathfrak{J})
& QConf(\mathfrak{J}) \\ \hline\hline
\mathbb{R} & Id & Id & SL(2,\mathbb{R}) & G_{2(2)} \\ \hline
\mathbb{R}\oplus \Gamma_{m,n} & SO(m)\times SO(n) & SO(m,n) & SL(2,%
\mathbb{R})\times SO(m+1,n+1) & SO(m+3,n+3) \\ \hline
J_{3}^{\mathbb{R}} & SO(3) & SL(3,\mathbb{R}) & Sp(6,\mathbb{R}) & F_{4(4)}
\\ \hline
J_{3}^{\mathbb{C}} & SU(3) & SL(3,\mathbb{C})_{\mathbb{R}} & SU(3,3) &
E_{6(2)} \\ \hline
J_{3}^{\mathbb{C}_{s}} & SL(3,\mathbb{R}) & SL(3,\mathbb{R})\times SL(3,%
\mathbb{R}) & SL(6,\mathbb{R}) & E_{6(6)} \\ \hline
J_{3}^{\mathbb{H}} & USp(6) & SU^{\ast }(6) & SO^{\ast }(12) & E_{7(-5)} \\
\hline
J_{3}^{\mathbb{H}_{s}} & Sp(6,\mathbb{R}) & SL(6,\mathbb{R}) & SO(6,6) &
E_{7(7)} \\ \hline
J_{3}^{\mathbb{O}} & F_{4(-52)} & E_{6(-26)} & E_{7(-25)} & E_{8(-24)} \\
\hline
J_{3}^{\mathbb{O}_{s}} & F_{4(4)} & E_{6(6)} & E_{7(7)} & E_{8(8)} \\ \hline
\end{array}%
\end{equation*}%
\caption{Invariance groups associated to cubic Jordan algebras. The notation
$G(\mathbb{C})_{\mathbb{R}}$ means the group $G(\mathbb{C})$ seen as a real
group}
\end{table*}

The following maximal Lie group embeddings hold (\textit{cfr. e.g.} \cite%
{Squaring-Magic}, and Refs. therein):%
\begin{eqnarray}
Str_{0} &\supset &Aut;  \label{s1} \\
Conf &\supset &Str_{0}\times SO(1,1);  \label{s2} \\
QConf &\supset &Conf\times Sp(2,\mathbb{R}),  \label{s3}
\end{eqnarray}%
and these actually hold for any of the (simple and semi-simple) cubic Jordan
algebras introduced above.

\subsection{\label{sec:F}Freudenthal Triple Systems}

Starting from a cubic Jordan algebra $\mathfrak{J}$, a \textit{Freudenthal
triple system} (FTS) (of reduced type) is defined as the vector space
\begin{equation}
\mathfrak{F}(\mathfrak{J}):=\mathbb{R}\oplus \mathbb{R}\oplus \mathfrak{%
J\oplus J}.
\end{equation}%
An element $\mathbf{x}\in \mathfrak{F}(\mathfrak{J})$ can formally be
written as a \textquotedblleft $2\times 2$ matrix\textquotedblright\ :
\begin{equation}
\mathbf{x}=%
\begin{pmatrix}
x & X \\
Y & y%
\end{pmatrix}%
,\text{\ }x,y\in \mathbb{R},~X,Y\in \mathfrak{J}.
\end{equation}%
An FTS is endowed\footnote{%
It is worth remarking that all the other necessary definitions, such as the
cubic and trace bilinear forms, are inherited from the underlying Jordan
algebra $\mathfrak{J}$.} with a non-degenerate bilinear antisymmetric
quadratic form, a quartic form and a trilinear triple product \cite%
{35,Brown:1969,36, 37, 9} :

\begin{enumerate}
\item Quadratic form $\{\bullet ,\mathbf{\bullet }\}$: $\mathfrak{F}(%
\mathfrak{J})\times \mathfrak{F}(\mathfrak{J})\rightarrow \mathbb{R}$,
defined as
\begin{equation}
\{\mathbf{x},\mathbf{y}\}:=\alpha \delta -\beta \gamma +Tr\left( A,D\right)
-Tr\left( B,C\right) ,  \label{eq:bilinearform}
\end{equation}%
where
\begin{equation}
\mathbf{x}=%
\begin{pmatrix}
\alpha & A \\
B & \beta%
\end{pmatrix}%
,~\mathbf{y}=%
\begin{pmatrix}
\gamma & C \\
D & \delta%
\end{pmatrix}%
.
\end{equation}

\item Quartic form $\Delta :\mathfrak{F}(\mathfrak{J})\rightarrow \mathbb{R}$%
, defined as
\begin{align}
\label{eq:quarticnorm}
\Delta (\mathbf{x}):&=-4\left( \alpha \mathcal{N}_{3}(A)+\beta \mathcal{N}%
_{3}(B)\right.\\ \notag
&\qquad\qquad\left.+\kappa (\mathbf{x})^{2}-Tr(A^{\sharp },B^{\sharp })\right) ,
\end{align}%
where
\begin{equation}
\kappa (\mathbf{x}):=\tfrac{1}{2}(\alpha \beta -Tr(A,B)).
\end{equation}

\item Triple product $T:\mathfrak{F}(\mathfrak{J})\times \mathfrak{F}(%
\mathfrak{J})\times \mathfrak{F}(\mathfrak{J})\rightarrow \mathfrak{F}(%
\mathfrak{J})$, defined as

\begin{equation}
\{T(\mathbf{x},\mathbf{y},\mathbf{w}),\mathbf{z}\}:=2\Delta (\mathbf{x},%
\mathbf{y},\mathbf{w},\mathbf{z}),
\end{equation}%
where $\Delta (\mathbf{x},\mathbf{y},\mathbf{w},\mathbf{z})$ is the full
linearization of $\Delta (\mathbf{x})$, such that $\Delta (\mathbf{x},%
\mathbf{x},\mathbf{x},\mathbf{x})=\Delta (\mathbf{x})$.
\end{enumerate}

The \textit{automorphism} group $Aut(\mathfrak{F}(\mathfrak{J}))$ is defined
as the set of all invertible $\mathbb{R}$-linear transformations which leave
both $\{\mathbf{x},\mathbf{y}\}$ and $\Delta (\mathbf{x},\mathbf{y},\mathbf{w%
},\mathbf{z})$ invariant \cite{Brown:1969}.

It can be proved \cite{Gunaydin:1975mp,Gunaydin:1989dq,Gunaydin:1992zh}
that, as anticipated above :

\begin{equation}
Aut(\mathfrak{F}(\mathfrak{J}))\cong Conf\left( \mathfrak{J}\right) .
\end{equation}

\subsection{\label{EFTS}Extended Freudenthal Triple Systems}

Every simple Lie algebra $\mathfrak{g}$ can be endowed with a $5$-grading,
determined by one of its generators $\Delta $, with one-dimensional $\pm 2$%
-graded subspaces :
\begin{equation}
\mathfrak{g}=\mathfrak{g}^{-2}\oplus \mathfrak{g}^{-1}\oplus \mathfrak{g}%
^{0}\oplus \mathfrak{g}^{+1}\oplus \mathfrak{g}^{+2}\,.
\end{equation}%
where
\begin{eqnarray}
\mathfrak{g}^{0} &=&\mathfrak{h}\oplus \Delta ; \\
\lbrack \Delta ,\mathfrak{t}] &=&m\mathfrak{t}\;\;\;\;\forall \mathfrak{t}%
\in \mathfrak{g}^{m}\;\;,\;m=0,\pm 1,\pm 2
\end{eqnarray}

As firstly discussed in \cite{Gunaydin:2000xr}, a $5$-graded\footnote{%
For $sl(2)$, the $5$-grading degenerates into a $3$-grading.} Lie algebra $%
\mathfrak{g}$ can geometrically be constructed as the \textit{quasi-conformal%
} Lie algebra $\mathfrak{QConf}\left( \mathfrak{J}\right) $ over a vector
space $\mathcal{E}\mathfrak{F}\left( \mathfrak{J}\right) $ coordinatized by $%
\mathcal{X}:=\left( \mathbf{x},\Phi \right) \in \mathcal{E}\mathfrak{F}%
\left( \mathfrak{J}\right) $, where $\mathbf{x}\in \mathfrak{F}\left(
\mathfrak{J}\right) $, and $\Phi $ is an extra real variable \cite%
{Gunaydin:2000xr,GP}:%
\begin{equation*}
\mathcal{E}\mathfrak{F}\left( \mathfrak{J}\right) :=\mathfrak{F}\left(
\mathfrak{J}\right) \oplus \mathbb{R}.
\end{equation*}

Remarkably, a norm $\mathcal{N}:\mathcal{E}\mathfrak{F}\left( \mathfrak{J}%
\right) \rightarrow \mathbb{R}$ can be defined by using the quartic form $%
\Delta $ previously introduced in $\mathfrak{F}\left( \mathfrak{J}\right) $,
as follows\footnote{%
Since the image of $\mathcal{\Delta }$ in $\mathfrak{F}\left( \mathfrak{J}%
\right) $ extends over the whole $\mathbb{R}$, for $\mathcal{\Delta }(%
\mathbf{x})<0$ the lightlike condition $\mathcal{N}(\mathcal{X})=0$ in $%
\mathcal{E}\mathfrak{F}\left( \mathfrak{J}\right) $ does not yield real
solutions for $\Phi $. However, as discussed in \cite{Gunaydin:2000xr}, this
problem can be solved by complexifying the whole $\mathcal{E}\mathfrak{F}%
\left( \mathfrak{J}\right) $ (\textit{i.e.}, by considering $\mathbb{F}=%
\mathbb{C}$ as ground field), thus obtaining a realization of the \textit{%
complexified} Lie algebra $\mathfrak{g}\left( \mathbb{C}\right) $ over $%
\left[ \mathcal{E}\mathfrak{F}\left( \mathfrak{J}\right) \right] _{\mathbb{C}%
}$.} :
\begin{equation}
\mathcal{N}(\mathcal{X}):=\mathcal{\Delta }(\mathbf{x})-\Phi ^{2}.
\label{N-call}
\end{equation}%
Furthermore, a \textquotedblleft quartic distance" $d_{4}:\mathbf{EF}%
\mathfrak{(J)\times }\mathbf{EF}\mathfrak{(J)}\rightarrow \mathbb{R}$
between any two points $\mathcal{X}=\left( \mathbf{x},\Phi \right) $ and $%
\mathcal{Y}:=\left( \mathbf{y},\Psi \right) $ in $\mathbf{EF}\left(
\mathfrak{J}\right) $ can be defined as
\begin{equation}
d_{4}(\mathcal{X},\mathcal{Y}):=\mathcal{\Delta }(\mathbf{x}-\mathbf{y}%
)-\left( \Phi -\Psi +\left\{ \mathbf{x},\mathbf{y}\right\} \right) ^{2},
\label{d4}
\end{equation}%
such that $\mathcal{N}(\mathcal{X})=d_{4}(\mathcal{X},\mathcal{Y}=0)$.

Then, the \textit{quasi-conformal} group $QConf\left( \mathfrak{J}\right) $
over $\mathbf{EF}\left( \mathfrak{J}\right) $ is defined as the set of all
invertible $\mathbb{R}$-linear transformations which leave invariant the
\textquotedblleft quartic light-cone", namely, the geometrical locus defined
by \cite{Gunaydin:2000xr}
\begin{equation}
d_{4}(\mathcal{X},\mathcal{Y})=0,~\forall \left( \mathcal{X},\mathcal{Y}%
\right) \in \left( \mathbf{EF}\left( \mathfrak{J}\right) \right) ^{2},
\end{equation}%
thus yielding that%
\begin{equation}
QConf\left( \mathfrak{J}\right) \cong Aut\left( \mathbf{EF}\left( \mathfrak{J%
}\right) \right) .
\end{equation}

Thus, every $5$-graded Lie algebra $\mathfrak{g}$, geometrically realized as
the \textit{quasi-conformal} Lie algebra $\mathfrak{QConf}\left( \mathfrak{J}%
\right) $ over a vector space $\mathbf{EF}\left( \mathfrak{J}\right) $,
admits a \textit{conformal invariant} given by the norm $\mathcal{N}_3$ (\ref{N-call}).

\section{\label{App-Frob}Frobenius norms and Lorentzian spin factors}

We will now prove that the enhancement from the rank-2 simple Jordan algebra
$\Gamma _{1,d-1}$ to the rank-3 semisimple Jordan algebra $\mathfrak{J}%
_{1,d-1}\equiv \mathbb{R}\oplus $ $\Gamma _{1,d-1}$ admits a Lorentzian
metric over the (d+1)-dimensional algebra itself, namely admits a quadratic
norm of signature $\left( n_{+},n_{-}\right) =(d,1)$ (in the `mostly plus'
choice, which we make here):%
\begin{equation}
\underset{\left( n_{+},n_{-}\right) =(d-1,1)}{\Gamma _{1,d-1}}%
\longrightarrow \underset{(n_{+},n_{-})=(d,1)}{\mathfrak{J}_{1,d-1}}.
\end{equation}
This ultimately explains the naming `Lorentzian spin factor' for $\mathfrak{J%
}_{1,d-1}$, and it has been denoted by `Frob:$\left( 1,d\right) $' in the
lowerscripts of formula (\ref{br}); indeed, the aforementioned Lorentzian
metric on $\mathfrak{J}_{1,d-1}$ is defined in terms of the \textit{%
Frobenius norm} of the (formal) matrix realization of $\mathfrak{J}_{1,d-1}$
itself as a $3\times 3$ matrix algebra (with bilinear product defined as the
symmetric part of the usual matrix product).

\vspace{-.6cm}
\subsection{$\Gamma _{1,d-1}$}

We start from the following (formal) parametrization of a generic element of
a simple, rank-2, formally real Jordan algebra $\Gamma _{1,d-1}$ :%
\begin{equation}
X_{ab}\equiv X:=\left(
\begin{array}{cc}
\phi +z & x^{\hat{I}} \\
x_{\hat{I}} & \phi -z%
\end{array}%
\right) \in \Gamma _{1,d-1},  \label{X}
\end{equation}%
where $\phi ,z\in \mathbb{R}$, and the index ranges are $a,b=1,2$, and $\hat{%
I}=1,...,d-2$. The (square of the) \textit{Frobenius norm} \cite{matrix_norms} of $X$ reads%
\begin{align}
\left\Vert X\right\Vert _{F}^{2} :&=\text{Tr}\left( XX\right) \equiv \left(
XX\right) _{a}^{a}=X_{ab}\delta ^{bc}X_{cd}\delta ^{da}  \notag \\
&=\left( \phi +z\right) ^{2}+\left( \phi -z\right) ^{2}+2\left\vert \vec{x}%
\right\vert ^{2} \notag \\
&=2\phi ^{2}+2z^{2}+2\left\vert \vec{x}\right\vert ^{2},
\label{F}
\end{align}%
where $\left\vert \vec{x}\right\vert ^{2}:=x_{\hat{I}}x^{\hat{I}}$ is the
Euclidean norm in $d-2$ (spatial) dimensions. Since Tr$\left( X\right)
=2\phi $, one can define the following metric over $\Gamma _{1,d-1}$ :%
\begin{align}
g\left( X,X\right) :&=\frac{1}{2}\left( \left\Vert X\right\Vert _{F}^{2}-%
\text{Tr}^{2}\left( X\right) \right)\notag \\
&=\frac{1}{2}\left[ 2\phi
^{2}+2z^{2}+2\left\vert \vec{x}\right\vert ^{2}-\text{Tr}^{2}\left( X\right) %
\right]\notag \\
& =-\phi ^{2}+z^{2}+\left\vert \vec{x}\right\vert ^{2},
\end{align}%
which has a Lorentzian (`mostly plus') signature $\left( n_{+},n_{-}\right)
=\left( d-1,1\right) $.

Since the quadratic norm $\mathcal{N}_{2}$ of $\Gamma _{1,d-1}$ is defined
as the (formal) determinant of the $2\times 2$ (formal) matrix realization
of $\Gamma _{1,d-1}$ itself,
\begin{equation}
\mathcal{N}_{2}\left( X\right) :=\text{det}\left( X\right) =\phi
^{2}-z^{2}-\left\vert \vec{x}\right\vert ^{2},
\end{equation}%
it follows that%
\begin{equation}
g_{-}\left( X,X\right) =-\mathcal{N}_{2}\left( X\right) .
\end{equation}%
In other words, the metric defined by the linearization of $\mathcal{N}_{2}$
is a legitimate Lorentzian metric on $\Gamma _{1,d-1}$, but in the `mostly minus'
choice :%
\begin{align}
g_{\mathcal{N}_{2}}\left( X,Y\right) :&=\frac{1}{2}\left[ \mathcal{N}%
_{2}(X+Y)-\mathcal{N}_{2}(X)-\mathcal{N}_{2}(Y)\right]\notag \\
& =-g_{-}(X,Y),~\forall
X,Y\in \Gamma _{1,d-1}.
\end{align}

\subsection{$\mathfrak{J}_{1,d-1}$}

We start from the following parametrization of the semi-simple, rank-3
Jordan algebra\footnote{%
Note that by setting $\tau =2y$ the treatment of this subsection reduces to
the treatment of the previous one.}:%
\begin{align}\label{X-11}
X_{AB}&\equiv X\qquad (\in \mathbb{R}\oplus \Gamma _{1,d-1}) \\
:&=\left(
\begin{array}{ccc}
\frac{1}{\sqrt{3}}\left( \tau +y\right) +z & x^{\hat{I}} & 0 \\
x_{\hat{I}} & \frac{1}{\sqrt{3}}\left( \tau +y\right) -z & 0 \\
0 & 0 & \frac{1}{\sqrt{3}}\left( \tau -2y\right)%
\end{array}%
\right), \notag
\end{align}%
where, again, $\tau ,y,z\in \mathbb{R}$, and the index ranges are $A,B=1,2,3$%
, and $\hat{I}=1,...,d-2$.

The (square of the) \textit{Frobenius norm} of $X$ (\ref{X-11}) is defined
as follows :%
\begin{align}
\left\Vert X\right\Vert _{F}^{2} :&=\text{Tr}\left( XX\right) \equiv \left(
XX\right) _{A}^{A}=X_{AB}\delta ^{BC}X_{CD}\delta ^{DA}  \notag \\
&=\left[ \frac{1}{\sqrt{3}}\left( \tau +y\right) +z\right] ^{2}+\left[
\frac{1}{\sqrt{3}}\left( \tau +y\right) -z\right] ^{2}\notag \\
&\qquad\qquad+\frac{1}{3}\left(
\tau -2y\right) ^{2}+2\left\vert \vec{x}\right\vert ^{2}\notag \\
&=\frac{2}{3}\left( \tau ^{2}+y^{2}+2\tau y\right) +2z^{2}+\frac{1}{3}\tau
^{2}+\frac{4}{3}y^{2}\notag \\
&\qquad\qquad-\frac{4}{3}\tau y+2\left\vert \vec{x}\right\vert ^{2}
\notag \\
&=\tau ^{2}+2y^{2}+2z^{2}+2\left\vert \vec{x}\right\vert ^{2}.
\end{align}%
Since Tr$\left( X\right) =\sqrt{3}y$, one can define the following metric
over $\mathbb{R}\oplus \Gamma _{1,d-1}$ :%
\begin{align}
g\left( X,X\right) :&=\frac{1}{2}\left( \left\Vert X\right\Vert _{F}^{2}-%
\text{Tr}^{2}\left( X\right) \right)  \notag \\
&=\frac{1}{2}\left[ \tau ^{2}+2y^{2}+2z^{2}+2\left\vert \vec{x}\right\vert
^{2}-\text{Tr}^{2}\left( X\right) \right]\notag \\
&=-\tau ^{2}+y^{2}+z^{2}+\left\vert
\vec{x}\right\vert ^{2},
\end{align}%
which has the desired Lorentzian (`mostly plus') signature $\left(
n_{+},n_{-}\right) =\left( d,1\right) $. This is denoted by `Frob:$\left(
1,d\right) $' in (\ref{br}). $\square $

\section{\label{App:I2}Infinitesimal transformations and the invariance of the $\mathcal{I}_2$ sign}

In this section, the whole set of infinitesimal transformations for each example is reported. These transformations are used to analyze the sign of $\mathcal{I}_2$. It is worth emphasizing that the latter is invariant only when all the \textit{positive} and \textit{negative} terms of $\mathcal{I}_2$ transform in the same way without flipping the sign. It is the case, for example, of the dilatation in the massless relativistic particle (Subsection \ref{Sec-rel}). Indeed,  $\mathcal{I}_2$ splits in a sum of a positive and negative part
\begin{align}
    \mathcal{I}_2=|\vec{x}|^2-x_0^2,
\end{align}
where $|\vec{x}|$ represents the modulus of the spatial part. As outlined in (\ref{I2Massless}), both parts transform in the same may under dilatations, without changing the sign (they are both multiplied by a positive factor $e^{-2\alpha}$). On the other hand, translations transform the negative part ($-x_0^2$) and the positive part ($|\vec{x}|^2$) differently, leading to a possible change in the sign (see equations (\ref{IIMasslessTTransl}) and (\ref{IIMasslessXTransl})).

Here below, the other examples are analyzed.

\subsection{\label{App:RMMS}Relativistic massless particle in maximally symmetric space}
The whole set of infinitesimal transformations of the phase space under $SO(2,d)$ reads
\begin{widetext}
\begin{equation}
\begin{split}
\delta_{\Lambda^{(\mu\nu)}} x^\rho&=\frac{1}{2}\Lambda_{\mu\nu}\{L^{\mu\nu},x^\rho\}=\frac{1}{2}\Lambda_{\mu\nu}(\eta^{\mu\rho}x^\nu-\eta^{\nu\rho}x^\mu);
\\
\delta_{\Lambda^{(+\mu)}} x^\nu&=b_\mu\{L^{+\mu},x^\nu\}=-(1+\sqrt{1-Kx^{2}})b^\nu+K\ b\cdot x\ x^\nu;\\
\delta_{\Lambda^{(-\mu)}} x^\nu&=c_\mu\{L^{-\mu},x^\nu\}=\frac{1}{1+\sqrt{1-Kx^{2}}}\left[-\frac{x^2}{2}c^\nu+c\cdot x x^\nu\right]-\frac{1}{2}\frac{Kx^2\ c\cdot x\ x^\nu}{(1+\sqrt{1-Kx^2})^2};\\
\delta_{\Lambda^{(+-)}} x^\mu&=\alpha\{L^{+-},x^\mu\}=-\alpha\ (1+\sqrt{1-Kx^{2}})\ x^\mu,
\end{split}
\label{CSTrasformationsRMMS}
\end{equation}
\begin{equation}
\begin{split}
\delta_{\Lambda^{(\mu\nu)}} p^\rho&=\frac{1}{2}\Lambda_{\mu\nu}\{L^{\mu\nu},p^\rho\}=\frac{1}{2}\Lambda_{\mu\nu}(\eta^{\mu\rho}p^\nu-\eta^{\nu\rho}p^\mu);\\
\delta_{\Lambda^{(+\mu)}} p^\nu&=b_\mu\{L^{+\mu},p^\nu\}=-\frac{K\ b\cdot p}{\sqrt{1-Kx^{2}}}x^\nu-K\ b\cdot x\ p^\nu-K\ b^\nu x\cdot p;\\
\delta_{\Lambda^{(-\mu)}} p^\nu&=c_\mu\{L^{-\mu},p^\nu\}=\frac{K\ x^\nu}{\sqrt{1-Kx^{2}}(1+\sqrt{1-Kx^{2}})^2}\left(\frac{x^2}{2}c\cdot p-x\cdot p c\cdot x+\frac{K\ x^2x\cdot p c\cdot x}{1+\sqrt{1-Kx^{2}}}\right)\cr
&+\frac{1}{1+\sqrt{1-Kx^{2}}}\left[c\cdot p x^\nu-c\cdot xp^\nu-x\cdot p b^\nu+\frac{2K\ c\cdot x x\cdot p x^\nu + K\ x^2c\cdot xp^\nu+ K\ x^2x\cdot p c^\nu}{2(1+\sqrt{1-Kx^{2}})}\right];\\
\delta_{\Lambda^{(+-)}} p^\mu&=\alpha\{L^{+-},p^\mu\}=\frac{\alpha}{\sqrt{1-Kx^2}}\left[p^\nu-K(x^2p^\nu+x\cdot p)\right],
\end{split}
\label{CSTrasformationsRMMS2}
\end{equation}
\end{widetext}
where we have renamed $b_\mu=\Lambda_{+\mu}$, $c_\mu=\Lambda_{-\mu}$ and $\alpha=\Lambda_{+-}$.

From the infinitesimal transformations, it is possible to deduce the behavior of the sign of $\mathcal{I}_2$. Let us report here the explicit formula of the latter:
\begin{align*}
    \mathcal{I}_2=\frac{x^2}{2}=\frac{-x_0^2+|\vec x|^2}{2},
\end{align*}
where $x_0^2$ pertains to the negative part of $\mathcal{I}_2$ and $|\vec x|^2$ to the positive part. We have already analyzed the rotations and dilatations.

If we consider the transformations generated by $L^{+\mu}$, we can observe that, for $b^2$ small enough, then
\begin{align}
    {x'}^\mu=\left(1+K\ b\cdot x\right)x^\nu-(1+\sqrt{1-Kx^{2}})b^\nu.
\end{align}
Iterating the transformation, one obtains that
\begin{align}
    {x'}^\mu=f_b(x)x^\nu+g_b(x)b^\nu.
\end{align}
Therefore, if $\mathcal{I}_2$ is for instance positive, it is possible to choose $b_0$ big enough in such a way to change the sign of $\mathcal{I}_2$.

The same is obtained from the transformations generated by $L^{-\mu}$.

\subsection{\label{App:RMassive}Relativistic massive particle in
Minkowski space}
The infinitesimal transformations of the phase space under $SO(2,d)$ is given by
\begin{equation}
\begin{split}
\delta_{\Lambda^{(\mu\nu)}} x^\rho&=\frac{1}{2}\Lambda_{\mu\nu}\{L^{\mu\nu},x^\rho\}=\frac{1}{2}\Lambda_{\mu\nu}(\eta^{\mu\rho}x^\nu-\eta^{\nu\rho}x^\mu);
\\
\delta_{\Lambda^{(+\mu)}} x^\nu&=b_\mu\{L^{+\mu},x^\nu\}\\
&=\left[\frac{1-a^2}{2a^3\ x\cdot p}\ b\cdot p+\frac{m^2}{2a^3(x\cdot p)^2}b\cdot x\right] x^\nu\\
&\qquad-\frac{1+a}{2a}b^\nu;
\\
\delta_{\Lambda^{(-\mu)}} x^\nu&=c_\mu\{L^{-\mu},x^\nu\}\\
&=\left[\frac{x^2}{(1+a)^2}\ c\cdot p+\frac{c\cdot x}{a}\right] x^\nu-\frac{ax^2}{1+a}c^\nu;
\\
\delta_{\Lambda^{(+-)}} x^\mu&=\alpha\{L^{+-},x^\mu\}=-\frac{\alpha x^\mu}{a},
\end{split}
\label{CSTrasformationsRMassless}
\end{equation}
\begin{equation}
\begin{split}
\delta_{\Lambda^{(\mu\nu)}} p^\rho&=\frac{1}{2}\Lambda_{\mu\nu}\{L^{\mu\nu},p^\rho\}=\frac{1}{2}\Lambda_{\mu\nu}(\eta^{\mu\rho}p^\nu-\eta^{\nu\rho}p^\mu);
\\
\delta_{\Lambda^{(+\mu)}} p^\nu&=b_\mu\{L^{+\mu},p^\nu\}
\\&=-\frac{m^2}{2a^3(x\cdot p)^2}\left(b\cdot p+\frac{m^2b\cdot x}{x\cdot p}\right)\left(x^\nu-p^\nu\frac{x^2}{x\cdot p}\right)\\
&\qquad-\frac{m^2}{2a(x\cdot p)^2}b\cdot xp^\nu;
\\
\delta_{\Lambda^{(-\mu)}} p^\nu&=c_\mu\{L^{-\mu},p^\nu\}\\
&=\frac{1}{a}\left(c\cdot p+\frac{m^2c\cdot x}{x\cdot p}\right)\left(x^\nu-p^\nu\frac{x^2}{x\cdot p}\right)-ac\cdot xp^\nu;
\\
\delta_{\Lambda^{(+-)}} p^\mu&=\alpha\{L^{+-},p^\mu\}=\alpha ax^\mu+\frac{\alpha m^2}{ax\cdot p}\left(x^\mu-\frac{x^2}{x\cdot p}p^\mu\right),
\end{split}
\label{CSTrasformationsRMassless2}
\end{equation}
where $b_\mu=\Lambda_{+\mu}$, $c_\mu=\Lambda_{-\mu}$ and $\alpha=\Lambda_{+-}$.

The analysis of the sign of $\mathcal{I}_2$ is very similar to the one in Subsection \ref{App:RMMS}. As already stated in Subsections \ref{Sec-relMax}, the only difference is represented by the presence of the mass, which is invariant under each $L^{MN}$. This \textit{breaks} also the invariance under dilatation. Indeed, if we consider $x^2-m^2>0$ (and thus $x^2>0$), then, under dilatation, $x^2\rightarrow e^{-2\alpha}x^2$ as for the massless particle. On the other hand, the negative term $-m^2$ remains invariant. We can choose $\alpha$ big enough to obtain a change in the sign of $\mathcal{I}_2$.

\subsection{\label{App:NRM}Non-relativistic massive particle}
It is immediate to realize that the $L^{IJ}$'s in (\ref{GenGalilei})
generate the Lie algebra $\mathfrak{so}(d-1)$ of rotations in $d-1$ spatial
dimensions\footnote{%
We will henceforth assume no summation on dummy indices, until different
notice.}%
\begin{equation}
\begin{array}{l}
\delta _{\Lambda^{(IJ)}}x^{K}:=\frac{1}{2}\Lambda _{IJ}\left\{ L^{IJ},x^{K}\right\}
=-\Lambda _{I}^{~K}x^{I}; \\
\delta_{\Lambda^{(IJ)}}p^{K}:=\frac{1}{2}\Lambda _{IJ}\left\{ L^{IJ},p^{K}\right\}
=\Lambda ^{K}_{~I}p^{I}; \\
\delta_{\Lambda^{(IJ)}}t:=\frac{1}{2}\Lambda _{IJ}\left\{ L^{IJ},t\right\} =0; \\
\delta_{\Lambda^{(IJ)}}H:=-\frac{1}{2}\Lambda _{IJ}\left\{ L^{IJ},H\right\} =0.%
\end{array}
\label{rotA}
\end{equation}%
Notice that the r.h.s. of (\ref{rotA}) is \textit{linear} in the
coordinates of $\mathcal{N}_{d}$: this implies that $\mathfrak{so}(d-1)$ is
\textit{linearly} realized within the coordinatization (\ref{par-4}) of $%
\mathfrak{F}\left( \mathfrak{J}_{1,d-1}\right) $ yielding to (\ref%
{GenGalilei}).


Analogously, by defining the $\left( d-1\right) $-dim. vector%
\begin{equation}
r^{I}:=x^{I}-\frac{t}{m}p^{I}  \label{r^IA}
\end{equation}%
and its versor\footnote{%
I.e., the corresponding unimodular vector, such that $\left\vert \hat{r}%
\right\vert ^{2}:=\hat{r}\cdot \hat{r}=1$.}%
\begin{equation}
\hat{r}^{I}:=\frac{x^{I}-\frac{t}{m}p^{I}}{\left\vert \vec{x}-\frac{t}{m}%
\vec{p}\right\vert }=\frac{x^{I}-\frac{t}{m}p^{I}}{\sqrt{\left( \vec{x}-%
\frac{t}{m}\vec{p}\right) \cdot \left( \vec{x}-\frac{t}{m}\vec{p}\right) }},
\label{versor rA}
\end{equation}%
one obtains%
\begin{eqnarray}
&&%
\begin{array}{l}
\delta_{\Lambda^{(+-)}}t:=\alpha\left\{ L^{+-},t\right\} = 2\alpha t\\
\delta_{\Lambda^{(+-)}}x^{I}:=\alpha\left\{ L^{+-},x^{I}\right\} = \alpha x^I\\
\delta_{\Lambda^{(+-)}}H:=\alpha\left\{ L^{+-},H\right\} = -2\alpha H\\
\delta_{\Lambda^{(+-)}}p^{I}:=\alpha\left\{ L^{+-},p^{I}\right\} = -\alpha p^I%
\end{array}
\label{+-A} \\
&&%
\begin{array}{l}
\delta_{\Lambda^{(+I)}}t:=b_I\left\{ L^{+I},t\right\} = 0\\
\delta_{\Lambda^{(+I)}}x^J:=b_I\left\{ L^{+I},x^J\right\} = -b^Jt\\
\delta_{\Lambda^{(+I)}}H:=b_I\left\{ L^{+I},H\right\} = -b\cdot p\\
\delta_{\Lambda^{(+I)}}p^{J}:=b_I\left\{ L^{+I},p^{J}\right\} = -mb^J%
\end{array}
\label{+IA} \\
&&%
\begin{array}{l}
\delta_{\Lambda^{(-I)}}t:=c_I\left\{ L^{-I},t\right\} = -\frac{\vec c\cdot\vec p}{m}t-\vec c\cdot\vec x\\
\delta_{\Lambda^{(-I)}}x^{J}:=c_I\left\{ L^{-I},x^{J}\right\} =-c^J\frac{\vec{x}\cdot\vec{p}-Ht}{m}-\frac{c\cdot xp^J}{m}\\
\delta_{\Lambda^{(-I)}}H:=c_I\left\{ L^{-I},H\right\} = \frac{\vec c\cdot\vec p}{m}H\\
\delta_{\Lambda^{(-I)}}p^{J}:=c_I\left\{ L^{-I},p^{J}\right\} =\frac{c\cdot pp^J}{m}-c^JH%
\end{array}
\label{-I} \\
&&%
\begin{array}{l}
\delta_{\Lambda^{(+0)}}t:=b_0\left\{ L^{+0},t\right\} =0; \\
\delta_{\Lambda^{(+0)}}x^{I}:=b_0\left\{ L^{+0},x^{I}\right\} =\mp b_0t\hat{r}^{I}; \\
\delta_{\Lambda^{(+0)}}H:=b_0\left\{ L^{+0},H\right\} =\mp b_0\vec{p%
}\cdot \hat{r}; \\
\delta_{\Lambda^{(+0)}}p^{I}:=b_0\left\{ L^{+0},p^{I}\right\} =\mp b_0m\hat{r}^{I};%
\end{array}
\label{+0} \\
&&%
\begin{array}{l}
\delta_{\Lambda^{(-0)}}t:=c_0\left\{ L^{-0},t\right\} =\mp c_0\left\vert \vec{r}\right\vert ; \\
\delta_{\Lambda^{(-0)}}x^{I}:=c_0\left\{ L^{-0},x^{I}\right\} =\mp c_0\frac{tH}{m}\hat{r}^{I}; \\
\delta_{\Lambda^{(-0)}}H:=c_0\left\{ L^{-0},H\right\} =\mp c_0\frac{%
H}{m}\vec{p}\cdot \hat{r}; \\
\delta_{\Lambda^{(-0)}}p^{I}:=c_0\left\{ L^{-0},p^{I}\right\} =\mp c_0H\hat{r}^{I};%
\end{array}
\label{-0} \\
&&%
\begin{array}{l}
\delta_{\Lambda^{(0I)}}t:=\Lambda _{0I}\left\{ L^{0I},t\right\} =0 \\
\delta_{\Lambda^{(0I)}}x^{J}:=\Lambda _{0I}\left\{ L^{0I},x^{J}\right\} =\mp\ d^{J}\left\vert \vec{r}\right\vert \pm\frac{t\ c\cdot p}{m}\hat{r}%
^{J} ; \\
\delta_{\Lambda^{(0I)}}H:=d_I\left\{ L^{0I},H\right\} =\pm\ \frac{%
d\cdot p}{m}\vec{p}\cdot \hat{r}; \\
\delta_{\Lambda^{(0I)}}p^{J}:=d_I\left\{ L^{0I},p^{J}\right\} =\pm\ d\cdot p\hat{r}^{J}.%
\end{array}
\label{0I}
\end{eqnarray}%
Here, we renamed $b_I=\Lambda_{+I}$, $c_I=\Lambda_{-I}$, $b_0=\Lambda_{+0}$, $c_0=\Lambda_{-0}$, $d_I=\Lambda_{0I}$ and $\alpha=\Lambda_{+-}$. We remark that the r.h.s. of the infinitesimal variations (\ref{rotA}), (%
\ref{+-A}) and (\ref{+IA}) under the action of $L^{IJ}$, $L^{+I}$ and $L^{+-}$  
is
linear in the coordinates of $\mathcal{N}_{d}$, whereas the r.h.s. of the
infinitesimal variations under all other generators of the conformal algebra
$\mathfrak{so}(2,d)$ realized on the coordinatization (\ref{par-4}) is
\textit{not}. This implies that $L^{IJ}$, $L^{+I}$ and $L^{+-}$
constitute the \textit{maximal} set of generators of $\mathfrak{so}(2,d)$
which are linearly realized within the coordinatization (\ref{par-4}) of $%
\mathfrak{F}\left( \mathfrak{J}_{1,d-1}\right) $ yielding to (\ref%
{GenGalilei}). Interestingly, these generators form the subalgebra $%
\widehat{\mathfrak{bars}}_{m}\left( 1,d-1\right) $ of $\mathfrak{so}(2,d)$, which we introduce in Subsec. \ref{Sec-bars}). Indeed, their Poisson brackets read
\begin{align}
    \begin{split}\label{Bars Gen}
        \{L^{IJ},L^{KL}\}& =\delta ^{IK}L^{JL}+\delta ^{JL}L^{IK}\\
        &\qquad-\delta
^{IL}L^{JK}-\delta ^{JK}L^{IL}; \\
        \{L^{IJ},L^{+K}\}&=-\eta^{JK}L^{+I};\\
        \{L^{+I},L^{+-}\}&=-L^{+I};\\
        \{L^{+-},L^{IJ}\}&=0.
    \end{split}
\end{align}

In order to study its sign, let us recall the expression of $\mathcal{I}_2$:
\begin{align*}
\mathcal{I}_{2}=\frac{t\vec{x}\cdot
\vec{p}+\left( m^{2}-t^{2}\right) H}{m} .
\end{align*}%

The transformations under $L^{+I}$ can be computed exactly, giving
\begin{align}
    \begin{split}
        x^I&\rightarrow x^I-b^It;\\
        p^I&\rightarrow p^I-b^Im;\\
        H&\rightarrow H-\vec b\cdot\vec p +mb^2\\
        t&\rightarrow t.
    \end{split}
\end{align}
Therefore,
\begin{align}
    \mathcal{I}_2\rightarrow\mathcal{I}_2-t\vec x\cdot\vec b-{m\vec b\cdot\vec p+m^2b^2},
\end{align}
which changes the sign of $\mathcal{I}_2$ for the right values of $\vec b$.

The other transformations cannot be computed explicitly. Nevertheless, it is possible to check that the two terms composing $\mathcal{I}_2$ transforms differently, leading to a possible change of sign.

Let us consider the transformations generated by $L^{-I}$. In the first order in $c^I$, the first term transforms as
\begin{align}
    t\vec x\cdot\vec p\rightarrow t\vec x\cdot\vec p &-t\left(H\ \vec c\cdot\vec r+\vec c\cdot\vec x\ \frac{ |\vec p|^2}{m}\right)\notag\\
    &-\vec x\cdot\vec p\left(\vec c\cdot\vec x+t\frac{\vec c\cdot\vec p}{m}\right),
\end{align}
whereas the second,
\begin{align}
    (m^2-t^2)H\rightarrow(m^2-t^2)H&+2tH\vec c\cdot\vec x\notag\\
    &+(m^2+t^2)\frac{\vec c\cdot\vec p}{m}H.
\end{align}
Iterating the process, we can conclude that
\begin{align}
    \begin{split}
        t\vec x\cdot\vec p&\rightarrow f_c(t,H,\vec x,\vec p)\ t\vec x\cdot\vec p+g_c(t,H,\vec x,\vec p),
        \\
        (m^2-t^2)H&\rightarrow [m^2-h_c^2(t,H,\vec x,\vec p)\ t^2]\ \hat{f}_c(t,H,\vec x,\vec p)\ H\\
        &\qquad\qquad+\hat{g}_c(t,H,\vec x,\vec p).
    \end{split}
\end{align}
Thus, it is possible to choose an appropriate value of $\vec c$ that changes the sign of $\mathcal{I}_2$.

With long but straightforward computations, one can verify that the same happens also for the other transformations.

\subsection{\label{App:Carroll}The massive Carroll particle}
The infinitesimal transformations of the Carroll particle's phase space under $SO(2,d)$ transformations are
\begin{eqnarray}
&&%
\begin{array}{l}
\delta _{\Lambda^{(IJ)}}x^{K}:=\frac{1}{2}\Lambda _{IJ}\left\{ L^{IJ},x^{K}\right\}
=-\Lambda _{I}^{~K}x^{I}; \\
\delta_{\Lambda^{(IJ)}}p^{K}:=\frac{1}{2}\Lambda _{IJ}\left\{ L^{IJ},p^{K}\right\}
=\Lambda ^{K}_{~I}p^{I}; \\
\delta_{\Lambda^{(IJ)}}t:=\frac{1}{2}\Lambda _{IJ}\left\{ L^{IJ},t\right\} =0; \\
\delta_{\Lambda^{(IJ)}}H:=-\frac{1}{2}\Lambda _{IJ}\left\{ L^{IJ},H\right\} =0.%
\end{array}
\label{IJcA}\\
&&%
\begin{array}{l}
\delta_{\Lambda^{(+-)}}t:=\alpha\left\{ L^{+-},t\right\} = 0\\
\delta_{\Lambda^{(+-)}}x^{I}:=\alpha\left\{ L^{+-},x^{I}\right\} = \alpha x^I\\
\delta_{\Lambda^{(+-)}}H:=\alpha\left\{ L^{+-},H\right\} = 0\\
\delta_{\Lambda^{(+-)}}p^{I}:=\alpha\left\{ L^{+-},p^{I}\right\} = -\alpha p^I%
\end{array}
\label{+-cA} \\
&&%
\begin{array}{l}
\delta_{\Lambda^{(+I)}}t:=b_{I}\left\{ L^{+I},t\right\} = b_{I}x^I\\
\delta_{\Lambda^{(+I)}}x^{J}:=b_{I}\left\{ L^{+I},x^{J}\right\} = 0\\
\delta_{\Lambda^{(+I)}}H:=b_{I}\left\{ L^{+I},H\right\} = 0\\
\delta_{\Lambda^{(+I)}}p^{J}:=b_{I}\left\{ L^{+I},p^{J}\right\} = -Hb^J%
\end{array}
\label{+IcA} \\
&&%
\begin{array}{l}
\delta_{\Lambda^{(-I)}}t:=c_{I}\left\{ L^{-I},t\right\} = c_{I}\left(-\frac{p^I}{2H}x^I+\frac{\vec x\cdot\vec p}{H^2}p^I\right)\\
\delta_{\Lambda^{(-I)}}x^{J}:=c_{I}\left\{ L^{-I},x^{J}\right\} = c_{I}\frac{x^Ip^J-x^Jp^I}{H}\\
\delta_{\Lambda^{(-I)}}H:=c_{I}\left\{ L^{-I},H\right\} = 0\\
\delta_{\Lambda^{(-I)}}p^{J}:=c_{I}\left\{ L^{-I},p^{J}\right\} =c_{I}\left(-\frac{p^2}{2H}\delta^{IJ}+\frac{p^Ip^J}{H^2}\right)%
\end{array}
\label{-Ic} \\
&&%
\begin{array}{l}
\delta_{\Lambda^{(+0)}}t:=b_{0}\left\{ L^{+0},t\right\} =b_{0}r; \\
\delta_{\Lambda^{(+0)}}x^{I}:=b_{0}\left\{ L^{+0},x^{I}\right\} =0; \\
\delta_{\Lambda^{(+0)}}H:=b_{0}\left\{ L^{+0},H\right\} =0; \\
\delta_{\Lambda^{(+0)}}p^{I}:=b_{0}\left\{ L^{+0},p^{I}\right\} =- b_{0}H\hat{r}^{I};%
\end{array}
\label{+0c} \\
&&%
\begin{array}{l}
\delta_{\Lambda^{(-0)}}t:=c_{0}\left\{ L^{-0},t\right\} = -c_{0}\frac{rp^2}{2H^2} ; \\
\delta_{\Lambda^{(-0)}}x^{I}:=c_{0}\left\{ L^{-0},x^{I}\right\} =c_{0}r\frac{p^I}{H}; \\
\delta_{\Lambda^{(-0)}}H:=c_{0}\left\{ L^{-0},H\right\} =0; \\
\delta_{\Lambda^{(-0)}}p^{I}:=c_{0}\left\{ L^{-0},p^{I}\right\} =-c_{0}H\hat{r}^{I};%
\end{array}
\label{-0c} \\
&&%
\begin{array}{l}
\delta_{\Lambda^{(0I)}}t:=d_{I}\left\{ L^{0I},t\right\} =0 \\
\delta_{\Lambda^{(0I)}}x^{J}:=d_{I}\left\{ L^{0I},x^{J}\right\} =-d_{I}\hat{r}\delta^{IJ}; \\
\delta_{\Lambda^{(0I)}}H:=d_{I}\left\{ L^{0I},H\right\} =0; \\
\delta_{\Lambda^{(0I)}}p^{J}:=d_{I}\left\{ L^{0I},p^{J}\right\} =d_{I}p^I\hat{r}^{J},%
\end{array}
\label{0Ic}
\end{eqnarray}%
where, once again, $b_I=\Lambda_{+I}$, $c_I=\Lambda_{-I}$, $b_0=\Lambda_{+0}$, $c_0=\Lambda_{-0}$, $d_I=\Lambda_{0I}$ and $\alpha=\Lambda_{+-}$. Here, $r=\sqrt{\vec x\cdot\vec x}$ and $\hat{r}^I={x^I}/{r}$. From these relations, it is evident that the linearly realized subgroup of $SO(2,d)$ is generated by $L^{IJ}$, $L^{+-}$ and $L^{+I}$. The Poisson brackets of these generators are always of the form (and can be calculated explicitly with) \eqref{conf}. In particular,
\begin{align*}\label{PoissonCarroll}
\begin{split}
    \{L^{IJ},L^{KL}\}& =\eta^{IK}L^{JL}+\eta^{JL}L^{IK}-\eta^{IL}L^{JK}-\eta^{JK}L^{IL}; \\
    \{L^{+I},L^{+-}\}& =L^{I+}= -L^{+I}; \\
    \{L^{+-},L^{IJ}\}& =0; \\
    \{L^{+I},L^{+J}\}& =0.
\end{split}%
\end{align*}

The analysis of this case in similar to the one of the non-relativistic particle of Subsection \ref{App:NRM}. We remember that
\begin{equation*}
\mathcal{I}_{2}=\left( \vec{p}\cdot \vec{x}\right) t+\frac{%
\left\vert \vec{p}\right\vert ^{2}}{2}\left( t^{2}+1\right) .
\end{equation*}%

The transformation generated by $L^{+I}$ can be computed exactly:
\begin{align}
    \begin{split}
        t&\rightarrow t+\vec b\cdot\vec x,\\
        p^I&\rightarrow p^I- b^IH.
    \end{split}
\end{align}
The other quantities are left invariant. Therefore,
\begin{equation}
    \begin{split}
        \mathcal{I}_{2}\rightarrow\mathcal{I}_{2}&+ (\vec b\cdot\vec x)\left(\vec x\cdot\vec p-Ht+|\vec p|^2\right)\\
        &-H(\vec b\cdot\vec p)\left(1+t^2\right)-(\vec b\cdot\vec x)^2\left(H-|\vec p|^2\right)\\
        &-2\vec b\cdot\vec x\ \vec b\cdot\vec p\ Ht+\frac{b^2H^2}{2}(1+t^2)\\
        &+b^2\ \vec b\cdot\vec x\ H^2t-(\vec b\cdot\vec x)^2\ \vec b\cdot\vec p\ H\\
        &+\frac{1}{2}b^2(\vec b\cdot\vec x)^2H^2.
    \end{split}
\end{equation}%
An exact result similar to the latter can also be easily obtained from the transformation generated by $L^{+0}$. For the other transformations, it is possible to follow the passages outlined in Subsection \ref{App:NRM} for the non-relativistic particle. In each case, the sign of $\mathcal{I}_2$ is not conserved.

\subsection{\label{App:HA}Hydrogen atom}
We can define the whole set of infinitesimal transformation given by $SO(2,d)$ on the hydrogen atom phase space. The transformations generated by $L^{IJ}$ represent the usual set of rotations:
\begin{eqnarray}
&&%
\begin{array}{l}
\delta _{\Lambda^{(IJ)}}x^{K}:=\frac{1}{2}\Lambda _{IJ}\left\{ L^{IJ},x^{K}\right\}
=-\Lambda _{I}^{~K}x^{I}; \\
\delta_{\Lambda^{(IJ)}}p^{K}:=\frac{1}{2}\Lambda _{IJ}\left\{ L^{IJ},p^{K}\right\}
=\Lambda ^{K}_{~I}p^{I}; \\
\delta_{\Lambda^{(IJ)}}t:=\frac{1}{2}\Lambda _{IJ}\left\{ L^{IJ},t\right\} =0; \\
\delta_{\Lambda^{(IJ)}}H:=-\frac{1}{2}\Lambda _{IJ}\left\{ L^{IJ},H\right\} =0.%
\end{array}
\label{IJH-2}
\end{eqnarray}
Then, $L^{00'}$ generates a transformation only along $t$:
\begin{eqnarray}
&&%
\begin{array}{l}
\delta_{\Lambda^{(00')}}t:=\Lambda _{00'}\left\{ L^{00'},t\right\} = \Lambda _{00'} \frac{m^2\alpha}{(-2mH)^{\frac 32}};\\
\delta_{\Lambda^{(00')}}x^{I}:=\Lambda _{00'}\left\{ L^{00'},x^{I}\right\} = 0;\\
\delta_{\Lambda^{(00')}}H:=\Lambda _{00'}\left\{ L^{00'},H\right\} = 0;\\
\delta_{\Lambda^{(00')}}p^{I}:=\Lambda _{00'}\left\{ L^{00'},p^{I}\right\} = 0;%
\end{array}
\label{00'}
\end{eqnarray}

The remaining transformations have a very complicated form in this parameterization, so we refrain from presenting them here. All in all, one obtains that only rotations are linearly realized.

In order to study the sign of $\mathcal{I}_2$ for the hydrogen atom, let us remind its expression:
\begin{align*}
\mathcal{I}_{2}&=\frac{1}{2} \left( r^{2}-\frac{m\alpha ^{2}}{2Hr%
^{2}}\right) \cos ^{2}u+\frac{2Hr^{2}}{2m\alpha ^{2}}\left( \vec{x}%
\cdot \vec{p}\right) ^{2}\\
&\qquad\qquad+\frac{\left\vert \vec{p}\right\vert ^{4}}{4mH}-%
\frac{\alpha \left\vert \vec{p}\right\vert ^{2}}{2Hr}.
\label{I2-HH}
\end{align*}%
It is manifestly invariant under $L^{IJ}$-transformations (namely, the only linearly realized transformations after the gauge fixing). Nevertheless, following the passages from Subsection \ref{App:RMMS} onwards, it is not invariant under the rest of the $SO(2,d)$ transformations and neither its sign is unchanged. A straightforward example is represented by the transformation generated by $L^{00'}$. It represents a \textit{translation} of the time $t$. Namely,
\begin{align}
    t\rightarrow t+\Lambda_{00'}\frac{m^2\alpha}{(-2mH)^{\frac 32}}.
\end{align}
The time $t$ only appear in the function $u$ (see the expression \eqref{u}). Thus, if, for instance, the first term of $\mathcal{I}_2$ is negative, leading to a negative expression of $\mathcal{I}_2$, then we can choose $\Lambda_{00'}$ in such a way that the first term is small enough to switch the $\mathcal{I}_2$ sign. With long computations, it is possible to verify that this situation is valid also for the rest of the transformations.
\vspace{2cm}


\bibliographystyle{apsrev4-2}
\bibliography{2TBib}

@article{Bars-Emergent,
    author = "Bars, Itzhak and Chen, Shih-Hung and Quelin, Guillaume",
    title = "{Dual field theories in (d-1)+1 emergent spacetimes from A unifying field theory in d+2 spacetime}",
    reportNumber = "USC-07-HEP-B4",
    doi = "10.1103/PhysRevD.76.065016",
    journal = "Phys. Rev. D",
    volume = "76",
    pages = "065016",
    year = "2007"
}

@article{KAC1980190,
title = {Some remarks on nilpotent orbits},
journal = {Journal of Algebra},
volume = {64},
number = {1},
pages = {190-213},
year = {1980},
issn = {0021-8693},
doi = {https://doi.org/10.1016/0021-8693(80)90141-6},
url = {https://www.sciencedirect.com/science/article/pii/0021869380901416},
author = {V.G. Kac}
}

@article{vinberg1976weyl,
  title={The Weyl group of a graded Lie algebra},
  author={Vinberg, E B},
  journal={Mathematics of the USSR-Izvestiya},
  volume={10},
  number={3},
  pages={463},
  year={1976},
  publisher={IOP Publishing}
}

@article{1,
  author  = {Jordan, P.},
  title   = {{\"U}ber die multiplikation quanten-mechanischer grossen},
  doi = "https://doi.org/10.1007/BF01333854",
  journal = {Zschr. f. Phys.},
  volume  = {80},
  year    = {1933},
  pages   = {285}
}

@article{Sato_Kimura_1977, 
title={A classification of irreducible prehomogeneous vector spaces and their relative invariants}, 
volume={65}, 
DOI={10.1017/S0027763000017633}, 
journal={Nagoya Mathematical Journal}, 
author={Sato, M. and Kimura, T.}, 
year={1977}, 
pages={1–155}
}

@article{DallAgata:2014tph,
    author = "Dall'Agata, Gianguido and Inverso, Gianluca and Marrani, Alessio",
    title = "{Symplectic Deformations of Gauged Maximal Supergravity}",
    eprint = "1405.2437",
    archivePrefix = "arXiv",
    primaryClass = "hep-th",
    reportNumber = "DFPD-2014-TH-06, NIKHEF-2014-010",
    doi = "10.1007/JHEP07(2014)133",
    journal = "JHEP",
    volume = "07",
    pages = "133",
    year = "2014"
}

@article{2,
  author  = {Jordan, P.},
  title   = {{\"U}ber verallgemeinerungsm{\"o}glichkeiten des formalismus der quantenmechanik},
  journal = {Nachr. Ges. Wiss. Gottingen},
  year    = {1933},
  pages   = {209--214}
}

@article{3,
    author = "Jordan, Pascual and von Neumann, J. and Wigner, Eugene P.",
    title = "{On an Algebraic generalization of the quantum mechanical formalism}",
    doi = "10.2307/1968117",
    journal = "Annals Math.",
    volume = "35",
    pages = "29--64",
    year = "1934"
}

@article{4,
  author  = {Jacobson, N.},
  title   = {Some groups of transformations defined by {Jordan} algebras},
  doi = "https://doi.org/10.1007/978-1-4612-3694-8_27",
  journal = {J. Reine Angew. Math.},
  volume  = {207},
  year    = {1961},
  pages   = {61--85}
}

@book{5,
  author    = {Jacobson, N.},
  title     = {Structure and Representations of {Jordan} Algebras},
  series    = {American Mathematical Society Colloquium Publications},
  volume    = {39},
  publisher = {American Mathematical Society},
  year      = {1968}
}

@article{small_orbits_maths,
    author = "Borsten, L. and Duff, M. J. and Ferrara, S. and Marrani, A. and Rubens, W.",
    title = "{Explicit Orbit Classification of Reducible Jordan Algebras and Freudenthal Triple Systems}",
    reportNumber = "IMPERIAL-TP-2011-MJD-3, CERN-PH-TH-2011-192",
    doi = "10.1007/s00220-013-1846-3",
    journal = "Commun. Math. Phys.",
    volume = "325",
    pages = "17--39",
    year = "2014"
}

@article{6,
  author  = {Elkies, N. and Gross, B. H.},
  title   = {The exceptional cone and the {Leech} lattice},
  doi = "https://doi.org/10.1155/S1073792896000426",
  journal = {Internat. Math. Res. Notices},
  volume  = {14},
  year    = {1996},
  pages   = {665--698}
}

@article{7,
  author  = {Gross, B. H.},
  title   = {Groups over $\mathbb{Z}$},
  doi = "https://doi.org/10.1007/s002220050053",
  journal = {Invent. Math.},
  volume  = {124},
  year    = {1996},
  pages   = {263--279}
}

@article{8,
  author  = {Krutelevich, S.},
  title   = {On a canonical form of a $3\times 3$ {Hermitian} matrix over the ring of integral split octonions},
  journal = {J. Algebra},
  volume  = {253},
  number  = {2},
  year    = {2002},
  pages   = {276--295}
}

@article{9,
  author  = {Krutelevich, S.},
  title   = {{Jordan} algebras, exceptional groups, and {Bhargava} composition},
  doi = "https://doi.org/10.1016/s0021-8693(02)00127-8",
  journal = {J. Algebra},
  volume  = {314},
  number  = {2},
  year    = {2007},
  pages   = {924--977}
}

@article{GP,
    author = "G{\"u}naydin, Murat and Pavlyk, Oleksandr",
    title = "{Generalized spacetimes defined by cubic forms and the minimal unitary realizations of their quasiconformal groups}",
    doi = "10.1088/1126-6708/2005/08/101",
    journal = "JHEP",
    volume = "08",
    pages = "101",
    year = "2005"
}

@book{McCrimmon,
  author    = {McCrimmon, K.},
  title     = {A Taste of {Jordan} Algebras},
  publisher = {Springer-Verlag New York Inc.},
  address   = {New York},
  year      = {2004}
}

@article{Brown,
  author  = {Brown, R. B.},
  title   = {Groups of type $E_7$},
  url = "http://eudml.org/doc/150932",
  journal = {J. Reine Angew. Math.},
  volume  = {236},
  year    = {1969},
  pages   = {79--102}
}

@article{35,
  author  = {Freudenthal, H.},
  title   = {Beziehungen der $E_7$ und $E_8$ zur oktavenebene I-II},
  doi = "https://doi.org/10.1016/S1385-7258%2854%2950032-6",
  journal = {Nederl. Akad. Wetensch. Proc. Ser.},
  volume  = {57},
  year    = {1954},
  pages   = {218--230}
}

@article{36,
  author  = {Faulkner, J. R.},
  title   = {A Construction of {Lie} Algebras from a Class of Ternary Algebras},
  doi = "https://doi.org/10.2307/1995694",
  journal = {Trans. Amer. Math. Soc.},
  volume  = {155},
  number  = {2},
  year    = {1971},
  pages   = {397--408}
}

@article{37,
  author  = {Ferrar, C. J.},
  title   = {Strictly Regular Elements in {Freudenthal} Triple Systems},
  doi = "https://www.jstor.org/stable/1996111",
  journal = {Trans. Amer. Math. Soc.},
  volume  = {174},
  year    = {1972},
  pages   = {313--331}
}

@article{ADM,
    author = "Cerchiai, Bianca L. and Ferrara, Sergio and Marrani, Alessio and Zumino, Bruno",
    title = "{Duality, Entropy and ADM Mass in Supergravity}",
    reportNumber = "UCB-PTH-09-05, LBNL-1936E, CERN-PH-TH-2009-001, SU-ITP-09-04",
    doi = "10.1103/PhysRevD.79.125010",
    journal = "Phys. Rev. D",
    volume = "79",
    pages = "125010",
    year = "2009"
}

@article{Bars-Gauged,
    author = "Bars, I. and Deliduman, C. and Andreev, O.",
    title = "{Gauged duality, conformal symmetry and space-time with two times}",
    doi = "10.1103/PhysRevD.58.066004",
    journal = "Phys. Rev. D",
    volume = "58",
    pages = "066004",
    year = "1998"
}

@article{Bars:DualityH,
  author  = {Bars, I. and Rosner, J. L.},
  title   = {Duality Between Hydrogen Atom and Oscillator Systems via Hidden $SO(d,2)$ Symmetry and {2T}-physics},
  journal = {J. Phys. A},
  volume  = {53},
  number  = {23},
  year    = {2020},
  pages   = {234001},
  doi     = {10.1088/1751-8121/ab87ba},
}

@book{Bars-Terning,
  author    = {Bars, I. and Terning, J.},
  title     = {Extra Dimensions in Space and Time},
  publisher = {Springer},
  address   = {New York},
  year      = {2010}
}

@article{Bars-Araya,
    author = "Araya, Ignacio J. and Bars, Itzhak",
    title = "{Generalized dualities in one-time physics as holographic predictions from two-time physics}",
    doi = "10.1103/PhysRevD.89.066011",
    journal = "Phys. Rev. D",
    volume = "89",
    number = "6",
    pages = "066011",
    year = "2014"
}

@article{Bars10,
    author = "Bars, Itzhak and Chen, Shih-Hung",
    title = "{Geometry and Symmetry Structures in 2T Gravity}",
    reportNumber = "USC-08-HEP-B5",
    doi = "10.1103/PhysRevD.79.085021",
    journal = "Phys. Rev. D",
    volume = "79",
    pages = "085021",
    year = "2009"
}

@article{Bars:Gravity,
    author = "Bars, Itzhak",
    title = "{Two time physics with gravitational and gauge field backgrounds}",
    reportNumber = "CITUSC-00-011",
    doi = "10.1103/PhysRevD.62.085015",
    journal = "Phys. Rev. D",
    volume = "62",
    pages = "085015",
    year = "2000"
}

@article{Bars:Super,
    author = "Bars, I. and Deliduman, C. and Minic, D.",
    title = "{Supersymmetric two time physics}",
    reportNumber = "USC-98-HEP-B6",
    doi = "10.1103/PhysRevD.59.125004",
    journal = "Phys. Rev. D",
    volume = "59",
    pages = "125004",
    year = "1999"
}

@article{Bars:Strings,
    author = "Bars, I. and Deliduman, C. and Minic, D.",
    title = "{Strings, branes and two time physics}",
    reportNumber = "USC-99-HEP-B2",
    doi = "10.1016/S0370-2693(99)01127-2",
    journal = "Phys. Lett. B",
    volume = "466",
    pages = "135--143",
    year = "1999"
}

@article{Bars:GravityII,
    author = "Bars, Itzhak",
    title = "{Gravity in 2T-Physics}",
    reportNumber = "USC-08-HEP-B2",
    doi = "10.1103/PhysRevD.77.125027",
    journal = "Phys. Rev. D",
    volume = "77",
    pages = "125027",
    year = "2008"
}

@article{Bars:QFT,
    author = "Bars, Itzhak",
    title = "{Two time physics in field theory}",
    reportNumber = "CITUSC-00-014",
    doi = "10.1103/PhysRevD.62.046007",
    journal = "Phys. Rev. D",
    volume = "62",
    pages = "046007",
    year = "2000"
}

@article{Bars:SM,
    author = "Bars, Itzhak",
    editor = "Feng, Jonathan L.",
    title = "{The Standard Model as a 2T-physics Theory}",
    reportNumber = "USC-06-HEP-B3",
    doi = "10.1063/1.2735245",
    journal = "AIP Conf. Proc.",
    volume = "903",
    number = "1",
    pages = "550--555",
    year = "2007"
}

@article{small_orbits,
    author = "Borsten, L. and Duff, M. J. and Ferrara, S. and Marrani, A. and Rubens, W.",
    title = "{Small Orbits}",
    doi = "10.1103/PhysRevD.85.086002",
    journal = "Phys. Rev. D",
    volume = "85",
    pages = "086002",
    year = "2012"
}

@article{CFMZ1-D=5,
    author = "Cerchiai, Bianca L. and Ferrara, Sergio and Marrani, Alessio and Zumino, Bruno",
    title = "{Charge Orbits of Extremal Black Holes in Five Dimensional Supergravity}",
    reportNumber = "CERN-PH-TH-2010-130, SU-ITP-10-08, UCB-PTH-10-12",
    doi = "10.1103/PhysRevD.82.085010",
    journal = "Phys. Rev. D",
    volume = "82",
    pages = "085010",
    year = "2010"
}

@article{KM,
  author  = {Kamenshchik, A. and Muscolino, F.},
  title   = {Looking for {Carroll} Particles in the Two-Time Spacetime},
  doi = "https://doi.org/10.1103/PhysRevD.109.025005",
  journal = {Phys. Rev. D},
  volume  = {109},
  year    = {2024},
  pages   = {025005},
  note    = {Also Ukr. J. Phys. 69 (2024) no.7, 448}
}

@article{FigueroaGalileiCarrollBargmann,
    author = "Figueroa-O'Farrill, Jos{\'e}",
    title = "{Lie algebraic Carroll/Galilei duality}",
    reportNumber = "EMPG-22-20",
    doi = "10.1063/5.0132661",
    journal = "J. Math. Phys.",
    volume = "64",
    number = "1",
    pages = "013503",
    year = "2023"
}

@article{Figueroa:Fractons,
    author = "Figueroa-O'Farrill, Jos{\'e} and P{\'e}rez, Alfredo and Prohazka, Stefan",
    title = "{Carroll/fracton particles and their correspondence}",
    reportNumber = "EMPG-23-09",
    doi = "10.1007/JHEP06(2023)207",
    journal = "JHEP",
    volume = "06",
    pages = "207",
    year = "2023"
}

@article{Figueroa:Galilei,
    author = "Figueroa-O'Farrill, Jos{\'e} Miguel and Pekar, Simon and P{\'e}rez, Alfredo and Prohazka, Stefan",
    title = "{Galilei particles revisited}",
    doi = "10.21468/SciPostPhysLectNotes.93",
    journal = "SciPost Phys. Lect. Notes",
    volume = "93",
    pages = "1",
    year = "2025"
}

@article{Springer:1962,
  author  = {Springer, T. A.},
  title   = {Characterization of a class of cubic forms},
  journal = {Nederl. Akad. Wetensch. Proc. Ser.},
  volume  = {A24},
  year    = {1962},
  pages   = {259--265}
}

@article{McCrimmon:1969,
  author  = {McCrimmon, K.},
  title   = {The {Freudenthal-Springer-Tits} construction of exceptional {Jordan} algebras},
  doi = "https://doi.org/10.2307/1995337",
  journal = {Trans. Amer. Math. Soc.},
  volume  = {139},
  year    = {1969},
  pages   = {495--510}
}

@article{Baez:2001dm,
    author = "Baez, John C.",
    title = "{The Octonions}",
    doi = "10.1090/S0273-0979-01-00934-X",
    journal = "Bull. Am. Math. Soc.",
    volume = "39",
    pages = "145--205",
    year = "2002",
    note = "[Erratum: Bull.Am.Math.Soc. 42, 213 (2005)]"
}

@article{F-Gimon-K,
    author = "Ferrara, Sergio and Gimon, Eric G. and Kallosh, Renata",
    title = "{Magic supergravities, N= 8 and black hole composites}",
    reportNumber = "CERN-PH-TH-2006-116, SU-ITP-2006-19, LBNL-60487, UCLA-06-TEP-19",
    doi = "10.1103/PhysRevD.74.125018",
    journal = "Phys. Rev. D",
    volume = "74",
    pages = "125018",
    year = "2006"
}

@article{Wissanji,
    author = "Dasgupta, Keshav and Hussin, Veronique and Wissanji, Alisha",
    title = "{Quaternionic Kahler Manifolds, Constrained Instantons and the Magic Square. I.}",
    doi = "10.1016/j.nuclphysb.2007.09.026",
    journal = "Nucl. Phys. B",
    volume = "793",
    pages = "34--82",
    year = "2008"
}

@book{Schafer:1966,
  author    = {Schafer, R.},
  title     = {Introduction to Nonassociative Algebras},
  url = "https://www.gutenberg.org/files/25156/25156-pdf.pdf",
  publisher = {Academic Press Inc.},
  address   = {New York},
  year      = {1966}
}

@article{Brown:1969,
  author  = {Brown, R. B.},
  title   = {Groups of type $E_7$},
  url = "https://projecteuclid.org/journals/bulletin-of-the-american-mathematical-society/volume-55/issue-8/Inner-derivations-of-non-associative-algebras/bams/1183514040.pdf",
  journal = {J. Reine Angew. Math.},
  volume  = {236},
  year    = {1969},
  pages   = {79--102}
}

@article{Gunaydin:2000xr,
    author = "G{\"u}naydin, M. and Koepsell, K. and Nicolai, H.",
    title = "{Conformal and quasiconformal realizations of exceptional Lie groups}",
    reportNumber = "AEI-2000-043, CERN-TH-2000-230",
    doi = "10.1007/PL00005574",
    journal = "Commun. Math. Phys.",
    volume = "221",
    pages = "57--76",
    year = "2001"
}

@book{Koecher,
  author    = {Koecher, M.},
  title     = {An elementary approach to bounded symmetric domains},
  publisher = {Rice Univ. lectures, Houston},
  year      = {1969}
}

@book{Loos,
  author    = {Loos, O.},
  title     = {Bounded symmetric domains and {Jordan} pairs},
  publisher = {Lectures, Univ. California, Irvine},
  year      = {1977}
}

@article{Gun-Bars,
    author = "Bars, I. and G{\"u}naydin, M.",
    title = "{Dynamical Theory Of Subconstituents Based On Ternary Algebras}",
    doi = "10.1103/PhysRevD.22.1403",
    journal = "Phys. Rev. D",
    volume = "22",
    pages = "1403--1413",
    year = "1980"
}

@article{GST,
    author = "G{\"u}naydin, M. and Sierra, G. and Townsend, P. K.",
    title = "{Exceptional Supergravity Theories and the MAGIC Square}",
    reportNumber = "LPTENS 83/31",
    doi = "10.1016/0370-2693(83)90108-9",
    journal = "Phys. Lett. B",
    volume = "133",
    pages = "72--76",
    year = "1983"
}

@article{GST2,
    author = "G{\"u}naydin, M. and Sierra, G. and Townsend, P. K.",
    title = "{The Geometry of N=2 Maxwell-Einstein Supergravity and Jordan Algebras}",
    reportNumber = "LPTENS 83/32",
    doi = "10.1016/0550-3213(84)90142-1",
    journal = "Nucl. Phys. B",
    volume = "242",
    pages = "244--268",
    year = "1984"
}

@article{Squaring-Magic,
    author = "Cacciatori, Sergio L. and Cerchiai, Bianca L. and Marrani, Alessio",
    title = "{Squaring the Magic}",
    reportNumber = "CERN-PH-TH-2012-229",
    doi = "10.4310/ATMP.2015.v19.n5.a1",
    journal = "Adv. Theor. Math. Phys.",
    volume = "19",
    pages = "923--954",
    year = "2015"
}

@article{Gunaydin:1975mp,
  author  = {G{\"u}naydin, M.},
  title   = {Exceptional realizations of {Lorentz} group: Supersymmetries and leptons},
  doi = "https://doi.org/10.1007/BF02734524",
  journal = {Nuovo Cim. A},
  volume  = {29},
  year    = {1975},
  pages   = {467}
}

@incollection{Gunaydin:1989dq,
  author    = {G{\"u}naydin, M.},
  title     = {The exceptional superspace and the quadratic {Jordan} formulation of quantum mechanics},
  doi = "https://doi.org/10.1017/CBO9780511563980.010",
  booktitle = {Elementary particles and the universe: Essays in honor of {Murray Gell-Mann}},
  publisher = {Cambridge University Press},
  year      = {1989},
  pages     = {99--119}
}

@article{Gunaydin:1992zh,
    author = "G{\"u}naydin, Murat",
    title = "{Generalized conformal and superconformal group actions and Jordan algebras}",
    reportNumber = "IASSNS-HEP-92-86",
    doi = "10.1142/S0217732393001124",
    journal = "Mod. Phys. Lett. A",
    volume = "8",
    pages = "1407--1416",
    year = "1993"
}

@article{j,
    author = "Cremmer, E. and Julia, B.",
    editor = "Salam, A. and Sezgin, E.",
    title = "{The N=8 Supergravity Theory. 1. The Lagrangian}",
    reportNumber = "LPTENS 78/23",
    doi = "10.1016/0370-2693(78)90303-9",
    journal = "Phys. Lett. B",
    volume = "80",
    pages = "48",
    year = "1978"
}

@article{Gun-2,
    author = "G{\"u}naydin, Murat and Pavlyk, Oleksandr",
    title = "{Spectrum Generating Conformal and Quasiconformal U-Duality Groups, Supergravity and Spherical Vectors}",
    doi = "10.1007/JHEP04(2010)070",
    journal = "JHEP",
    volume = "04",
    pages = "070",
    year = "2010"
}

@article{matrix_norms,
    author = "Ferrara, Sergio and Marrani, Alessio",
    title = "{Matrix Norms, BPS Bounds and Marginal Stability in N=8 Supergravity}",
    reportNumber = "CERN-PH-TH-2010-204, SU-ITP-10-27",
    doi = "10.1007/JHEP12(2010)038",
    journal = "JHEP",
    volume = "12",
    pages = "038",
    year = "2010"
}

@article{Maclay:DynamicalSymmetries,
    author = "Maclay, G. Jordan",
    title = "{Dynamical Symmetries of the H Atom, One of the Most Important Tools Of Modern Physics: SO(4) to SO(4,2), Background, Theory, and Use in Calculating Radiative Shifts}",
    doi = "https://doi.org/10.3390/sym12081323",
    journal = "Symmetry",
    volume = "12",
    pages = "1323",
    year = "2020"
}

@article{LL,
    author = "L{\'e}vy-Leblond, Jean Marc",
    title = "{Une nouvelle limite non-relativiste du groupe de Poincar{\'e}}",
    journal = "Ann. Inst. H. Poincare Phys. Theor. A",
    volume = "3",
    number = "1",
    pages = "1--12",
    year = "1965"
}

@article{WI,
    author = "Inonu, E. and Wigner, Eugene P.",
    title = "{On the Contraction of groups and their represenations}",
    doi = "10.1073/pnas.39.6.510",
    journal = "Proc. Nat. Acad. Sci.",
    volume = "39",
    pages = "510--524",
    year = "1953"
}

@article{Duval0,
    author = "Duval, C. and Gibbons, G. W. and Horvathy, P. A.",
    title = "{Conformal Carroll groups}",
    doi = "10.1088/1751-8113/47/33/335204",
    journal = "J. Phys. A",
    volume = "47",
    number = "33",
    pages = "335204",
    year = "2014"
}

@article{Duval,
    author = "Duval, C. and Gibbons, G. W. and Horvathy, P. A. and Zhang, P. M.",
    title = "{Carroll versus Newton and Galilei: two dual non-Einsteinian concepts of time}",
    doi = "10.1088/0264-9381/31/8/085016",
    journal = "Class. Quant. Grav.",
    volume = "31",
    pages = "085016",
    year = "2014"
}

@article{Duval1,
    author = "Duval, C. and Gibbons, G. W. and Horvathy, P. A.",
    title = "{Conformal Carroll groups and BMS symmetry}",
    doi = "10.1088/0264-9381/31/9/092001",
    journal = "Class. Quant. Grav.",
    volume = "31",
    pages = "092001",
    year = "2014"
}

@article{Bergshoeff,
    author = "Bergshoeff, Eric and Gomis, Joaquim and Longhi, Giorgio",
    title = "{Dynamics of Carroll Particles}",
    reportNumber = "UG-2014-31, ICCUB-14-050",
    doi = "10.1088/0264-9381/31/20/205009",
    journal = "Class. Quant. Grav.",
    volume = "31",
    number = "20",
    pages = "205009",
    year = "2014"
}

@article{Ciambelli,
    author = "Ciambelli, Luca and Leigh, Robert G. and Marteau, Charles and Petropoulos, P. Marios",
    title = "{Carroll Structures, Null Geometry and Conformal Isometries}",
    reportNumber = "CPHT-RR025.052019, CPHT-RR010.022019",
    doi = "10.1103/PhysRevD.100.046010",
    journal = "Phys. Rev. D",
    volume = "100",
    number = "4",
    pages = "046010",
    year = "2019"
}

@article{Donnay,
    author = "Donnay, Laura and Marteau, Charles",
    title = "{Carrollian Physics at the Black Hole Horizon}",
    doi = "10.1088/1361-6382/ab2fd5",
    journal = "Class. Quant. Grav.",
    volume = "36",
    number = "16",
    pages = "165002",
    year = "2019"
}

@article{Henneaux,
    author = "Henneaux, Marc and Salgado-Rebolledo, Patricio",
    title = "{Carroll contractions of Lorentz-invariant theories}",
    doi = "10.1007/JHEP11(2021)180",
    journal = "JHEP",
    volume = "11",
    pages = "180",
    year = "2021"
}

@article{Boer,
    author = "de Boer, Jan and Hartong, Jelle and Obers, Niels A. and Sybesma, Watse and Vandoren, Stefan",
    title = "{Carroll Symmetry, Dark Energy and Inflation}",
    reportNumber = "NORDITA 2021-086",
    doi = "10.3389/fphy.2022.810405",
    journal = "Front. in Phys.",
    volume = "10",
    pages = "810405",
    year = "2022"
}

@article{Gomis,
    author = "Gomis, Joaquim and Kleinschmidt, Axel",
    title = "{Infinite-Dimensional Algebras as Extensions of Kinematic Algebras}",
    doi = "10.3389/fphy.2022.892812",
    journal = "Front. in Phys.",
    volume = "10",
    pages = "892812",
    year = "2022"
}

@article{Hansen,
    author = "Hansen, Dennis and Obers, Niels A. and Oling, Gerben and S{\o}gaard, Benjamin T.",
    title = "{Carroll Expansion of General Relativity}",
    reportNumber = "NORDITA 2021-156",
    doi = "10.21468/SciPostPhys.13.3.055",
    journal = "SciPost Phys.",
    volume = "13",
    number = "3",
    pages = "055",
    year = "2022"
}

@article{Campoleoni,
    author = "Campoleoni, Andrea and Henneaux, Marc and Pekar, Simon and P{\'e}rez, Alfredo and Salgado-Rebolledo, Patricio",
    title = "{Magnetic Carrollian gravity from the Carroll algebra}",
    doi = "10.1007/JHEP09(2022)127",
    journal = "JHEP",
    volume = "09",
    pages = "127",
    year = "2022"
}

@article{Boer1,
    author = "de Boer, Jan and Hartong, Jelle and Obers, Niels A. and Sybesma, Watse and Vandoren, Stefan",
    title = "{Carroll stories}",
    reportNumber = "NORDITA-2023-036",
    doi = "10.1007/JHEP09(2023)148",
    journal = "JHEP",
    volume = "09",
    pages = "148",
    year = "2023"
}

@article{LL1,
    author = "L{\'e}vy-Leblond, Jean Marc",
    title = "{On the unexpected fate of scientific ideas: An archeology of the Carroll group}",
    doi = "10.21468/SciPostPhysProc.14.006",
    journal = "SciPost Phys. Proc.",
    volume = "14",
    pages = "006",
    year = "2023"
}

@article{Bargmann:UnitaryReps,
    author = "Bargmann, V.",
    title = "{On Unitary ray representations of continuous groups}",
    doi = "10.2307/1969831",
    journal = "Annals Math.",
    volume = "59",
    pages = "1--46",
    year = "1954"
}

@article{SenGupta:Carroll,
    author = "Sen Gupta, N. D.",
    title = "{On an analogue of the Galilei group}",
    doi = "10.1007/BF02740871",
    journal = "Nuovo Cim. A",
    volume = "44",
    number = "2",
    pages = "512--517",
    year = "1966"
}

\end{document}